\documentclass{amsart}
\usepackage{amssymb,amsmath,amscd,amsthm}
\usepackage{amsfonts,epsfig,latexsym,graphicx,amssymb,pict2e}

\newtheorem{Lm}{Lemma}[section]
\newtheorem{Thm}[Lm]{Theorem}
\newtheorem{Prop}[Lm]{Proposition}

\newtheorem{Rem}[Lm]{Remark}
\newtheorem{Cor}[Lm]{Corollary}
\newtheorem{Def}[Lm]{Definition}
\newtheorem{problem}[Lm]{Problem}

\newcommand{\CalW}{{\mathcal{W}}}
\newcommand{\CalA}{{\mathcal{A}}}

\def\R{\mathbb{R}}

\def\Z{\mathbb{Z}}
\def\N{\mathbb{N}}
\def\Q{\mathbb{Q}}
\def\T{\mathbb{T}}

\def\bdef{\begin{Def}}
\def\endef{\end{Def}}
\def\bthm{\begin{Thm}}
\def\ethm{\end{Thm}}
\def\bprop{\begin{Prop}}
\def\enprop{\end{Prop}}
\def\blm{\begin{Lm}}
\def\elm{\end{Lm}}
\def\bcor{\begin{Cor}}
\def\ecor{\end{Cor}}
\def\brm{\begin{Rem}}
\def\erm{\end{Rem}}
\def\bfig{\begin{picture}}
\def\efig{\end{picture}}
\def\beq{\begin{eqnarray}}
\def\eneq{\end{eqnarray}}
\def\beal{\begin{aligned}}
\def\enal{\end{aligned}}

\title[Schr\"odinger Operators Arising in the Study of Quasicrystals]{Spectral Properties of Schr\"odinger Operators Arising in the Study of Quasicrystals}

\author{David Damanik}

\address{Department of Mathematics, Rice University, Houston, TX~77005, USA}

\email{damanik@rice.edu}

\thanks{D.\ D.\ was supported in part by a Simons Fellowship and NSF grants DMS--0800100 and DMS--1067988.}

\author{Mark Embree}

\address{Department of Computational and Applied Mathematics, Rice University, Houston, TX~77005, USA}

\email{embree@rice.edu}

\thanks{M.\ E.\ was supported in part by NSF grant DMS--CAREER--0449973.}

\author{Anton Gorodetski}

\address{Department of Mathematics, University of California, Irvine CA 92697, USA}

\email{asgor@math.uci.edu}

\thanks{A.\ G.\ was supported in part by NSF grants DMS--0901627 and IIS--1018433.}

\date{\today}

\begin{document}

\begin{abstract}
We survey results that have been obtained for self-adjoint operators, and especially Schr\"odinger operators, associated with mathematical models of quasicrystals. After presenting general results that hold in arbitrary dimensions, we focus our attention on the one-dimensional case, and in particular on several key examples. The most prominent of these is the Fibonacci Hamiltonian, for which much is known by now and to which an entire section is devoted here. Other examples that are discussed in detail are given by the more general class of Schr\"odinger operators with Sturmian potentials. We put some emphasis on the methods that have been introduced quite recently in the study of these operators, many of them coming from hyperbolic dynamics. We conclude with a multitude of numerical calculations that illustrate the validity of the known rigorous results and suggest conjectures for further exploration.
\end{abstract}

\maketitle

\tableofcontents

\section{Introduction}

The area of mathematical quasicrystals is fascinating due to its richness and very broad scope. One perspective that leads to a rich theory is the question of how well quantum wave packets can travel in a quasicrystalline medium. This, in turn, leads to a study of the time-dependent Schr\"odinger equation governed by a potential that reflects the aperiodic order of the environment to which the quantum state is exposed.

This survey paper describes the current ``state of the art'' of mathematical results concerning quantum transport in mathematical quasicrystal models. We will describe the (Schr\"odinger) operators that are typically considered in this context, as well as the spectral and quantum dynamical properties of these operators that are relevant to our basic motivating question. Few results hold without further assumptions. After presenting these general results, we will therefore specialize our class of operators in various ways. On the one hand, additional general results follow from passing to a one-dimensional setting, most notably some specific consequences of Kotani theory. On the other hand, there is much interest in certain central examples, such as the Fibonacci model, and more generally the Sturmian case or potentials generated by primitive substitutions. For these special cases, many more results are known, most of which will be described in detail here. Another purpose of this survey is to highlight new tools that have recently been introduced in the study of these models that have led to very fine quantitative results in the Fibonacci case and beyond.

The structure of the paper is as follows. In Section~\ref{s.2} we describe the operators that are typically considered in the case of general dimension. In Section~\ref{s.3} we present the results known to hold in this very general setting. Next, in Section~\ref{s.4}, we pass to the one-dimensional situation, where the operators are slightly redefined to conform with the bulk of the literature. Additional general results are described in this scenario. These range from Kotani theory, via proofs of zero-measure spectrum and hence absence of absolutely continuous spectrum based on Kotani theory, to proofs of absence of point spectrum and hence purely singular continuous spectrum based on Gordon's lemma. We also discuss how these general results can be applied to several classes of examples. Section~\ref{s.5} is devoted to the special case of the Fibonacci Hamiltonian. This is the most prominent one-dimensional quasicrystal model; in addition to the general results mentioned before, there are fine estimates of the local and global dimensions of the spectrum and the density of states measure, as well as the optimal H\"older exponent of the integrated density of states and the upper transport exponent. For this model the question about quantum transport behavior also has very satisfactory answers. All these results, along with comments on their proofs, are presented in this section. Then, in Section~\ref{s.6}, we discuss how the approach and the results may be extended from the Fibonacci case to the more general Sturmian case. We present and discuss numerical results in Section~\ref{s.7}. Finally, Section~\ref{s.8} lists and discusses a number of open problems that are suggested by the existing results.

\bigskip

Related matters have been surveyed earlier in \cite{B92, BG95, D00, D07a, S95}.

\section{Schr\"odinger Operators Arising in the Study of Quasicrystals}\label{s.2}

In this section we describe self-adjoint operators that have been studied in the context of quasicrystals. There is a clear distinction between the case of one space dimension and the case of higher space dimensions. While the geometry of the quasicrystal model plays an important role in higher dimensions, this is not the case for one-dimensional models. In the former case, this leads to a dependence of the underlying Hilbert space on the realization of the model. In particular, as one typically embeds any such realization in a family of realizations, this leads to some technical issues that need to be addressed mathematically. In the latter case, on the other hand, one usually works in a universal Hilbert space and the aperiodic order features are solely reflected by the potential of the operator. This allows one to invoke the standard theory of ergodic Schr\"odinger operators with a fixed Hilbert space. In this survey we will present the known results in the settings in which they have been obtained, which we now describe.

Quasicrystals are commonly modeled either by Delone sets in Euclidean space or by tilings of Euclidean space. In fact, any such Delone set or tiling is embedded in a Delone or tiling dynamical system, which is obtained by considering the set of translates and then taking the closure of this set with respect to a suitable topology. This orbit closure is called the hull of the initial Delone set or tiling. The dynamics are then given by the natural action of the Euclidean space on the hull through translations. For this dynamical system, one then identifies ergodic measures, and they are typically unique. While each element of the hull gives rise to a Schr\"odinger operator, it is the ergodic framework that allows one to prove statements that hold for almost all such operators, as opposed to results for a single such operator. Occasionally, it is then even possible to extend results that hold for almost all elements of the hull to all elements of the hull by approximation.

Delone sets and tilings are in some sense dual and hence equivalent to each other.\footnote{One can go back and forth between these two settings by decorations and the Voronoi construction.} For definiteness, we will consider a framework based on Delone sets. Let us fix a dimension $d \ge 1$. The Euclidean norm on $\R^d$ will be denoted by $|\cdot|$. We denote by $B(x,r)$ the closed ball in $\R^d$ that is centered at $x$ and has radius $r$.

\begin{Def}
A set $\Lambda \subset \R^d$ is called a Delone set if there are $r,R > 0$ such that
$$
B(x,r) \cap \Lambda = \{ x \} \quad \forall x \in \Lambda
$$
and
$$
B(x,R) \cap \Lambda \not= \emptyset \quad \kern15pt \forall x \in \R^d.
$$
\end{Def}

Thus for a Delone set we have a lower bound on the separation between points of the set, and an upper bound for the size of ``holes'' in the set. One also says that a Delone set is uniformly discrete and relatively dense. Since Delone sets are closed and we want to take orbit closures, let us now define the underlying topology on $\mathcal{F}(\R^d)$, the closed subsets of $\R^d$, following \cite{LS1}.

\begin{Def}
For $k \in \Z_+$ and $F,G \in \mathcal{F}(\R^d)$, we define
$$
d_k(F,G) := \inf \left( \{ \varepsilon > 0 : F \cap B(0,k) \subset U_\varepsilon(G) \text{ and } G \cap B(0,k) \subset U_\varepsilon(F) \} \cup \{ 1 \} \right),
$$
where $U_\varepsilon(\cdot)$ is an open neighborhood.
For $k \in \Z_+$, $\varepsilon > 0$, and $F \in \mathcal{F}(\R^d)$, we define
$$
U_{\varepsilon,k}(F) : = \{ G \in \mathcal{F}(\R^d) : d_k(F,G) < \varepsilon \}.
$$
The topology on $\mathcal{F}(\R^d)$ with neighborhood basis $\{ U_{\varepsilon,k}(F) \}$ will be called the natural topology and denoted by $\tau_\mathrm{nat}$.
\end{Def}

\begin{Prop}
{\rm (a)} Translations are continuous with respect to $\tau_\mathrm{nat}$. \\
{\rm (b)} $\mathcal{F}(\R^d)$ endowed with $\tau_\mathrm{nat}$ is compact. \\
{\rm (c)} $\tau_\mathrm{nat}$ is metrizable.
\end{Prop}

See \cite{LS1} for a metric that induces $\tau_\mathrm{nat}$.

\begin{Def}
Let $\Omega$ be a set of Delone sets and denote by $T$ the translation action of $\R^d$, that is, $T_t x = x + t$. The pair $(\Omega,T)$ is a \emph{Delone dynamical system} if $\Omega$ is invariant under $T$ and closed in the natural topology.
\end{Def}

\begin{Def}
A Delone dynamical system is said to be of \emph{finite local complexity} if for every radius $s > 0$, there is a uniform upper bound on the number of different patterns one can observe in $\omega$ intersected with a ball of radius $s$. Here, $\omega$ ranges over $\Omega$, the center of the ball ranges over $\R^d$. {\rm (}A pattern that appears in $\omega$ is any finite subset of $\omega$ modulo translations{\rm )}.
\end{Def}

\begin{Def}
Suppose $(\Omega,T)$ is a Delone dynamical system of finite local complexity. A family $\{ A_\omega \}_{\omega \in \Omega}$ of bounded operators $A_\omega : \ell^2(\omega) \to \ell^2(\omega)$ is said to have \emph{finite range} if there is $s > 0$ such that for all $\omega \in \Omega$, $A_\omega(x,y)$ only depends on the pattern of $\omega$ in the $s$-neighborhood of $x$ and $y$, and $A_\omega(x,y) = 0$ whenever $|x-y| \ge s$.
\end{Def}

The class of operators so defined encompasses all discrete operators that are usually considered as quantum Hamiltonians in the context of multi-dimensional quasicrystals. In one dimension, however, it is customary to realign points as a lattice (i.e., $\Z$) and to encode the geometry in the matrix elements of the operator. Even more specifically, one focuses on nearest neighbor interactions and hence obtains a tridiagonal matrix in the standard basis of $\ell^2(\Z)$. Much of the mathematical literature focuses on the discrete Schr\"odinger case, where the terms on the first off-diagonals are all equal to one and the quasicrystalline structure of the environment is reflected in the terms on the diagonal. That is, the family of discrete Schr\"odinger operators one then considers is of the form $\{ H_\omega \}_{\omega \in \Omega}$, where $\Omega$ is typically a subshift over a finite alphabet, and for $\omega \in \Omega$, $H_\omega$ acts on vectors from $\ell^2(\Z)$ as
$$
[H_\omega \psi](n) = \psi(n+1) + \psi(n-1) + V_\omega(n) \psi(n).
$$
Here the potential $V_\omega$ is given by $V_\omega(n) = f(T^n \omega)$, where $T : \Omega \to \Omega$ is the standard shift transformation and $f$ is at least continuous, usually locally constant, and often just depends on a single entry of the sequence $\omega$.

\section{General Results in Arbitrary Dimension}\label{s.3}

\subsection{Spectrum and Spectral Types}

One of the fundamental results for the families of operators introduced above is the almost sure constancy of the spectrum and the spectral type with respect to an ergodic measure $\mu$ associated with $(\Omega,T)$. This follows from the covariance condition
$$
A_{\omega + t} = U_t A_\omega U_t^*, \quad \omega \in \Omega, \; t \in \R^d,
$$
where $U_t : \ell^2(\omega) \to \ell^2(\omega + t)$ is the unitary operator induced by translation by $t$, along with the definition of ergodicity applied to traces of spectral projections in the usual way. This establishes the following result; compare \cite{LS1}.

\begin{Thm} \label{thm:as_spec}
Suppose $(\Omega,T)$ is a Delone dynamical system of finite local complexity and $\mu$ is an ergodic measure. Let $\{ A_\omega \}_{\omega \in \Omega}$ be a family of bounded self-adjoint operators of finite range. Then there exist $\Sigma, \Sigma_\mathrm{pp}, \Sigma_\mathrm{sc}, \Sigma_\mathrm{ac}$ and a subset $\Omega_0 \subseteq \Omega$ of full $\mu$-measure such that for every $\omega \in \Omega_0$, we have $\sigma(A_\omega) = \Sigma$, $\sigma_\mathrm{pp}(A_\omega) = \Sigma_\mathrm{pp}$, $\sigma_\mathrm{sc}(A_\omega) = \Sigma_\mathrm{sc}$, and $\sigma_\mathrm{ac}(A_\omega) = \Sigma_\mathrm{ac}$.
\end{Thm}

\subsection{Existence of the Integrated Density of States}\label{ss.IDS}

Suppose $\{ A_\omega \}_{\omega \in \Omega}$ is a family of bounded self-adjoint operators of finite range. For $Q \subset \R^d$ bounded, the restriction $A_\omega|_Q$ defined on $\ell^2(Q \cap \omega)$ has finite rank. Therefore,
$$
n(A_\omega,Q)(E) := \# \{ \text{eigenvalues of } A_\omega|Q \text{ that are } \le E\}
$$
is finite and $E \mapsto \frac{1}{|Q|} n(A_\omega,Q)(E)$ is the distribution function of the measure $\rho^{A_\omega}_Q$ defined by
$$
\int \varphi \, d\rho^{A_\omega}_Q = \frac{1}{|Q|} \mathrm{Tr} (\varphi(A_\omega|Q)), \quad \varphi \in C_b(\R).
$$

For $s > 0$ and $Q \subseteq \R^d$, denote by $\partial_s Q$ the set of points in $\R^d$ whose distance from the boundary of $Q$ is bounded by $s$. A sequence $\{ Q_k \}$ of bounded subsets of $\R^d$ is called a van Hove sequence if $\frac{\mathrm{vol}(\partial_s Q_k)}{\mathrm{vol}(Q_k)} \to 0$ as $k \to \infty$ for every $s > 0$.

The following result was shown in \cite{LS2} (compare also the earlier papers \cite{H93, H95}).

\begin{Thm} \label{thm:idsconv}
Suppose $(\Omega,T)$ is a strictly ergodic\footnote{Strict ergodicity means that all orbits are dense and that there is a unique invariant Borel probability measure.} Delone dynamical system of finite local complexity. Let $\{ A_\omega \}_{\omega \in \Omega}$ be a family of bounded self-adjoint operators of finite range and let $\{Q_k\}$ be a van Hove sequence. Then, as $k \to \infty$, the distributions of $E \mapsto \rho_{Q_k}^{A_\omega} ((-\infty,E])$ converge to the distribution $E \mapsto \rho^A ((-\infty,E])$ with respect to $\|\cdot\|_\infty$, and the convergence is uniform in $\omega \in \Omega$.
\end{Thm}

In fact, the limiting distribution can be given in closed form; see \cite{LS2}. It is called the integrated density of states, and the associated measure is called the density of states measure. The remarkable feature of this result is the strength of the convergence, in that the distribution functions converge uniformly in the $\|\cdot\|_\infty$ topology. This is of particular interest in cases when the limiting distribution function has jumps. The next subsection shows that the latter phenomenon may actually happen.

\subsection{Locally Supported Eigenfunctions and Discontinuities of the IDS}

In dimensions strictly greater than one, the local structure of a Delone set may be chosen such that suitable finite range operators have finitely supported eigenfunctions at a suitable energy. If these local configurations occur sufficiently regularly, it follows that the energy in question will be a point of discontinuity of the integrated density of states. This observation may now be supplemented in two ways. On the one hand, a given Delone dynamical system may be transformed into one that is equivalent to the original one in the sense of mutual local derivability, which does have the required local configurations. On the other hand, any discontinuity of the integrated density of states must arise in this way, that is, through the regular occurrence of finitely supported eigenfunctions at the energy in question. These issues were discussed in the paper \cite{KLS} (see that paper for the definition of mutual local derivability). Let us state the results from that paper precisely.

\begin{Thm} \label{thm:kls1}
Suppose $(\Omega,T)$ is a strictly ergodic Delone dynamical system of finite local complexity. Let $\{ A_\omega \}_{\omega \in \Omega}$ be a family of bounded self-adjoint operators of finite range. Then there exist $(\Omega',T)$ and $\{ A_\omega' \}_{\omega \in \Omega'}$ such that  $(\Omega,T)$ and  $(\Omega',T)$ are mutually locally derivable and $A_\omega'$ has locally supported eigenfunctions with the same eigenvalue for every $\omega \in \Omega'$. Moreover, $A_\omega'$ can be chosen to be the nearest neighbor Laplacian of a suitable graph.
\end{Thm}

\begin{Thm} \label{thm:kls2}
Suppose $(\Omega,T)$ is a strictly ergodic Delone dynamical system of finite local complexity. Let $\{ A_\omega \}_{\omega \in \Omega}$ be a family of bounded self-adjoint operators of finite range. Then $E \in \R$ is a point of discontinuity of $\rho^A$ if and only if there exists a locally supported eigenfunction of $A_\omega$ for $E$ for one {\rm (}\kern-1pt equivalently, all\kern1.5pt{\rm )} $\omega \in \Omega$.
\end{Thm}

\section{General Results in One Dimension}\label{s.4}

Starting with this section, we will focus on the case of one space dimension.
Far more rigorous results are known for this special case than for the general case.
In particular, much is known about the structure of the spectrum as a set, as well as the type of the spectral measures.

As mentioned above, in the one dimensional setting one typically passes to a somewhat different choice of the model.
Thus, for definiteness, we will restrict our attention in much of the remainder of this paper to the following scenario. We consider Schr\"odinger operators in $\ell^2(\Z)$,
\begin{equation}\label{e.operom}
[H_\omega \psi](n) = \psi(n+1) + \psi(n-1) + V_\omega(n) \psi(n),
\end{equation}
where the potentials are of the form $V_\omega(n) = f(T^n \omega)$, with $\omega$ in a compact metric space $\Omega$, a homeomorphism $T : \Omega \to \Omega$, and $f \in C(\Omega,\R)$. We also fix an ergodic measure $\mu$.

Notice that the Hilbert space in which $H_\omega$ acts is now $\omega$-independent, and the aperiodic order features of the medium that is being modeled are completely subsumed in the potential $V_\omega$ of the operator $H_\omega$. In this we follow the standard convention, for this class of operators has been commonly studied.  One could consider operators that are formally more akin to the operators considered above in general dimensions. However, this would not lead to any significant mathematical difference. Loosely speaking, the aperiodically-ordered Delone set in $\R$ is just being reconfigured as $\Z$, and the local properties of the operator that depend on the pattern near a point in the general setting affect the value of the potential at the point in question accordingly in our present setting.

In fact, this scenario is more general than considered in the previous section. The better analog would be the case where $(T \omega)_n = \omega_{n+1}$ is the shift on $\mathcal{A}^\Z$ for some finite set $\mathcal{A}$, $\Omega \subseteq \mathcal{A}^\Z$ is $T$-invariant and closed, and $f : \Omega \to \R$ is locally constant, that is, it only depends on the values of $\omega_n$ for some finite set of $n$ values. However, some of the results below hold in the more general setting, and we will impose further restrictions when they are needed.

Consequently, Theorems~\ref{thm:as_spec} and~\ref{thm:idsconv} now take the following form.

\begin{Thm}\label{t.4.1}
There are sets $\Sigma$, $\Sigma_\mathrm{ac}$, $\Sigma_\mathrm{sc}$, and $\Sigma_\mathrm{pp}$ such that for $\mu$-almost every $\omega \in \Omega$, we have
$\sigma(H_\omega) = \Sigma$, $\sigma_\mathrm{ac}(H_\omega) = \Sigma_\mathrm{ac}$, $\sigma_\mathrm{sc}(H_\omega) = \Sigma_\mathrm{sc}$, and $\sigma_\mathrm{pp}(H_\omega) = \Sigma_\mathrm{pp}$.
\end{Thm}

\begin{Thm}\label{t.4.2}
The measures $\int \varphi \, dN_k^\omega = \frac1k \mathrm{Tr} (\varphi(H_\omega|_{[1,k]}))$ converge weakly to the measure $\int \varphi \, dN = \int \langle \delta_0 , \varphi(H_\omega) \delta_0 \rangle \, d\mu(\omega)$ as $k \to \infty$.
\end{Thm}

The second result uses a weaker notion of convergence than in Theorem~\ref{thm:idsconv},
the price we have to pay for casting this problem in the more general setting.
However, this is fine after all, due to the following result.

\begin{Thm}\label{t.4.3}
The measure $dN$ is continuous.
\end{Thm}

Moreover, we have:

\begin{Thm}\label{t.4.4}
The topological support of the measure $dN$ is equal to $\Sigma$.
\end{Thm}

Theorems~\ref{t.4.1}--\ref{t.4.4} are standard results from the theory of ergodic Schr\"odinger operators on $\ell^2(\Z)$; compare \cite{CL}.

\subsection{Spectrum and the Absence of Uniform Hyperbolicity}

Let us consider solutions to the difference equation
\begin{equation}\label{e.eveom}
u(n+1) + u(n-1) + V_\omega(n) u(n) = E u(n), \quad n \in \Z,
\end{equation}
for some energy $E \in \R$. Clearly, $u$ solves \eqref{e.eveom} if and only if
\begin{equation}\label{e.ostmom}
\begin{pmatrix} u(n+1) \\ u(n) \end{pmatrix} = \begin{pmatrix} E - V_\omega(n) & -1 \\ 1 & 0 \end{pmatrix} \begin{pmatrix} u(n) \\ u(n-1) \end{pmatrix}, \quad n \in \Z.
\end{equation}
Since $V_\omega(n) = f(T^n \omega)$, one naturally defines
$$
A_E(\omega) = \begin{pmatrix} E - f(\omega) & -1 \\ 1 & 0 \end{pmatrix},
$$
so that \eqref{e.ostmom} implies
$$
\begin{pmatrix} u(n+1) \\ u(n) \end{pmatrix} = A_E(T^n \omega) \times \cdots \times A_E(T \omega) \begin{pmatrix} u(1) \\ u(0) \end{pmatrix}
$$
for $n \ge 1$ and solutions $u$ to \eqref{e.eveom}. We set $M_{E,\omega}(n) = A_E(T^n \omega) \times \cdots \times A_E(T \omega)$.

\begin{Def}
We let
$$
\mathcal{UH} = \{ E \in \R : \exists \, c>1 \text{ such that } \forall \omega \in \Omega, \; n \ge 1 \text{ we have } \|M_{E,\omega}(n)\| \ge c^n \}.
$$
\end{Def}

Johnson showed in \cite{J86} that the set $\mathcal{UH}$ is equal to the resolvent set of $H_\omega$ for any $\omega$ that has a dense $T$-orbit. As a particular consequence, we may state the following:

\begin{Thm}\label{t.specuh}
Suppose $T$ is minimal. Then for every $\omega \in \Omega$, we have $\sigma(H_\omega) = \R \setminus \mathcal{UH}$. In particular, for any ergodic measure $\mu$, we have $\Sigma = \R \setminus \mathcal{UH}$.
\end{Thm}

\subsection{Kotani Theory}

In the previous subsection, we saw that the partition $\R = \mathcal{UH} \sqcup \R \setminus \mathcal{UH}$ yields the partition of the energy axis into resolvent set and spectrum. Let us partition the energy axis even further by introducing the Lyapunov exponent
$$
L_\mu(E) = \lim_{n \to \infty} \frac1n \int \log \|M_{E,\omega}(n)\| \; d\mu(\omega).
$$
The existence of the limit follows quickly by subadditivity. Moreover, due to the ergodicity of $\mu$ and Kingman's Subadditive Ergodic Theorem, we have
$$
L_\mu(E) = \lim_{n \to \infty} \frac1n \log \|M_{E,\omega}(n)\| \quad \mbox{ for $\mu$-a.e.\ $\omega \in \Omega$}.
$$
Obviously, we have $L_\mu(E) > 0$ for every $E \in \mathcal{UH}$. We let
$$
\mathcal{Z}_\mu = \{ E : L_\mu(E) = 0 \}
$$
and
$$
\mathcal{NUH}_\mu = \{ E \in \R : L_\mu(E) > 0 \} \setminus \mathcal{UH},
$$
so that our final partition is $\R = \mathcal{UH} \sqcup \mathcal{NUH}_\mu \sqcup \mathcal{Z}_\mu$. Notice that in these definitions, the dependence of the partition of $\R \setminus \mathcal{UH}$ into $\mathcal{NUH}_\mu \sqcup \mathcal{Z}_\mu$ on the choice of the ergodic measure $\mu$ is emphasized through the subscript for the $\mu$-dependent sets. It is customary to drop this explicit subscript from $L$, $ \mathcal{Z}$, and $\mathcal{NUH}$, and we will henceforth do so as well.

Recall that the essential closure of a set $S \subseteq \R$ is given by
$$
\overline{S}^\mathrm{ess} = \{ E \in \R : \forall \, \varepsilon > 0 \text{ we have } \mathrm{Leb}(S \cap (E- \varepsilon, E + \varepsilon)) > 0 \}.
$$
Then, the following result is the celebrated Ishii-Pastur-Kotani theorem; see, for example, \cite{D07b, I73, K84, P80, S83}.
\begin{Thm}
We have $\Sigma_\mathrm{ac} = \overline{\mathcal{Z}}^\mathrm{ess}$.
\end{Thm}

Since the essential closure of a set $S$ is empty if and only if $S$ has zero Lebesgue measure, this result yields a characterization of the almost sure purely singular spectrum. In fact, this is the typical situation for our models of interest \cite{K89}.

\begin{Thm}\label{t.kotanifv}
Suppose the range of $f : \Omega \to \R$ is finite and $\mathrm{Leb}(\mathcal{Z}) > 0$. Denote the push-forward of $\mu$ under $\Omega \ni \omega \mapsto V_\omega \in (\mathrm{Ran} \, f)^\Z$ by $\mu^*$. Then, $\mathrm{supp} \, \mu^*$ is finite.
\end{Thm}

Since $\mathrm{supp} \, \mu$ is $T$-invariant, this means that every element of $\mathrm{supp} \, \mu^*$ is periodic. In other words, if the potentials are ergodic, aperiodic and take finitely many values, then $\mathcal{Z}$ has zero Lebesgue measure and the almost sure absolutely continuous spectrum is empty.

\subsection{Zero-Measure Spectrum}

The realization that $\mathrm{Leb}(\mathcal{Z}) = 0$ for potentials that are ergodic, aperiodic and take finitely many values is at the heart of proofs of zero-measure spectrum in these cases. Whenever Theorem~\ref{t.kotanifv} applies, it suffices to show that the spectrum is contained in (and hence coincides with) $\mathcal{Z}$ in order to establish zero-measure spectrum. There are two approaches that establish this identity. One shows that the Lyapunov exponent must vanish for every energy in the spectrum. The other is based on Theorem~\ref{t.specuh}, which says that $\Sigma = \mathcal{Z} \cup \mathcal{NUH}$. Hence, it suffices to prove that if $E$ is such that the Lyapunov exponent is positive, the convergence of $\frac1n \log \|M_{E,\omega}(n)\|$ to $L(E)$ must be uniform. The second approach can be made to work in greater generality, whereas the first approach often gives additional information that has other interesting consequences.

We will discuss how the first approach is implemented when we discuss Sturmian models and, specifically, the Fibonacci Hamiltonian in later sections. Here we therefore discuss in some detail how the second approach works. We first state the main result, then explain how it is a natural consequence of the line of reasoning employed. We emphasize that this is a result that holds in the symbolic setting. That is, for a finite set $\CalA$, called the \textit{alphabet}, we consider the shift transformation $T : \CalA^\Z \to \CalA^\Z$ given by $(T \omega)(n) = \omega(n+1)$ and closed, $T$-invariant subsets $\Omega$ of $\CalA^\Z$, which are called \textit{subshifts} (\textit{over} $\CalA$). We denote by $\mathcal{W}_\Omega$ the set of all finite words over the alphabet $\CalA$ that occur in elements of $\Omega$, and we write $\mathcal{W}_\Omega(n)$ for those elements of $\mathcal{W}_\Omega$ that have length $n$. A function on $\Omega$ is called \textit{locally constant} if $f(\omega)$ only depends on finitely many entries. More formally, there exists $k \in \Z_+$ such that $f(\omega) = f(\omega')$ for all $\omega, \omega' \in \Omega$ with $\omega_{-k} \ldots \omega_k = \omega_{-k}' \ldots \omega_k'$. Note that $f$ is locally constant if and only if it is continuous and takes only finitely many values.

As usual, a subshift $\Omega$ is called \textit{minimal} if the topological dynamical system $(\Omega,T)$ is minimal (i.e., the $T$-orbit of every $\omega \in \Omega$ is dense in $\Omega$), it is called \textit{uniquely ergodic} if $(\Omega,T)$ has a unique invariant Borel probability measure, and it is called \textit{strictly ergodic} if it is both minimal and uniquely ergodic. In the uniquely ergodic case, the unique invariant Borel probability measure must necessarily be ergodic.

\begin{Def}
Let $\Omega$ be a strictly ergodic subshift with unique $T$-invariant measure $\mu$. It satisfies the Boshernitzan condition {\rm (B)} if
$$
\limsup_{n \to \infty} \left( \min_{w \in \mathcal{W}_\Omega(n)} n \cdot \mu \left( [w] \right) \right) > 0.
$$
\end{Def}

Here is the result from \cite{DL06a} showing that (B) is a sufficient condition for zero-measure spectrum:

\begin{Thm}\label{t.boshthmdl}
Suppose $\Omega$ is a strictly ergodic subshift that satisfies condition {\rm (B)} and $f : \Omega \to \R$ is locally constant. Then the convergence of $\frac1n \log \|M_{E,\omega}(n)\|$ to $L(E)$ is uniform for every $E \in \R$. In particular, $\mathcal{NUH}$ is empty and $\Sigma = \mathcal{Z}$. Thus, if $\Omega$ and $f$ are such that the $V_\omega$ are aperiodic, then $\mathrm{Leb}(\Sigma) = 0$. In the latter case, $\Sigma_\mathrm{ac} = \emptyset$.
\end{Thm}

The proof of Theorem~\ref{t.boshthmdl} is more easily understood if one imposes a stronger assumption. Indeed, let us assume for the time being that $f(\omega)$ depends only on $\omega_0$ and $\min_{w \in \mathcal{W}_\Omega(n)} n \cdot \mu \left( [w] \right)$ is uniformly large for all $n$, not merely for a subsequence. That is, suppose there is $\delta > 0$ such that $|w| \cdot \mu \left( [w] \right) \ge \delta$ for every $w \in \mathcal{W}_\Omega$.

By our stronger assumption on $f$, we may view $\log \|M_{E,\omega}(n)\|$ as a quantity associated with the word $w = \omega_1 \ldots \omega_n \in \mathcal{W}_\Omega(n)$ that we will denote by $F(w)$. If we do so, then the goal is to prove that $|w|^{-1} F(w)$ converges uniformly as $|w| \to \infty$ and each $w$ belongs to $\mathcal{W}_\Omega$. It is a known consequence of unique ergodicity of $(\Omega,T)$ that the convergence in
$$
F^+ := \limsup_{w \in \mathcal{W}_\Omega, \, |w| \to \infty} |w|^{-1} F(w)
$$
is uniform. By the uniform upper bound and the frequent occurrence of any word due to the strengthened assumption, one can derive a similar uniform result for the $\liminf$, and hence establish uniform convergence. If one only has condition (B), then a similar way of reasoning can be carried out for the sequence of length scales from (B). In this way, one can establish uniform convergence along this sequence. To interpolate, one can employ the Avalanche Principle from \cite{GS01}.

The paper \cite{DL06b} contains numerous applications of Theorem~\ref{t.boshthmdl} to specific classes of subshifts, some of which will be described in Subsection~\ref{ss.examples} below.

When \cite{DL06a,DL06b} appeared, all known zero-measure results for Schr\"odinger operators defined by strictly ergodic subshifts (see, e.g., \cite{B, BBG91, BIST89, BG93, L02a, L02b, LTWW, s5}) were covered by this approach, that is, they all held for subshifts that satisfy condition (B). Recently, Liu and Qu constructed examples of strictly ergodic subshifts that do not satisfy (B) but for which the associated Schr\"odinger operators do have zero-measure spectrum (and in fact the convergence of $\frac1n \log \|M_{E,\omega}(n)\|$ to $L(E)$ is uniform for every $E \in \R$) \cite{LQ11}.

Naturally, once one knows that the spectrum has zero Lebesgue measure, one would like to determine its fractal (e.g., Hausdorff, lower and upper box counting) dimensions, as well as similar quantities such as thickness and denseness. These more delicate questions have been studied for a rather small number of examples, which will be discussed in subsequent sections. Zero-measure spectrum, on the other hand, is known in much greater generality, and this is the topic of the next subsection.

\subsection{Examples}\label{ss.examples}

In this subsection, we present several classes of popular examples of potentials that are ergodic, aperiodic, and take finitely many values (so that Kotani's central result applies) and discuss the validity of condition (B) (so that the associated Schr\"odinger operators have zero-measure spectrum). For more details, we refer the reader to \cite{DL06b}.

\subsubsection{Linearly Recurrent Subshifts and Subshifts Generated by Primitive Substitutions}\label{ss.subst}

A subshift $\Omega$ over $\CalA$ is called \textit{linearly recurrent} (or \textit{linearly repetitive}) if there exists a constant $K$ such that if $v,w \in \CalW (\Omega)$ with $|w| \ge K |v|$, then $v$ is a subword of $w$. Clearly, every linearly recurrent subshift $\Omega$ satisfies (B). A popular way to generate linearly recurrent subshifts is via primitive substitutions. A substitution $S : \CalA \to \CalA^*$ is called \textit{primitive} if there exists $k \in \N$ such that for every $a,b \in \CalA$, $S^k(a)$ contains $b$. Such a substitution generates a subshift $\Omega$ as follows. It is easy to see that there are $m \in \N$ and $a \in \CalA$ such that $S^m(a)$ begins with $a$. If we iterate $S^m$ on the symbol $a$, we obtain a one-sided infinite limit, $u$, called a \textit{substitution sequence}. $\Omega$ then consists of all two-sided sequences for which all subwords are also subwords of $u$. One can verify that this construction is in fact independent of the choice of $u$, and hence $\Omega$ is uniquely determined by $S$. Prominent examples\footnote{These examples appear explicitly in many papers in the physics literature on Schr\"odinger operators generated by primitive substitution; compare \cite{AS, BG, Be, BG95, GA, Ho, KST, L89}.} are given by
$$
\begin{array}{|l|l|}
\hline
a \mapsto ab, \; b \mapsto a & \text{ Fibonacci}\\
\hline
a \mapsto ab, \; b \mapsto ba & \text{ Thue-Morse}\\
\hline
a \mapsto ab, \; b \mapsto aa & \text{ Period Doubling}\\
\hline a \mapsto ab, \; b \mapsto ac, \; c \mapsto db, \; d
\mapsto dc & \text{ Rudin-Shapiro}\\
\hline
\end{array}
$$

The following was shown in \cite{DHS} (and independently in \cite{DZ00}):

\begin{Prop}\label{psimplieslr}
If the subshift $\Omega$ is generated by a primitive substitution, then it is linearly recurrent and hence satisfies condition {\rm (B)}.
\end{Prop}

It may happen that a non-primitive substitution generates a linearly recurrent subshift. An example is given by $a \mapsto aaba$, $b \mapsto b$. In fact, the class of linearly recurrent subshifts generated by substitutions was characterized in \cite{DL06c}.\footnote{See also \cite{dOL00, dOL02, LdO03} for results for Schr\"odinger operators arising from a specific class of non-primitive substitutions.} In particular, it turns out that a subshift generated by a substitution is linearly recurrent if and only if it is minimal.

\subsubsection{Sturmian and Quasi-Sturmian Subshifts}

Consider a minimal subshift $\Omega$ over $\CalA$. The (factor) complexity function $p : \Z_+ \to \Z_+$ is defined by $p(n) = \# \CalW_\Omega(n)$. Hedlund and Morse showed in \cite{HM} that $\Omega$ is aperiodic if and only if $p(n) \ge n+1$ for every $n \in \Z_+$. Aperiodic minimal subshifts of minimal complexity, $p(n) = n+1$ for every $n \in \N$, exist and they are called Sturmian. If the complexity function satisfies $p(n) = n + k$ for $n \ge n_0$, $k,n_0 \in \N$, the subshift is called quasi-Sturmian. It is known that quasi-Sturmian subshifts are exactly those subshifts that are a morphic image of a Sturmian subshift; compare \cite{C,C2,P}.

There are a large number of equivalent characterizations of Sturmian subshifts; compare \cite{B96}. We are mainly interested in their geometric description in terms of an irrational rotation. Let $\alpha \in (0,1)$ be irrational and consider the rotation by $\alpha$ on the circle, $R_\alpha : [0,1) \to [0,1), \;\; R_\alpha \theta = \{\theta + \alpha\}$, where $\{x\}$ denotes the fractional part of $x$, $\{x\} = x \mod 1$. The coding of the rotation $R_\alpha$ according to a partition of the circle into two half-open intervals of length $\alpha$ and $1-\alpha$, respectively, is given by the sequences $v_n(\alpha,\theta) = \chi_{[0,\alpha)}(R_\alpha^n \theta)$. We obtain a subshift
$$
\Omega_\alpha = \overline{ \{ v(\alpha,\theta) : \theta \in [0,1) \} } = \{ v(\alpha,\theta) : \theta \in [0,1) \} \cup \{ \tilde{v}^{(k)}(\alpha) : k \in \Z \} \subset\{0,1\}^{\Z},
$$
which can be shown to be Sturmian. Here, $\tilde{v}^{(k)}_n(\alpha) = \chi_{(0,\alpha]}(R_\alpha^{n+k} 0)$. Conversely, every Sturmian subshift is essentially of this form, that is, if $\Omega$ is minimal and has complexity function $p(n) = n+1$, then, up to a one-to-one morphism, $\Omega = \Omega_\alpha$ for some irrational $\alpha \in (0,1)$.

Using this description and some classical results in diophantine approximation, the following result was shown in \cite{DL06b}.

\begin{Thm}\label{sturmbosh}
Every Sturmian subshift obeys the Boshernitzan condition {\rm (B)}.
\end{Thm}

Moreover, establishing stability of (B) under morphic images, one obtains the following consequence.

\begin{Cor}\label{quasisturmbosh}
Every quasi-Sturmian subshift obeys {\rm (B)}.
\end{Cor}

\subsubsection{Circle Maps}\label{Circle}

Let $\alpha \in (0,1)$ be irrational and $\beta \in (0,1)$ arbitrary. The coding of the rotation $R_\alpha$ according to a
partition into two half-open intervals of length $\beta$ and $1-\beta$, respectively, is given by the sequences $v_n(\alpha,\beta,\theta) = \chi_{[0,\beta)}(R_\alpha^n \theta)$. We obtain a subshift
$$
\Omega_{\alpha,\beta} = \overline{ \{ v(\alpha,\beta,\theta) : \theta \in [0,1) \} } \subset\{0,1\}^{\Z}.
$$
Subshifts generated this way are usually called circle map subshifts or subshifts generated by the coding of a rotation.

The paper \cite{DL06b} established the following results for circle map subshifts in connection with property (B):

\begin{Thm}\label{cmpos}
Let $\alpha \in (0,1)$ be irrational. Then the subshift $\Omega_{\alpha,\beta}$ satisfies {\rm (B)} for Lebesgue almost every $\beta \in (0,1)$.
\end{Thm}

\begin{Thm}
Let $\alpha \in (0,1)$ be irrational with bounded continued fraction coefficients, that is, $a_n \le C$. Then $\Omega_{\alpha,\beta}$ satisfies {\rm (B)} for every $\beta \in (0,1)$.
\end{Thm}

\begin{Thm}
Let $\alpha \in (0,1)$ be irrational with unbounded continued fraction coefficients. Then there exists $\beta \in (0,1)$ such that $\Omega_{\alpha,\beta}$ does not satisfy {\rm (B)}.
\end{Thm}

\subsubsection{Interval Exchange Transformations}\label{IET}

Subshifts generated by interval exchange transformations (IETs) are natural generalizations of Sturmian subshifts. IETs are defined as follows. Given a probability vector $\lambda = (\lambda_1,\ldots,\lambda_m)$ with $\lambda_i > 0$ for $1 \le i \le m$, we let $\mu_0 = 0$, $\mu_i = \sum_{j = 1}^i \lambda_j$,
and $I_i = [\mu_{i-1},\mu_i)$. Let $\tau$ be a permutation of $\CalA_m = \{1,\ldots,m\}$, that is, $\tau \in S_m$, the symmetric group. Then $\lambda^\tau = (\lambda_{\tau^{-1}(1)}, \ldots, \lambda_{\tau^{-1}(m)})$ is also a probability vector, and we can form the corresponding $\mu_i^\tau$ and $I_i^\tau$. Denote the unit interval $[0,1)$ by $I$. The $(\lambda,\tau)$ interval exchange transformation is then defined by
$$
T : I \to I, \; \; T(x) = x - \mu_{i-1} + \mu_{\tau(i) - 1}^\tau \text{ for } x \in I_i, \; 1 \le i \le m.
$$
It exchanges the intervals $I_i$ according to the permutation $\tau$.

The transformation $T$ is invertible, and its inverse is given by the $(\lambda^\tau,\tau^{-1})$ interval exchange transformation.

The symbolic coding of $x \in I$ is $\omega_n(x) = i$ if $T^n(x) \in I_i$. This induces a subshift over the alphabet $\CalA_m$: $\Omega_{\lambda,\tau} = \overline{ \{ \omega(x) : x \in I \} }$.

Sturmian subshifts correspond to the case of two intervals, as a first return map construction shows.

Keane \cite{Ke1} proved that if the orbits of the discontinuities $\mu_i$ of $T$ are all infinite and pairwise distinct, then $T$ is minimal. In this case, the coding is one-to-one and the subshift is minimal and aperiodic.  This holds in particular if $\tau$ is irreducible and $\lambda$ is irrational. Here, $\tau$ is called irreducible if $\tau(\{1,\ldots,k\}) \not= \{1,\ldots,k\}$ for every $k < m$ and $\lambda$ is called irrational if the $\lambda_i$ are rationally independent.

Regarding property (B), Boshernitzan has proved two results. First, in \cite{Bosh4} the following is shown:

\begin{Thm}
For every irreducible $\tau \in S_m$ and for Lebesgue almost every $\lambda$, the subshift $\Omega_{\lambda,\tau}$ satisfies {\rm (B)}.
\end{Thm}

In fact, Boshernitzan shows that for every irreducible $\tau \in S_m$ and for Lebesgue almost every $\lambda$, the subshift $\Omega_{\lambda,\tau}$ satisfies a stronger condition where the sequence of $n$ values for which $\eta(n)$ is large cannot be too sparse. This condition is easily seen to imply (B), and hence the theorem above.

In a different paper, \cite{Bosh3}, Boshernitzan singles out an explicit class of subshifts arising from interval exchange transformations that satisfy (B). The transformation $T$ is said to be of (rational) rank $k$ if the $\mu_i$ span a $k$-dimensional space over $\Q$ (the field of rational numbers).

\begin{Thm}
If $T$ has rank $2$, the subshift $\Omega_{\lambda,\tau}$ satisfies {\rm (B)}.
\end{Thm}

\subsection{Singular Continuous Spectrum}

As seen in the previous section, the spectrum has zero Lebesgue measure whenever condition (B) holds. This condition is satisfied by a wide class of models, in particular by all typical quasicrystal models. As pointed out in Theorem~\ref{t.boshthmdl}, a consequence of zero-measure spectrum is the absence of absolutely continuous spectrum. That is, if $\sigma(H_\omega)$ has zero Lebesgue measure, then $\sigma_\mathrm{ac}(H_\omega) = \emptyset$, since all spectral measures are supported by the spectrum, and any measure supported by a set of zero Lebesgue measure must be purely singular by definition.

To complement this, one can often show the absence of point spectrum. That is, there are a variety of tools that allow one to show that $H_\omega$ has no eigenvalues, and hence $\sigma_\mathrm{pp}(H_\omega) = \emptyset$ as well. Putting the two results together, one obtains that $H_\omega$ has purely singular continuous spectrum.

The primary tool that allows one to exclude eigenvalues is based on the Gordon lemma, which assumes that the potential has infinitely many suitably aligned local periodicities. Overall, this nicely implements the philosophy that aperiodic order is intermediate between periodic and random. The aperiodicity implies the absence of absolutely continuous spectrum via Kotani's theorem (and hence one does not have the spectral type that appears for a periodic medium), while the order feature implies the absence of point spectrum via a fingerprint of local periodicity (and hence one does not have the spectral type that appears for a random medium).\footnote{In a random model, the values of the potential at the various sites are given by independent identically distributed random variables. This model is usually called the Anderson model.}

A potential $V : \Z \to \R$ is called a \textit{Gordon potential} if there are $q_k \to \infty$ such that for every $k$, we have $V(n) = V(n+q_k) = V(n-q_k)$ for $1 \le n \le q_k$. That is, $V$ looks like a periodic potential around the origin, as one sees at least three suitably aligned periodic unit cells there, and the period may be chosen arbitrarily large. The following Gordon lemma is based in spirit on \cite{G76}. In this particular form it was shown in \cite{DP86}.

\begin{Lm}\label{l.gordon3}
Suppose $V$ is a Gordon potential. Then, for every $E$, the difference equation
$$
u(n+1) + u(n-1) + V(n) u(n) = E u(n)
$$
has no non-trivial square-summable solutions. In particular, the associated Schr\"odinger operator $H$ in $\ell^2(\Z)$, given by
$$
[H\psi](n) = \psi(n+1) + \psi(n-1) + V(n) \psi(n),
$$
has empty point spectrum.
\end{Lm}

By ergodicity, $T$-invariance, and the Gordon lemma, if one can show that
$$
\mu \left( \{ \omega \in \Omega : V_\omega \text{ is a Gordon potential} \} \right) > 0,
$$
then
$$
\mu \left( \{ \omega \in \Omega : H_\omega \text{ has empty point spectrum} \} \right) = 1.
$$

On the other hand, for any aperiodic minimal subshift, at least one of its elements fails to have the required Gordon three-block symmetries \cite{D00b}. Thus, one cannot use this appraoch to show uniform absence of eigenvalues for all $\omega \in \Omega$. Nevertheless, results to this effect are known, established with the following variant of the Gordon lemma.

\begin{Lm}\label{l.gordon2}
Suppose $V : \Z \to \R$ is such that there are $q_k \to \infty$ such that $V(n) = V(n+q_k)$ for $1 \le n \le q_k$. Suppose further that $E$ is such that
\begin{equation}\label{e.2bgass}
\sup_k \left| \mathrm{Tr} \left( \begin{pmatrix} E - V(q_k) & -1 \\ 1 & 0 \end{pmatrix} \times \cdots \times \begin{pmatrix} E - V(1) & -1 \\ 1 & 0 \end{pmatrix} \right) \right| < \infty.
\end{equation}
Then, the difference equation
$$
u(n+1) + u(n-1) + V(n) u(n) = E u(n)
$$
has no non-trivial solutions that are square-summable on $\Z_+$ and hence $E$ is not an eigenvalue of the associated Schr\"odinger operator $H$ in $\ell^2(\Z)$. In particular, if the assumption \eqref{e.2bgass} holds for every $E$ in the spectrum of $H$, then $H$ has empty point spectrum.
\end{Lm}

In many quasicrystal models, the existence of hierarchical structures gives rise to a so-called trace map, which in turn can often be used to ensure that \eqref{e.2bgass} holds for all energies in the spectrum. Thus, the analysis then reduces to finding suitable squares of arbitrary length starting at the origin.

Thus, in the symbolic setting at hand, the observations above give rise to problems that concern the subword structure of the potentials, and hence fall in the general area of combinatorics on words.

Let us describe the results that have been obtained in this way for the examples discussed above. In all these results, the choice of the sampling function $f$ is more restricted than above. Namely, one usually assumes that $f(\omega) = g(\omega_0)$ with an injective map $g : \CalA \to \R$.

\subsubsection{Subshifts Generated by Primitive Substitutions}

Suppose $S$ is a primitive substitution over the alphabet $\CalA$ and let $\Omega \subseteq \CalA^\Z$ be the subshift associated with it. Recall that it is strictly ergodic and denote the unique invariant probability measure by $\mu$. The \textit{index} of $\Omega$ is given by the largest fractional power occurring in an (and hence any) element of $\Omega$. Formally, the index is defined as follows. Given $w \in \mathcal{W}_\Omega$ and any prefix $v$ of $w$, we denote the word $w^kv$ with $k \in \Z_+$ by $w^r$, where $r = k + \frac{|v|}{|w|}$. Then, $\mathrm{ind}_\Omega(w) = \sup \{ r \in \Q \cap [1,\infty) : w^r \in \mathcal{W}_\Omega \}$ and $\mathrm{ind}(\Omega) = \sup \{ \mathrm{ind}_\Omega(w) : w \in \mathcal{W}_\Omega \}$.

The following result was shown in \cite{D00b}.\footnote{See \cite{D98b} for a precursor dealing with the period doubling case. In this special case it was later shown that the absence of eigenvalues even holds for all $\omega \in \Omega$ \cite{D01}.}

\begin{Thm}
If $S$ is a primitive substitution and the associated subshift $\Omega$ satisfies $\mathrm{ind}(\Omega) > 3$, then $\mu \left( \{ \omega \in \Omega : V_\omega \text{ is a Gordon potential} \} \right) > 0$. Consequently, $\mu \left( \{ \omega \in \Omega : H_\omega \text{ has empty point spectrum} \} \right) = 1$.
\end{Thm}

The underlying idea is simple. Since the subshift is invariant under $S$, any word appearing with index strictly greater than $3$ generates by iteration of $S$ a sequence of words whose lengths go to infinity and whose index is bounded away from $3$. This allows one to bound from below the frequency with which third powers occur and hence yields measure estimates on the Gordon three-block conditions that are good enough to show that the $\limsup$ of these sets must have positive measure. Since the elements of the $\limsup$ of these sets give rise to Gordon potentials, the result follows.

Applications of this theorem include the Fibonacci substitution (since $\mathrm{ind}(\Omega) \ge \mathrm{ind}_\Omega(abaab) \ge 3.2$), the period doubling substitution (since $\mathrm{ind}(\Omega) \ge \mathrm{ind}_\Omega(ab) \ge 3.5$), and many others. Of course, the result does not apply to the Thue-Morse substitution, which is famous mainly because $\mathrm{ind}(\Omega) = 2$. Unfortunately, it is still open whether the point spectrum is almost surely empty in the Thue-Morse case. The Gordon approach fails due to small index, and no other methods are known that yield an almost sure result.\footnote{The absence of eigenvalues for a dense $G_\delta$ set of $\omega$'s can be established in this example and many others using palindromes instead of powers \cite{HKS}. However, using palindromes one cannot prove the absence of eigenvalues for a full measure set of $\omega$'s \cite{DZ00}.}

\subsubsection{Sturmian and Quasi-Sturmian Subshifts}\label{ss.standqs}

Damanik, Killip, and Lenz showed the following result in \cite{DKL} (see also \cite{DL99a} for a uniform result for almost every Sturmian subshift).

\begin{Thm}\label{t.sturmscspec}
For every Sturmian subshift $\Omega$, $H_\omega$ has empty point spectrum for every $\omega \in \Omega$.
\end{Thm}

This result was the culmination of a sequence of partial results. Among those, we single out S\"ut\H{o} \cite{S87}, who proved empty point spectrum for one $\alpha$ and one $\omega$, Bellissard et al.\ \cite{BIST89}, who proved it for all $\alpha$ and one $\omega$, Delyon-Petritis \cite{DP86}, who proved it for almost all $\alpha$ and almost all $\omega$, and Kaminaga \cite{K96}, who proved it for all $\alpha$ and almost all $\omega$. Here, $\alpha \in (0,1) \setminus \Q$ denotes the slope associated with a Sturmian subshift. Recall that Sturmian subshifts are in one-to-one correspondence with $(0,1) \setminus \Q$. Here, \cite{BIST89, DKL, DL99a, S87} used Lemma~\ref{l.gordon2}, whereas \cite{DP86,K96} used Lemma~\ref{l.gordon3}.

The extension of Theorem~\ref{t.sturmscspec} to the quasi-Sturmian case was obtained by Damanik and Lenz in \cite{DL03}.

\begin{Thm}\label{t.quasisturmscspec}
For every quasi-Sturmian subshift $\Omega$, $H_\omega$ has empty point spectrum for every $\omega \in \Omega$.
\end{Thm}

\subsubsection{Circle Maps}

Recall that a circle map subshift is determined by the parameters $\alpha \in (0,1) \setminus \Q$ and $\beta \in (0,1)$. It is strictly ergodic and we denote the unique ergodic measure by $\mu$. Delyon and Petritis proved the following in \cite{DP86}.

\begin{Thm}
For almost every $\alpha$ and every $\beta$, the corresponding circle map subshift $\Omega$ is such that $\mu \left( \{ \omega \in \Omega : V_\omega \text{ is a Gordon potential} \} \right) = 1$. Consequently, $\mu \left( \{ \omega \in \Omega : H_\omega \text{ has empty point spectrum} \} \right) = 1$.
\end{Thm}

In fact, the full measure set of $\alpha$ values is explicitly described in terms of the continued fraction expansion. The condition was weakened by Kaminaga in \cite{K96}, still however excluding an explicit zero measure set. This weaker condition was only shown to imply $\mu \left( \{ \omega \in \Omega : V_\omega \text{ is a Gordon potential} \} \right) > 0$, which of course is still sufficient to allow one to deduce $\mu \left( \{ \omega \in \Omega : H_\omega \text{ has empty point spectrum} \} \right) = 1$.

\subsubsection{Interval Exchange Transformations}

Recall that an IET subshift is determined by an irreducible permutation $\tau$ and a probability vector $\lambda$. Cobo, Gutierrez, and de Oliveira showed the following result in \cite{CGO08} (see also \cite{dOG03}).

\begin{Thm}
For every irreducible permutation $\tau$ and almost every $\lambda$, the associated IET subshift $\Omega$ is such that $\mu \left( \{ \omega \in \Omega : V_\omega \text{ is a Gordon potential} \} \right) = 1$. Consequently, $\mu \left( \{ \omega \in \Omega : H_\omega \text{ has empty point spectrum} \} \right) = 1$.
\end{Thm}

\subsection{Transport Properties}\label{s.transport}

Quasicrystal models have behavior that is markedly different from the periodic and random cases in many different respects. In the previous subsections we have seen that the spectrum is typically a zero-measure Cantor set, while for periodic and random potentials it is always given by a finite union of non-degenerate compact intervals. Moreover, the spectral type is typically singular continuous, while it is always absolutely continuous in the periodic case and
almost surely pure point in the (one-dimensional) random case.

In this subsection we consider yet another perspective from which the quasicrystal model behavior is expected to differ from the behavior of the periodic and random cases. Namely, we consider the spreading of wave packets under the time-dependent Schr\"odinger equation. That is, given a Schr\"odinger operator $H_\omega$ and a normalized element $\psi$ of $\ell^2(\Z)$, we consider $\psi(t) = e^{-itH_\omega} \psi$, where $e^{-itH_\omega}$ is defined via the spectral theorem. Then, $\psi(\cdot)$ satisfies the time-dependent Schr\"odinger equation $i \partial_t \psi(t) = H_\omega \psi(t)$ with initial condition $\psi(0) = \psi$. The quantum mechanical interpretation is that the probability of finding the quantum particle at site $n \in \Z$ at time $t \in \R$ is given by $a(n,t) := | \langle \delta_n , \psi(t) \rangle |^2$. The initial state is naturally localized in some fixed compact set, up to a small error, since it belongs to $\ell^2(\Z)$. More specifically, one is often interested in the initial state $\psi = \delta_0$ (or some $\delta_n$), which is completely localized. After having fixed the initial state, one is then interested in how fast $\psi(t)$ spreads out in space, or more specifically, how long one has to wait until $a(n,t)$ is no longer negligibly small at some distant site $n$. In general, this is a difficult problem. Questions of this kind are easier to study for compound quantities; that is, some averaging in $n$ and/or $t$ helps one generate quantities for which interesting statements can be proven.

A popular way to average in time is to consider Ces\`aro averages,
$$
\tilde a(n,T) = \frac1T \int_0^T a(n,t) \, dt = \frac1T \int_0^T | \langle \delta_n , \psi(t) \rangle |^2 \, dt.
$$
Let also
$$
M_p(t) = \sum_{n \in \Z} (1 + |n|^p) a(n,t), \quad \tilde M_p(T) = \sum_{n \in \Z} (1 + |n|^p) \tilde a(n,T), \quad p > 0.
$$
Notice that for $t$ (resp., $T$) fixed, $a(\cdot,t)$ and $\tilde a(\cdot,T)$ are probability distributions on $\Z$, and hence the quantities above are ($1$ plus) the $p$-th moment of the respective probability distribution. Here we assume that the initial state is either a Dirac delta function or at least sufficiently well localized so that these moments are finite.

Wave packet spreading then is reflected by growth in time of these moments. To detect power-law growth, one introduces the so-called transport exponents
\begin{align*}
\beta^+(p) &= \limsup_{t \to \infty} \frac{\log M_p(t)}{p \log t}, \quad\qquad
\beta^-(p) = \liminf_{t \to \infty} \frac{\log M_p(t)}{p \log t}, \\[.25em]
\tilde \beta^+(p) & = \limsup_{T \to \infty} \frac{\log \tilde M_p(T)}{p \log T}, \quad\qquad\kern-3pt
\tilde \beta^-(p)  = \liminf_{T \to \infty} \frac{\log \tilde M_p(T)}{p \log T}.
\end{align*}
Each of these four functions of $p$ is non-decreasing in $p$ and takes values in the interval $[0,1]$.

In view of the monotonicity of the transport exponents, it is natural to consider their limiting values for small and large values of $p$. Thus, denote
\begin{align*}
\alpha^\pm_\ell &= \lim_{p \downarrow 0} \beta^\pm (p), \quad\qquad
\alpha^\pm_u = \lim_{p \uparrow \infty} \beta^\pm (p), \\[.25em]
\tilde \alpha^\pm_\ell &= \lim_{p \downarrow 0} \tilde \beta^\pm (p), \quad\qquad
\tilde \alpha^\pm_u = \lim_{p \uparrow \infty} \tilde \beta^\pm (p).
\end{align*}
We note that there are other useful ways of capturing wave packet spreading, and refer the reader to \cite{DT10} for a comprehensive survey.

The transport exponents take the constant value $0$ for random potentials and (at least the time-averaged quantities) the constant value $1$ for periodic potentials. Thus if one is able to prove the occurrence of fractional values of the transport exponents, one exhibits wave packet spreading that is strictly intermediate between the periodic and random cases. Results of this kind are notoriously difficult to establish. The few known results for quasicrystal models will be described in detail in later sections on Fibonacci and Sturmian potentials.

\section{The Fibonacci Hamiltonian}\label{s.5}

The Fibonacci Hamiltonian is the most prominent model in the study of electronic properties of quasicrystals. It is given by the discrete one-dimensional Schr\"odinger operator
\beq\label{e.fib}
[H_{\lambda,\omega} u](n) = u(n+1) + u(n-1) + \lambda \chi_{[1-\alpha,1)}(n \alpha + \omega \!\!\!\! \mod 1) u(n),
\eneq
where $\lambda > 0$ is the coupling constant, $\alpha = \frac{\sqrt{5}-1}{2}$ is the frequency, and $\omega \in [0,1)$ is the phase. An alternative way to obtain the same potential is via the Fibonacci substitution; see Section~\ref{ss.subst} above. This operator family has been studied in many papers since the early 1980's (see, e.g., \cite{AS, GA, Ho, Ho1, IRT, KKT, Koh, KST, Lev, OPRSS, OP, W} for early works in the physics literature), and numerous fundamental results are known. In this section we describe the current ``state of the art'' for this model.

\subsection{Trace Map Formalism}

Even the earliest papers on the Fibonacci Hamiltonian realized the importance of a certain renormalization procedure in its study, see \cite{KKT, OPRSS}. This led in particular to the consideration of a certain dynamical system, the so-called trace map, whose properties are closely related to many spectral properties of the operator (\ref{e.fib}). The existence of the trace map and its connection to spectral properties is a consequence of the invariance of the potential under a substitution rule. This works in great generality; see \cite{AB93, D00} and references therein.

The one-step transfer matrices associated with the difference equation $H_{\lambda,\omega} u = E u$ are given by
$$
T_{\lambda,\omega}(m,E) = \begin{pmatrix} E - \lambda \chi_{[1-\alpha,1)}(m \alpha + \omega \!\!\!\! \mod 1) & -1 \\ 1 & 0 \end{pmatrix}.
$$
Denote the Fibonacci numbers by $\{F_k\}$, that is, $F_0 = F_1 = 1$ and $F_{k+1} = F_k + F_{k-1}$ for $k \ge 1$. Then the fact that the potential for zero phase is invariant under the Fibonacci substitution implies that the matrices
$$
M_{-1}(E) = \begin{pmatrix} 1 & -\lambda \\ 0 & 1 \end{pmatrix} , \quad M_0(E) = \begin{pmatrix} E & -1 \\ 1 & 0 \end{pmatrix}
$$
and
$$
M_k(E) = T_{\lambda,0}(F_k,E) \times \cdots \times T_{\lambda,0}(1,E) \quad \text{ for } k \ge 1
$$
obey the recurrence relations
$$
M_{k+1}(E) = M_{k-1}(E) M_k(E)
$$
for $k \ge 0$. Passing to the variables
$$
x_k(E) = \frac12 \mathrm{Tr} M_k(E),
$$
this in turn implies
\begin{equation}\label{e.tracerec}
x_{k+1}(E) = 2 x_k(E) x_{k-1}(E) - x_{k-2}(E)
\end{equation}
for $k \ge 1$, with $x_{-1}(E)=1$, $x_0(E)=E/2$, and $x_1=(E-\lambda)/2$.
The recursion relation~\eqref{e.tracerec} exhibits a conserved quantity; namely, we have
\begin{equation}\label{e.traceinvariant}
x_{k+1}(E)^2+x_k(E)^2+x_{k-1}(E)^2 - 2 x_{k+1}(E) x_k(E) x_{k-1}(E) -1 = \frac{\lambda^2}{4}
\end{equation}
for every $k \ge 0$.

Given these observations, it is then convenient to introduce the \textit{trace map}
\beq\label{e.T}
T : \Bbb{R}^3 \to \Bbb{R}^3, \; T(x,y,z)=(2xy-z,x,y).
\eneq
Aside from the context described here, this map appears in a natural way in problems related to dynamics of mapping classes \cite{Go}, Fuchsian groups \cite{Bo1}, number theory \cite{Bo2}, Painlev\'e sixth equations \cite{CaL, IU}, the Ising model for quasicrystals \cite{Be, HGB, Y1, Y2}, the Fibonacci quantum walk \cite{RMM, R}, among others \cite{BGJ, DMY, Suth, Y3}. See \cite{Can} or \cite{Br} for an algebraic explanation of this universality. We refer the reader also to \cite{HM, R96, RB94} for further reading on the Fibonacci trace map.

\begin{figure}\begin{minipage}[t]{5cm} \includegraphics[width=1\textwidth]{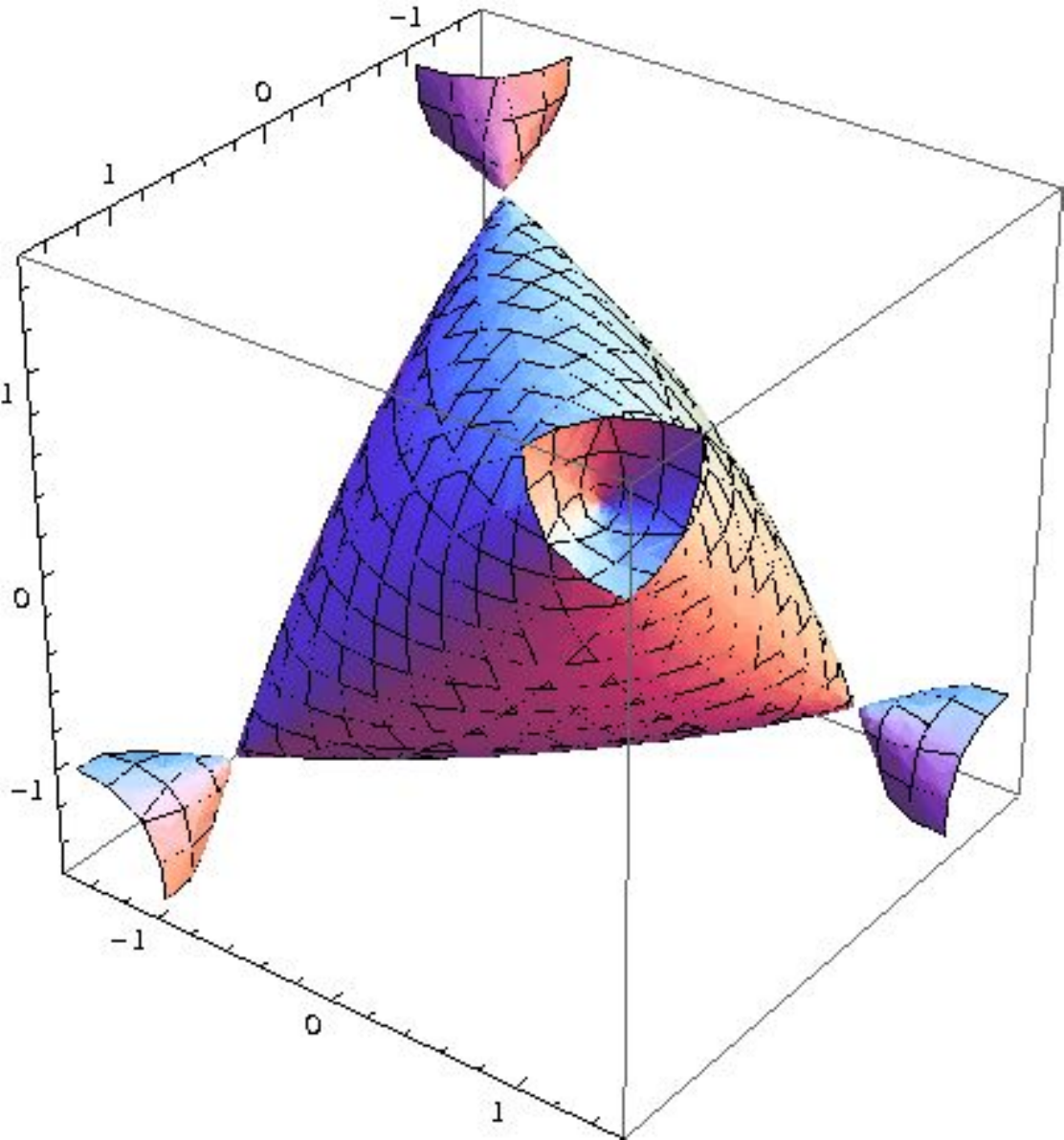} \par
\caption{The surface $S_{0.01}$.} \label{fig:s0.01}
\end{minipage} \hfill 
%
\begin{minipage}[t]{5cm} \includegraphics[width=1\textwidth]{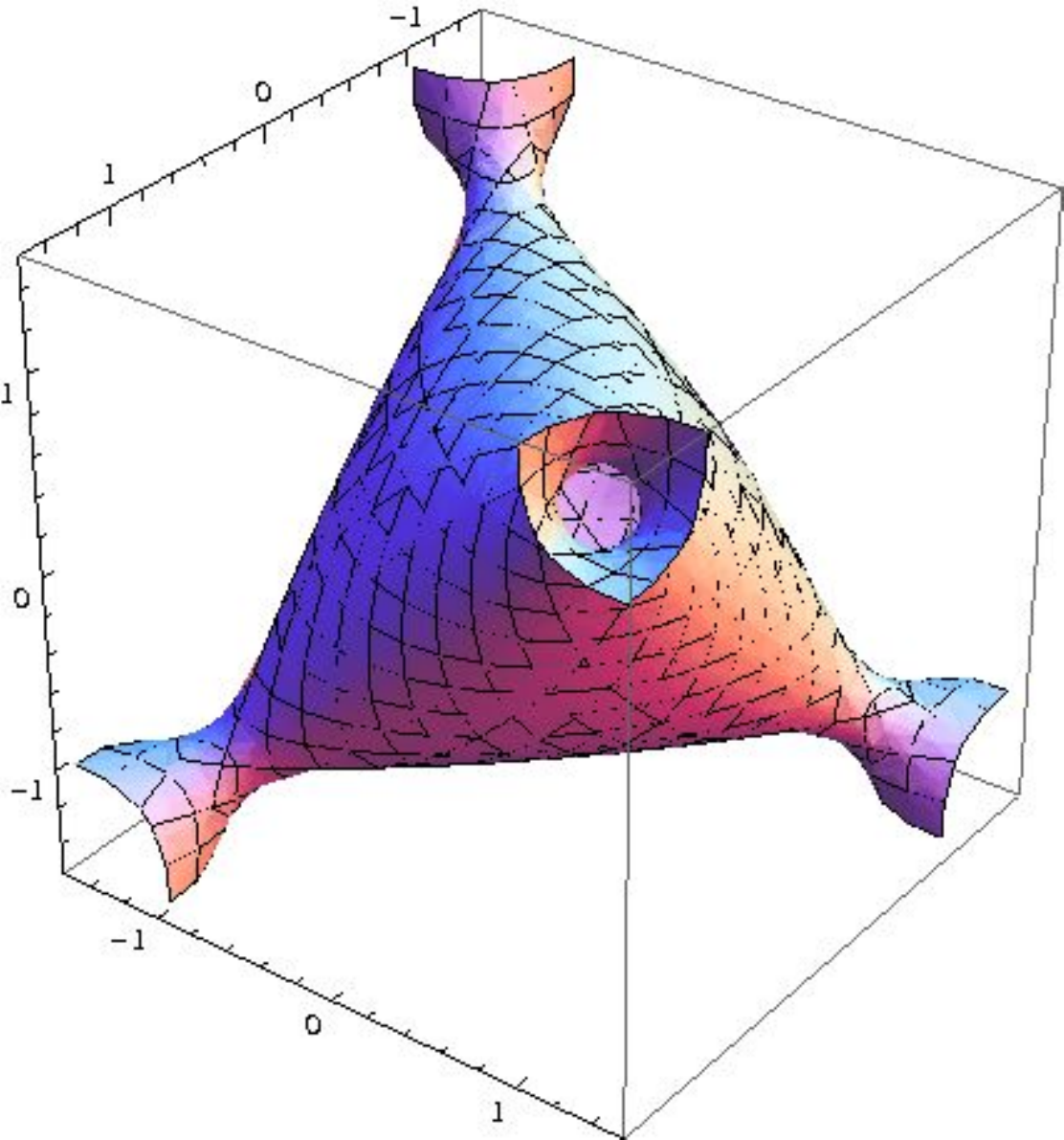}
\par \caption{The surface $S_{0.5}$.} \label{fig:s0.5} \end{minipage}
\end{figure}

The function
$$
G(x,y,z) = x^2+y^2+z^2-2xyz-1
$$
is invariant under the action of $T$\footnote{The function $G(x,y,z)$ is usually called the \textit{Fricke character} or \textit{Fricke-Vogt invariant}.} (which explains \eqref{e.traceinvariant}), and hence $T$ preserves the family of cubic surfaces\footnote{The surface $S_0$ is called the \textit{Cayley cubic}.}
\begin{equation} \label{Slam}
S_\lambda = \left\{(x,y,z)\in \Bbb{R}^3 : x^2 + y^2 + z^2 - 2xyz = 1 + \frac{\lambda^2}{4} \right\}.
\end{equation}
Plots of the surfaces $S_{0.01}$
and $S_{0.5}$ are given in Figures~\ref{fig:s0.01} and \ref{fig:s0.5}, respectively.


Denote by $\ell_\lambda$ the line
$$
\ell_\lambda = \left\{ \left(\frac{E-\lambda}{2}, \frac{E}{2}, 1 \right) : E \in \Bbb{R} \right\}.
$$
It is easy to check that $\ell_\lambda \subset S_\lambda$.

S\"ut\H{o} proved the following central result in \cite{S87}.

\bthm\label{spectrum}
An energy $E$ belongs to the spectrum of $H_{\lambda,\omega}$ if and only if the positive semiorbit of the point $\left(\frac{E-\lambda}{2}, \frac{E}{2}, 1 \right)$ under iterates of the trace map $T$ is bounded.
\ethm

To obtain this theorem, S\"ut\H{o} argued as follows. Denote
$$
\sigma_k = \{ E \in \R : |x_k(E)| \le 1 \}
$$
and
$$
\Sigma_k = \sigma_k \cup \sigma_{k+1}.
$$
These sets depend on the coupling constant $\lambda$, and whenever we want to make this dependence explicit, we will write $\sigma_{k,\lambda}$ and $\Sigma_{k,\lambda}$. An analysis of the trace recursion \eqref{e.tracerec} shows that the sets $\Sigma_k$ are decreasing, and hence it is natural to consider their limit $\tilde \Sigma = \bigcap \Sigma_k$. Clearly, if $E \in \tilde \Sigma$, then $\{ x_n(E) \}$ remains bounded due to \eqref{e.traceinvariant}. On the other hand, the analysis of the trace recursion \eqref{e.tracerec} also yields that whenever $E \notin \Sigma_k$ for some $k$, then $|x_{n-k}(E)|$ obeys an explicit super-exponentially growing lower bound. That is, the sequence $\{ x_n(E) \}$ remains bounded if and only if $E \in \tilde \Sigma$. Notice that the point $(\frac{E-\lambda}{2}, \frac{E}{2}, 1)$ is just $(x_1(E) , x_0(E) , x_{-1}(E))$, so that Theorem~\ref{spectrum} follows as soon as $\Sigma = \tilde \Sigma$ is established. The inclusion $\Sigma \subseteq \tilde \Sigma$ holds since $\sigma_k$ is precisely the spectrum of the canonical periodic approximant of period $F_k$ and the fact that these periodic approximants converge strongly. The inclusion $\tilde \Sigma \subseteq \Sigma$ holds since one can use the boundedness of $\{ x_n(E) \}$ for $E \in \tilde \Sigma$ along with the Gordon lemma to show that no solution for this energy is square-summable at $+\infty$, which implies that $E$ must be in the spectrum.

\subsection{Hyperbolicity of the Trace Map}

Let $f : M \to M$ be a diffeomorphism of a Riemannian manifold $M$. Let us recall that an invariant closed set $\Lambda$ of the diffeomorphism $f$ is \textit{hyperbolic} if there exists a splitting of a tangent space $T_x M = E^s_x \oplus E^u_x$ at every point $x \in \Lambda$ such that this splitting is invariant under $Df$, and the differential $Df$ exponentially contracts vectors from stable subspaces $\{E^s_x\}$ and exponentially expands vectors from unstable subspaces $\{E^u_x\}$. A hyperbolic set $\Lambda$ of a diffeomorphism $f : M \to M$ is \textit{locally maximal} if there exists a neighborhood $U(\Lambda)$ such that
$$
\Lambda = \bigcap_{n \in \Bbb{Z}} f^n(U).
$$

We will consider diffeomorphisms of a surface, $\text{\rm dim } M = 2$, and  hyperbolic sets of topological dimension zero. In this case a locally maximal hyperbolic set $\Lambda$ can be locally represented as a product of ``stable" and ``unstable" Cantor sets $C^s$ and $C^u$. Both Cantor sets $C^s$ and $C^u$ are {\it dynamically defined}. Dynamically defined Cantor sets have strong self-similar structure and exhibit many nice properties. The  formal definition in the general case (when the underlying symbolic system is a general topological Markov chain) is somewhat tedious, and can be found, for example, in \cite{PT}. To provide some intuition to the reader, we give here the definition of a dynamically defined Cantor set in the simplest case when the corresponding symbolic dynamical system is a full shift.

\bdef
Let $I\subset \mathbb{R}^1$ be a closed interval. A Cantor set $C\subset I$ is {\rm dynamically defined} if there are strictly monotone contracting maps $\psi_1,  \psi_2, \ldots, \psi_k: I\to I, \psi_i(I)\cap \psi_j(I) = \emptyset$ if $i \ne j$, such that $C = \bigcap_{n \in \mathbb{N}}I_n$, where $I_1 = \psi_1(I) \cup \cdots \cup \psi_k(I)$ and $I_{n+1} = \psi_1(I_n) \cup \cdots \cup \psi_k(I_n)$.
\endef

If $\psi_1, \psi_2, \ldots, \psi_k$ are $C^{1+\varepsilon}$-functions, then the Cantor set has zero measure, depends continuously on $\psi_1, \ldots, \psi_k$, and is ``regular'' in many other ways. We will be interested in the {\it Hausdorff dimension} and the {\it thickness} of the Cantor sets $C^s$ and $C^u$. Denote the Hausdorff dimension of the set $C$ by $\dim_H C$.

In our case, $\dim_H \Lambda = \dim_H C^s + \dim_H C^u$; see \cite{MM, PV}. Moreover, if $f$ depends $C^r$-smoothly on a parameter, then $\dim_H \Lambda$ is also a smooth function of the parameter; see \cite{Ma}.

\bdef
Let $C \subset \mathbb{R}$ now be an arbitrary Cantor set and denote by $I$ its convex hull. Any connected component of $I \backslash C$ is called a \emph{gap} of $C$. A \emph{presentation} of $C$ is given by an ordering $\mathcal{U} = \{U_n\}_{n \ge 1}$ of the gaps of $C$. If $u \in C$ is a boundary point of a gap $U$ of $C$, we denote by $K$ the connected component of $I\backslash (U_1\cup U_2\cup\cdots \cup U_n)$ {\rm (}with $n$ chosen so that $U_n = U${\rm )} that contains $u$ and write
$$
\tau(C, \mathcal{U}, u) = \frac{|K|}{|U|}.
$$

With this notation, the \emph{thickness} $\tau(C)$ and the \emph{denseness} $\theta(C)$ of $C$ are given by
\begin{equation} \label{eq:thick}
\tau(C) = \sup_{\mathcal{U}} \inf_{u} \tau(C, \mathcal{U}, u), \qquad \theta(C) = \inf_{\mathcal{U}} \sup_{u} \tau(C, \mathcal{U}, u).
\end{equation}
\endef

The thickness and the denseness of a Cantor set $C$ are related to the Hausdorff dimension of $C$ by the inequalities (cf.~\cite[Section~4.2]{PT})
\beq\label{e.thickness}
\frac{\log 2}{\log(2+ \frac{1}{\tau(C)})}\le \dim_H C \le \frac{\log 2}{\log(2 + \frac{1}{\theta(C)})}.
\eneq
For more details on thickness, see \cite{d2, MMR, PT}. An important property of thickness was discovered by Newhouse \cite{n1}:

\bthm\label{t.sumofcantorsets}
If $C_1$ and $C_2$ are two Cantor sets and $\tau(C_1) \cdot \tau(C_2) \ge 1$, then the sum $C_1 + C_2$ contains an interval. In the special case $C_1 = C_2 =: C$, we have that $\tau(C) \ge 1$ implies that $C+C$ is an interval.
\ethm

Consider the restriction $T_{\lambda} : S_\lambda \to S_\lambda$ of the trace map $T$ from \eqref{e.T} to the invariant surface $S_\lambda$, $T_{\lambda} = T|_{S_\lambda}$. Denote by $\Omega_{\lambda}$ the set of points in $S_\lambda$ whose full orbits under $T_{\lambda}$ are bounded.

\bthm\label{Casdagli}
For every $\lambda > 0$, the set $\Omega_{\lambda}$ is a locally maximal hyperbolic set of $T_{\lambda} : S_\lambda \to S_\lambda$. It is homeomorphic to a Cantor set.
\ethm

Theorem~\ref{Casdagli} was proved for $\lambda \ge 16$ by Casdagli \cite{Cas}, for small values of $\lambda$ by Damanik and Gorodetski \cite{DG1}, and finally for all $\lambda > 0$ by Cantat \cite{Can}.

Since $\ell_\lambda \subset S_\lambda$ the set of points on $\ell_\lambda$ whose forward semiorbits are bounded is exactly equal to $\ell_\lambda\cap W^s(\Omega_\lambda)$. Then the spectrum  $\Sigma_\lambda$ is affine equivalent to the set $\ell_\lambda \cap W^s(\Omega_\lambda)$.

\bthm\label{transversality}
For every $\lambda > 0$, the line $\ell_\lambda$ intersects the leaves of $W^s(\Omega_\lambda)$ transversally.
\ethm

This transversality statement was proved for $\lambda \ge 16$ by Casdagli \cite[Section~2]{C}, and for sufficiently small $\lambda > 0$ by Damanik and Gorodetski \cite{DG1}. A proof that works for all values of the coupling constant $\lambda > 0$ was given by Damanik, Gorodetski, and Yessen in \cite{DGY14}.

Theorem~\ref{transversality} allows one to consider the spectrum $\Sigma_\lambda$ as a dynamically defined Cantor set. Therefore the following holds.

\bcor\label{c.dyndefcantorset}
For every $\lambda > 0$, the spectrum $\Sigma_\lambda$ is a dynamically defined Cantor set, and hence:

\begin{itemize}

\item[{\rm (i)}] For every small $\varepsilon > 0$ and every $x \in \sigma(H_{\lambda,\omega})$, we have
\begin{align*}
\dim_H \left( (x-\varepsilon, x+\varepsilon) \cap \sigma(H_{\lambda,\omega}) \right) & = \dim_B \left( (x-\varepsilon, x+\varepsilon) \cap \sigma(H_{\lambda,\omega}) \right) \\
& = \dim_H \sigma(H_{\lambda,\omega}) \\
& =\dim_B \sigma(H_{\lambda,\omega}).
\end{align*}

\item[{\rm (ii)}] The Hausdorff dimension $\dim_H \sigma(H_{\lambda,\omega})$ is an analytic function of $\lambda$, and is strictly between zero and one.

\end{itemize}
\ecor

\subsection{Hausdorff Dimension of the Spectrum at Large Coupling}\label{sec 5.3}

The fact that the box counting dimension of the spectrum exists and coincides with its Hausdorff dimension allows one to determine the asymptotic behavior of this $\lambda$-dependent quantity in the large coupling limit. In fact, Damanik, Embree, Gorodetski, and Tcheremchantsev proved the following in \cite{DEGT}.

\begin{Thm}\label{t.DEGT}
We have
$$
\lim_{\lambda \to \infty} \left( \dim \Sigma_\lambda \right) \cdot \log \lambda = \log(1+\sqrt{2}).
$$
\end{Thm}

Let us briefly explain how this result is obtained. Recall that the spectrum is related to the spectra of the canonical periodic approximants by
$$
\Sigma_\lambda = \bigcap_{k \ge 1} \Sigma_{k,\lambda} = \bigcap_{k \ge 1} \sigma_{k,\lambda} \cup \sigma_{k+1,\lambda}.
$$
Since each periodic spectrum $\sigma_{k,\lambda}$ is a finite union of non-degenerate compact intervals and the lengths of these intervals can be shown to be decaying, it is natural to use $\Sigma_{k,\lambda}$ as one possible cover of $\Sigma_\lambda$ and estimate the Hausdorff dimension of $\Sigma_\lambda$ from above in this way. On the other hand, since each interval of $\sigma_{k,\lambda}$ can be shown to have non-empty intersection with $\Sigma_\lambda$, one can estimate the box counting dimension of $\Sigma_\lambda$ from below in this way. We observe how crucial it is that these dimensions coincide here. Thus, the analysis of the participating intervals comes down to proving good estimates for their length.

To estimate the length, one makes use of the following basic fact from one-dimensional Floquet theory. The preimage of the \emph{open} interval $(-1,1)$ under $x_k$ consists of exactly $F_k$ disjoint open intervals, on which $x_k$ is strictly monotone. In fact, in this particular case, the same statement is true for the corresponding \emph{closed} intervals (i.e., the periodic spectra in question have all their gaps open). Thus, the length of one of these intervals (say $I = [a,b]$) can be estimated as follows. Since
$$
2 = |x_k(a) - x_k(b)| = \int_a^b |x_k'(E)| \, dE,
$$
we have
$$
\frac{2}{\max_{E \in I} |x_k'(E)|} \le |I| \le \frac{2}{\min_{E \in I} |x_k'(E)|}.
$$
In order to prove estimates for $|x_k'(E)|$, one differentiates the trace recursion \eqref{e.tracerec} and proceeds inductively, making use of the trace invariant \eqref{e.traceinvariant}. This approach was pioneered by Raymond \cite{Ra} and then used in many subsequent papers. In this inductive approach, it turns out to be important to determine, for a given energy $E$ in one of the intervals of $\sigma_{k,\lambda}$, in how many of the earlier sets $\sigma_{k',\lambda}$, $k' < k$, the energy $E$ in question lies. This gives rise to a combinatorial question that was completely answered in \cite{DEGT}. Combining these combinatorial results with the length estimates one can prove in this way for the intervals in question, the overall strategy above yields the following specific estimates:
\begin{align}
\label{eq:bcdup} \dim_H \Sigma_\lambda & \le \frac{\log (1 + \sqrt{2})}{\log \left( \frac12 \left[ (\lambda - 4) + \sqrt{(\lambda - 4)^2 - 12} \, \right] \right)} & \text{ for } \lambda \ge 8, \\
\label{eq:bcdlow} \dim_B^- \Sigma_\lambda & \ge \frac{\log (1 + \sqrt{2})}{\log \left( 2 \lambda + 22 \right)} & \text{ for }  \lambda > 4.
\end{align}
Theorem~\ref{t.DEGT} is then a direct consequence of these estimates and the fact that the Hausdorff dimension and the box counting dimension of $\Sigma_\lambda$ are equal.

\subsection{Quantitative Characteristics of the Spectrum at Small Coupling}

Fractal properties of $\Sigma_\lambda$ for small $\lambda$ were studied in \cite{DG2}. Among many other things, that paper established the following pair of theorems.

\begin{Thm}\label{t.1}
We have
$$
\lim_{\lambda \to 0} \dim \Sigma_\lambda = 1.
$$
More precisely, there are constants $C_1, C_2 > 0$ such that
$$
1 - C_1 \lambda \le \dim \Sigma_\lambda \le 1 - C_2 \lambda
$$
for $\lambda > 0$ sufficiently small.
\end{Thm}

\begin{Thm}\label{t.2}
We have
$$
\lim_{\lambda \to 0} \tau(\Sigma_\lambda) = \infty.
$$
More precisely, there are constants $C_3, C_4 > 0$ such that
$$
C_3 \lambda^{-1} \le \tau(\Sigma_\lambda)\le \theta(\Sigma_\lambda) \le  C_4 \lambda^{-1}
$$
for $\lambda > 0$ sufficiently small.
\end{Thm}

Theorem~\ref{t.1} is a consequence of the connection \eqref{e.thickness} between the Hausdorff dimension of a Cantor set and its denseness and thickness, along with the estimates for the latter quantities provided by Theorem \ref{t.2}.

Let us briefly explain how Theorem \ref{t.2} can be obtained.  The Cayley cubic $S_0$ (cf.~\eqref{Slam}) has four conic singularities and can be represented as a union of a two dimensional sphere (with four conic singularities) and four unbounded components. The restriction of the trace map to the sphere
is a pseudo-Anosov map (a factor of a hyperbolic map of a two-torus), and its Markov partition can be presented explicitly (see \cite{Cas} or \cite{DG1, DG2}). For small values of $\lambda$, the map $T : S_{\lambda}\to S_\lambda$ ``inherits'' the hyperbolicity of this pseudo-Anosov map everywhere away from the singularities. The dynamics near the singularities must be considered separately. Consider the dynamics of $T$ near one of the singularities, say, near the point $p = (1, 1, 1)$. The set $\text{Per}_2(T)$ of periodic orbits of period two is a smooth curve that contains the point $p$ and intersects $S_\lambda$ at two points (denote them by $p_1(\lambda)$ and $p_2(\lambda)$) for $\lambda > 0$. Finite pieces of stable and unstable manifolds of $p_1(\lambda)$ and $p_2(\lambda)$ are a distance of order $\lambda$ from each other. In order to estimate the thickness (and the denseness) of the spectrum $\Sigma_\lambda$, we notice first that the Markov partition for $T : S_0 \to S_0$ can be continuously extended to a Markov partition for $T : S_\lambda \to S_\lambda$. The extended Markov partition is formed by finite parts of the stable and unstable manifolds of $p_1(\lambda)$, $p_2(\lambda)$, and the other six periodic points that are continuations of the three remaining singularities. Therefore the size of the elements of these Markov partitions remains bounded, and the size of the distance between them is of order $\lambda$. The natural approach now is to use the distortion property (see, e.g., \cite{PT}) to show that for the iterated Markov partition, the ratio of the distance between the elements to the size of an element is of the same order. The main technical problem here is again the dynamics of the trace map near the singularities, since the curvature of $S_\lambda$ is very large there for small $\lambda$. Nevertheless, one can still estimate the distortion that is obtained during a transition through a neighborhood of a singularity and prove boundedness of the distortion for arbitrarily large iterates of the trace map. This implies Theorem \ref{t.2}.

\subsection{The Density of States Measure}\label{sec:fibids}

Let us now turn to the formulation of results involving the integrated density of states, a quantity of fundamental importance associated with an ergodic family of Schr\"odinger operators. The integrated density of states (IDS) was introduced in Section \ref{ss.IDS} in a more general context, and represents the distribution function of a density of states measure -- a measure supported on the spectrum and, in particular, reflecting the asymptotic distribution of eigenvalues of finite dimensional approximations.

Denote the density of states measure of the Fibonacci Hamiltonian for a given coupling constant $\lambda$ by $dN_{\lambda}$. Repeating the definition from Section \ref{ss.IDS} in this particular case, we have
\begin{equation}\label{eq:idsN}
N_{\lambda}(E) = \lim_{n \to \infty} \frac{\#\{\text{eigenvalues of } H_{\lambda, \omega, [1,n]} \text{ that are } \le E \}}{n},
\end{equation}
where $H_{\lambda, \omega, [1,n]}$ is the restriction of $H_{\lambda, \omega}$ to the interval $[1,n]$ with Dirichlet boundary conditions, and the limit does not actually depend on the phase $\omega$.

It is interesting to analyze the regularity of the density of states measure. This question was studied for general potentials \cite{c,cs,cs2,GK11,l}, random potentials \cite{ck,st}, and analytic quasi-periodic potentials \cite{AJ10,b,b2,bgs,GS01,GS08,H09,s4}. In the case of Fibonacci Hamiltonian, the IDS is H\"older continuous.

\begin{Thm}\label{global}
For every $\lambda > 0$, there exist $C_\lambda < \infty$ and $\gamma_\lambda > 0$ such that
$$
| N_\lambda(E_1) - N_\lambda(E_2) | \le C_\lambda |E_1 - E_2|^{\gamma_\lambda}
$$
for every $E_1,E_2$ with $|E_1 - E_2| < 1$.
\end{Thm}

This follows directly from \cite{DKL}; see also \cite{D98, DL99b, IRT, IT, JL2} for some previous related results.

It is also interesting to obtain the asymptotics of the optimal H\"older exponent for large and small couplings.
In the large coupling regime, we have the following \cite{DG4} (recall that $\alpha = \frac{\sqrt{5}-1}{2}$).

\begin{Thm}\label{main}
{\rm (a)} Suppose $\lambda > 4$. Then for every
$$
\gamma < \frac{3\log(\alpha^{-1})}{2\log(2\lambda + 22)},
$$
there is some $\delta > 0$ such that the IDS associated with the family of Fibonacci Hamiltonians satisfies
$$
| N_\lambda(E_1) - N_\lambda(E_2) | \le |E_1 - E_2|^{\gamma}
$$
for every $E_1,E_2$ with $|E_1 - E_2| < \delta$.
\\[1mm]
{\rm (b)} Suppose $\lambda \ge 8$. Then for every
$$
\tilde \gamma > \frac{3\log(\alpha^{-1})}{2\log \left(\frac{1}{2} \left( (\lambda - 4) + \sqrt{(\lambda - 4)^2 - 12} \right)\right)}
$$
and every $0 < \delta < 1$, there are $E_1,E_2$ with $0< |E_1 - E_2| < \delta$ such that
$$
| N_\lambda(E_1) - N_\lambda(E_2) | \ge |E_1 - E_2|^{\tilde \gamma}.
$$
\end{Thm}
\begin{Cor}
The optimal H\"older exponent $\gamma$ behaves asymptotically as $\frac{3\log(\alpha^{-1})}{2\log \lambda}$ in the large coupling regime.
\end{Cor}

The proof is based on the self-similarity of the spectrum and an analysis of the periodic approximants (in the spirit of the proof of Theorem \ref{t.DEGT}).

In the small coupling regime, we have the following \cite{DG4}:

\begin{Thm}\label{t.idsholdersmall}
The integrated density of states $N_{\lambda}(\cdot)$ is H\"older continuous with H\"older exponent $\gamma_{\lambda}$, where $\gamma_{\lambda}\to \frac{1}{2}$ as $\lambda\to 0$, and $\gamma_{\lambda}<\frac{1}{2}$ for small $\lambda>0$.

More precisely:

{\rm (a)} For any $\gamma \in (0, \frac{1}{2})$, there exists $\lambda_0 > 0$ such that for any $\lambda \in (0, \lambda_0)$, there exists $\delta>0$ such that
$$
|N_{\lambda}(E_1) - N_{\lambda}(E_2)| \le |E_1 - E_2|^{\gamma}
$$
for every $E_1, E_2$ with $|E_1-E_2| < \delta$;

{\rm (b)} For any sufficiently small $\lambda > 0$, there exists $\tilde \gamma = \tilde \gamma (\lambda) < \frac{1}{2}$ such that for every $\delta > 0$, there are $E_1, E_2$ with $0 < |E_1 - E_2| < \delta$ and
$$
|N_{\lambda}(E_1) - N_{\lambda}(E_2)| \ge |E_1 - E_2|^{\tilde \gamma}.
$$
\end{Thm}

The proof uses the trace map formalism and a relation between the IDS of $H_{\lambda, \omega}$ and the measure of maximal entropy for the trace map $T_{\lambda}$. Namely, the density of states measure is proportional to the projection (along the stable manifolds) to $\ell_\lambda$ of the normalized restriction of the measure of maximal entropy $\mu_{\max}(T_\lambda)$ to an element of the Markov partition. After that, the proof uses a comparison of expansion rates of $T_\lambda$ and $T_0$ (and is reminiscent of the proof of H\"older continuity of conjugacies between two hyperbolic dynamical systems).

Another interesting feature of the Fibonacci Hamiltonian is the uniform scaling of the density of states measure. Namely, the following result (that summarizes the results from  \cite{DGY14},  \cite{DG3}, and \cite{P14}) holds.


\begin{Thm}\label{t.main}
For every $\lambda > 0$, there is $d_\lambda \in (0,1)$ so that the density of states measure $dN_\lambda$ is of exact dimension $d_\lambda$, that is, for $dN_\lambda$-almost every $E \in \R$, we have
$$
\lim_{\varepsilon \downarrow 0} \frac{\log N_\lambda(E - \varepsilon , E + \varepsilon)}{\log \varepsilon} = d_\lambda.
$$
Moreover, in $(0,\lambda_0)$,  $d_\lambda$ is an analytic function of $\lambda$, and
$$
\lim_{\lambda \downarrow 0} d_\lambda = 1.
$$
\end{Thm}

The proof is based on the relation between $dN_\lambda$ and $\mu_{\max}(T_\lambda)$, and the exact dimensionality of hyperbolic measures \cite{BPS, LY,  Pe}.

The Hausdorff dimension of the spectrum is an upper bound for $d_\lambda$, but a priori it is not clear whether these numbers must coincide. Barry Simon conjectured that for a large class of models these quantities must be different.\footnote{The conjecture does not appear anywhere in print, but it was popularized by Barry Simon in many talks given by him in the past four years.} The next result by Damanik, Gorodetski, and Yessen \cite{DGY14} shows that this conjecture is true (see also \cite{DG3} for an earlier partial result).

\begin{Thm}\label{t.main2}
For every $\lambda > 0$, we have $d_\lambda < \dim_H \Sigma_\lambda$.
\end{Thm}

The proof is based on the comparison of the measure of maximal entropy for $T_\lambda$ (which is ``responsible'' for $d_\lambda$) and the equilibrium measure for the potential given by minus the log of the expansion rate. The Hausdorff dimension of the unstable projection of the latter is equal to $\dim_H \Sigma_\lambda$, and the thermodynamical description of this measure (see \cite{MM}) implies that for any other ergodic invariant measure, the dimension of its unstable projection is strictly smaller. In order to prove that those two measures are actually different, one uses the fact that the measure of maximal entropy is an equilibrium measure that corresponds to zero potential. Therefore it is enough to show that the two potentials under consideration are not cohomological, which can be done using a comparison of multipliers of different periodic orbits of~$T_\lambda$.

\subsection{Gap Opening and Gap Labeling}

The spectrum $\Sigma_\lambda$ jumps from being an interval for $\lambda = 0$ to being a zero-measure Cantor set for $\lambda > 0$. Hence, as the potential is turned on, a dense set of gaps opens immediately. It is natural to ask about the size of these gaps; see \cite{BBG92}. These gap openings were studied in \cite{B} for the Thue-Morse potential (where the gaps open as a power of $\lambda$) and in \cite{BBG91} for the period doubling potential (where some gaps open linearly, and some others are superexponentially small in $\lambda$). In the Fibonacci case, all gaps open linearly \cite{DG2, DGY14}:

\begin{Thm}\label{t.3}
The boundary points of a gap in the spectrum $\Sigma_\lambda$ depend smoothly on the coupling constant $\lambda$. Moreover, given any one-parameter continuous family $\{U_\lambda\}_{\lambda > 0}$ of gaps of $\Sigma_\lambda$,\footnote{By a continuous family $\{U_\lambda\}_{\lambda > 0}$ of gaps of $\Sigma_\lambda$ we mean that $U_\lambda$ is a bounded connected component of $\R \setminus \Sigma_\lambda$ and the left endpoint and the right endpoint of $U_\lambda$ each depend continuously on $\lambda$.} we have that
$$
\lim_{\lambda\to 0} \frac{|U_\lambda|}{|\lambda|}
$$
exists and belongs to $(0,\infty)$.
\end{Thm}

Theorem~\ref{t.3} follows again from dynamical properties of the trace map. Namely, each singularity of the Cayley cubic $S_0$ gives birth to two periodic points on the surface $S_\lambda$, $\lambda > 0$. The distance between the periodic points is of order $\lambda$. The stable manifolds of these periodic points ``cut'' gaps in $\ell_\lambda$ that correspond to gaps in the spectrum. The curves formed by the families of the periodic points are normally hyperbolic manifolds of the trace map, and hence (see \cite{HPS, PSW}) their strong stable manifolds form a $C^1$ foliation. This implies that the size of each gap is also of order $\lambda$ (as $\lambda \to 0$), and Theorem \ref{t.3} follows.

The limit in Theorem~\ref{t.3} certainly depends on the family of gaps chosen. In order to study this dependence, one needs to use some labeling of the gaps. As is well known, the density of states produces such a gap labeling. That is, one can identify a canonical set of gap labels, which is only associated with the underlying dynamics (in this case, an irrational rotation of the circle or the shift-transformation on a substitution-generated subshift over two symbols), in such a way that the value of $N(E,\lambda)$ for $E \in \R \setminus \Sigma_\lambda$ must belong to this canonical set. In the Fibonacci case, this set is well-known (see, e.g., \cite[Eq.~(6.7)]{BBG92}) and the general gap labeling theorem specializes to the following statement:
\begin{equation}\label{f.fibgaplabels}
\{ N(E,\lambda) : E \in \R \setminus \Sigma_\lambda \} \subseteq \{ \{ m \alpha \} : m \in \Z \} \cup \{ 1 \}
\end{equation}
for every $\lambda \not= 0$. Here $\{ m \alpha \}$ denotes the fractional part of $m \alpha$, that is, $\{ m \alpha \} = m \alpha - \lfloor m \alpha \rfloor$.
Notice that the set of gap labels is indeed $\lambda$-independent and only depends on the value of $\alpha$ from the underlying circle rotation. Since $\alpha$ is irrational, the set of gap labels is dense. In general, a dense set of gap labels is indicative of a Cantor spectrum and hence a common (and attractive) stronger version of proving Cantor spectrum is to show that the operator ``has all its gaps open.'' For example, the Ten Martini Problem for the almost Mathieu operator is to show Cantor spectrum, while the Dry Ten Martini Problem is to show that all labels correspond to gaps in the spectrum. The former problem has been completely
solved \cite{AJ}, while the latter has not yet been completely settled. Indeed, it is in general a hard problem to show that all labels given by the gap labeling theorem correspond to gaps, and there are only few results of this kind. It turns out that the stronger (or ``dry'') form of Cantor spectrum holds for the  Fibonacci Hamiltonian \cite{DGY14}:

\begin{Thm}\label{t.completegaplabeling}
For every $\lambda > 0$, all gaps allowed by the gap labeling theorem are open. That is,
\begin{equation}\label{f.completelabeling}
\{ N(E,\lambda) : E \in \R \setminus \Sigma_\lambda \} = \{ \{ m \alpha \} : m \in \Z \} \cup \{ 1 \}.
\end{equation}
\end{Thm}

Earlier, \eqref{f.completelabeling} was shown for $\lambda > 4$ by Raymond \cite{Ra}, and for $\lambda > 0$ sufficiently small by Damanik and Gorodetski \cite{DG2}.

Using the gap labeling, we can refine the statement of Theorem \ref{t.3}. For $m \in \Z \setminus \{ 0 \}$, denote by $U_m(\lambda)$ the gap of $\Sigma_\lambda$ where the integrated density of states takes the value $\{ m \alpha \}$. Then, the following result from \cite{DG2} holds:

\begin{Thm}\label{t.gaplimit}
There is a finite constant $C^*$ such that for every $m \in \Z \setminus \{ 0 \}$,
$$
\lim_{\lambda \to 0} \frac{|U_m(\lambda)|}{|\lambda|} = \frac{C_m}{|m|}
$$
for a suitable $C_m \in (0, C^*)$.
\end{Thm}

To see why Theorem \ref{t.gaplimit} holds, notice that each family of gaps converges (as $\lambda \to 0$) to a point of intersection of $\ell_0$ with a stable manifold of one of the singularities. The intersections that have larger labels are in a sense ``produced'' from intersections with smaller labels by the action of the inverse of the trace map. For gaps with small labels, we know from Theorem~\ref{t.3} that $\lim_{\lambda \to 0} \frac{|U_m(\lambda)|}{|\lambda|} < {C^*}$  for some constant $C^*>0$. The length (in coordinates on the two-torus covering $S_0$) of the piece of the stable manifold from the singularity to the point of intersection after $k$ applications of the map is of order $\left( \frac{1 + \sqrt{5}}{2} \right)^k \sim |m|$, and the contraction that will be applied to the gap is of order
$$
\left( \frac{\sqrt{5} - 1}{2} \right)^k \sim \left( \frac{\sqrt{5} - 1}{2} \right)^{\frac{\log |m|}{\log \left( \frac{1 + \sqrt{5}}{2} \right)}} = \frac{1}{|m|}.
$$

\subsection{Transport Properties}

There is a substantial number of papers that investigate the transport exponents associated with the Fibonacci Hamiltonian; see, for example, \cite{BLS, D98, D05, DEGT, DKL, DST, DT03, DT05, DT07, DT08, JL2, KKL}. While we won't describe all the known results, we want to at least highlight some of them and put them in perspective.  As pointed out earlier, one of the fascinating features of quasicrystal models is that the intermediate nature of their aperiodic order between periodic and random is reflected in a number of ways, be it through the spectrum (by spectral measures being purely singular continuous) or through transport behavior. Here we want to address the latter point. All the papers listed above have the goal of proving estimates that show that the transport properties of the Fibonacci Hamiltonian are markedly different from those of periodic or random media.

Since there is ballistic transport (all transport exponents are equal to one) in the periodic case and no transport (all transport exponents are equal to zero) in the random case, one therefore wants to show that the transport exponents take values in the open interval $(0,1)$. Proving non-trivial lower bounds turns out to be comparatively easier and was accomplished in the late 1990's \cite{D98, JL2} for zero phase. Several subsequent papers then went on to extend the lower bound to all phases and improved the estimates \cite{DEGT, DKL, DST, DT03, DT05, DT08}. Upper bounds for transport exponents, on the other hand, proved to be elusive for some time. Note a key difference here: to bound transport exponents from below, one ``only'' has to show that some portion of the wave packet moves sufficiently fast. On the other hand, to bound transport exponents from above, one essentially has to control the entire wave packet and show that it does not move too fast (i.e., ballistically). Thus, it is potentially easier to prove upper bounds on transport that are dual to the type of lower bound that had been established, and this indeed turned out to be the case. The papers \cite{D05, KKL} showed that at least some non-trivial portion of the wave packet moves slowly. Full control and hence genuine upper bounds for transport exponents were finally obtained in 2007 and later \cite{BLS, DT07, DT08}.

Let us now state some of the transport results explicitly. Some general remarks that should be made are the following:
\begin{itemize}

\item[(a)] Almost all results concern time-averaged quantities (i.e., the exponents $\tilde \beta^\pm(p)$ defined in  Section \ref{s.transport}).

\item[(b)] Most papers focus on the case $\psi(0) = \delta_0$. We will limit our attention here to this case as well.

\item[(c)] The optimality of the known estimates improves when $p$ and/or $\lambda$ are large. In particular, the bounds are known to be tight in the limit $\lambda,p \uparrow \infty$.

\item[(d)] For finite values of $\lambda$ and $p$, the method of choice to obtain the best known bound varies.

\item[(e)] For $\lambda$ and $p$ large enough, the transport exponent may exceed the dimension of the spectrum.

\end{itemize}

Here is a result from \cite{DT05} that establishes the best known estimates for zero phase and given $\lambda$ and $p$:

\begin{Thm}\label{t.fiblowerdynbound}
Suppose $\lambda > 0$ and set
$$
\gamma = D \log (2 + \sqrt{8 + \lambda^2})
$$
{\rm (}where $D$ is some universal constant{\rm )} and
$$
\kappa = \log \left[ \frac{\sqrt{17}}{20 \log (1 + \alpha)} \right].
$$
Then, the time-averaged transport exponent corresponding to the initial state $\psi(0) = \delta_0$ and zero-phase Fibonacci Hamiltonian $H_{\lambda,0}$ obey
\begin{equation}\label{e.fibbetabound}
\tilde \beta^\pm (p) \ge \begin{cases} \frac{p + 2 \kappa}{(p + 1) (\gamma + \kappa + 1/2)}, & p \le 2 \gamma + 1; \\[.25em] \frac{1}{\gamma + 1}, & p > 2 \gamma + 1. \end{cases}
\end{equation}
\end{Thm}

Here is a result from \cite{DT07, DT08} that concerns the regime of large $\lambda$ and $p$:

\begin{Thm}
Consider the Fibonacci Hamiltonian $H_{\lambda,\omega}$ and the initial state $\psi(0) = \delta_0$. For $\lambda > \sqrt{24}$, we have
$$
\tilde \alpha_u^\pm \ge \frac{2 \log (1+\alpha)}{\log (2\lambda + 22)},
$$
and for $\lambda \ge 8$, we have
$$
\tilde \alpha_u^\pm \le \frac{2 \log (1+\alpha)}{\log \left( \frac12 \left[ (\lambda - 4) + \sqrt{(\lambda - 4)^2 - 12} \, \right] \right)}.
$$
Both estimates holds uniformly in $\omega$. In particular,
$$
\lim_{\lambda \to \infty} \tilde \alpha_u^\pm \cdot \log \lambda = 2 \log (1 + \alpha),
$$
and convergence is uniform in $\omega$.
\end{Thm}

In fact, the upper bound can be proved also for the non-time-averaged quantities, as shown in \cite{DT08}.

\begin{Thm}
Consider the Fibonacci Hamiltonian $H_{\lambda,\omega}$ and the initial state $\psi(0) = \delta_0$. For $\lambda \ge 8$ and uniformly in $\omega$, we have
$$
\alpha_u^\pm \le \frac{2 \log (1+\alpha)}{\log \left( \frac12 \left[ (\lambda - 4) + \sqrt{(\lambda - 4)^2 - 12} \, \right] \right)}.
$$
\end{Thm}

Some other estimates on transport exponents were obtained recently using different methods in \cite{DGY14}.

\subsection{Connections between Spectral Characteristics and Dynamical Quantities}

In \cite{DGY14} explicit relations between spectral quantities for the Fibonacci Hamiltonian and the dynamical characteristics of the Fibonacci trace map were obtained. In the next theorem, $\mu_{\lambda,\mathrm{max}}$ denotes the measure of maximal entropy of $T_\lambda|_{\Omega_\lambda}$ and $\mu_\lambda$ denotes the equilibrium measure of $T_\lambda|_{\Omega_\lambda}$ that corresponds to the potential $- \dim_H \Sigma_\lambda \cdot \log \|DT_\lambda|_{E^u}\|$. Recall that $\alpha$ denotes the inverse of the golden ratio.

\begin{Thm}\label{t.identities}
For every $\lambda > 0$, we have
\begin{align}
\tilde \alpha^\pm_u(\lambda) & = \frac{\log (1+\alpha)}{\inf_{p \in {\rm Per}(T_\lambda)} \mathrm{Lyap}^u(p)}, \label{e.transportexponentidentity} \\
\dim_H \Sigma_\lambda & = \frac{h_{\mu_\lambda}}{\mathrm{Lyap}^u \mu_\lambda}, \label{e.spectrumidentity} \\
\dim_H \nu_\lambda & = \dim_H \mu_{\lambda,\mathrm{max}} = \frac{h_\mathrm{top}(T_\lambda)}{\mathrm{Lyap}^u \mu_{\lambda,\mathrm{max}}} = \frac{\log (1+\alpha)}{\mathrm{Lyap}^u \mu_{\lambda,\mathrm{max}}}, \label{e.doesmeasureidentity} \\
\gamma_\lambda & = \frac{\log (1+\alpha)}{\sup_{p \in {\rm Per}(T_\lambda)} \mathrm{Lyap}^u(p)}. \label{e.hoelderexponentidentity}
\end{align}
\end{Thm}

The following theorem from \cite{DGY14} shows that for the Fibonacci Hamiltonian and every value of the coupling constant, the four quantities satisfy strict inequalities.

\begin{Thm}\label{t.strictinequalities}
For every $\lambda > 0$, we have
\begin{equation}\label{e.inequalities}
\gamma_\lambda < \dim_H \nu_\lambda < \dim_H \Sigma_\lambda < \tilde \alpha^\pm_u(\lambda).
\end{equation}
\end{Thm}

The particular inequality $\dim_H \nu_\lambda < \dim_H \Sigma_\lambda$ in \eqref{e.inequalities} establishes a conjecture of Barry Simon,\footnote{The conjecture does not appear anywhere in print, but it was popularized by Barry Simon in many talks given by him in the past four years.} which was made based on an analogy with work of Makarov and Volberg \cite{Mak, MV, V}; see \cite{DG3} for a more detailed discussion.
The inequality
\begin{equation}\label{e.dimspectranspexp}
\dim_H \Sigma_\lambda < \tilde \alpha^\pm_u(\lambda)
\end{equation}
in \eqref{e.inequalities} is related to a question of Yoram Last. He asked in \cite{L96} whether in general $\dim_H \Sigma_\lambda$ bounds $\tilde \alpha^\pm_u(\lambda)$ from above and conjectured that the answer is no. The inequality \eqref{e.dimspectranspexp} confirms this. See  \cite{DT08} and  \cite{DG5} for earlier partial results.

\bigskip

The identities in Theorem~\ref{t.identities} are instrumental in the proof of Theorem~\ref{t.strictinequalities}. Indeed, once the identities \eqref{e.transportexponentidentity}--\eqref{e.hoelderexponentidentity} are established, Theorem~\ref{t.strictinequalities} can be proved using the thermodynamic formalism, which we will describe next. Define $\phi : \Omega_\lambda \to \R$ by $\phi(x) = -\log \|DT_\lambda (x)|_{E^u}\|$ and consider the pressure function (sometimes called the Bowen function) $P : t \mapsto P(t\phi)$, where $P(\psi)$ is the topological pressure.\footnote{There are many classical books on the thermodynamical formalism; for example, \cite{Bow, Ru, W}. We also refer the reader to the recent introductory texts \cite{B11, Iommi, Sa}.} This function has been heavily studied; the next statement summarizes some known results; compare \cite{Bow, Kel, PP, Ru, Wal, W1}.

\begin{Prop}\label{p.thermodynamic}
Suppose that $\sigma_A : \Sigma_A \to \Sigma_A$ is a topological Markov chain defined by a transitive $0$--$1$ matrix $A$, and $\phi : \Sigma_A \to \R$ is a H\"older continuous function.
Denote by $\frak{M}$ the space of $\sigma_A$-invariant Borel probability measures.
Then, the following statements hold.
\begin{itemize}

\item[{\rm (1)}] Variational principle: $P(t\phi) = \sup_{\mu \in \frak{M}} \left\{ h_\mu + t \int \phi \, d\mu \right\}$.

\item[{\rm (2)}] For every $t \in \R$, there exists a unique invariant measure $\mu_t \in \frak{M}$ {\rm (}the equilibrium state{\rm )} such that $P(t\phi) = h_{\mu_t} + t \int \phi \, d\mu_t$.

\item[{\rm (3)}] $P(t\phi)$ is a real analytic function of $t$.

\item[{\rm (4)}] If $\phi$ is cohomological to a constant, then $P(t\phi)$ is a linear function; if $\phi$ is not cohomological to a constant, then $P(t\phi)$ is strictly convex and decreasing.

\item[{\rm (5)}] For every $t_0 \in \R$, the line $h_{\mu_{t_0}} + t \int \phi \, d\mu_{t_0}$ is tangent to the graph of the function $P(t\phi)$ at the point $(t_0, P(t_0\phi))$.

\item[{\rm (6)}] The following limits exist:
$$
\lim_{t \to \infty} \int \phi \, d\mu_t = \sup_{\mu \in \frak{M}} \int \phi \, d\mu, \ \ \ \  \lim_{t \to -\infty} \int \phi \, d\mu_t = \inf_{\mu \in \frak{M}} \int \phi \, d\mu.
$$
The graph of the function $t \mapsto P(t\phi)$ lies strictly above each of the lines $t\cdot \sup_{\mu \in \frak{M}} \int \phi \, d\mu$ and $t\cdot \inf_{\mu \in \frak{M}} \int \phi \, d\mu$.

\end{itemize}
\end{Prop}

\begin{figure}[t]
\includegraphics[scale=.7]{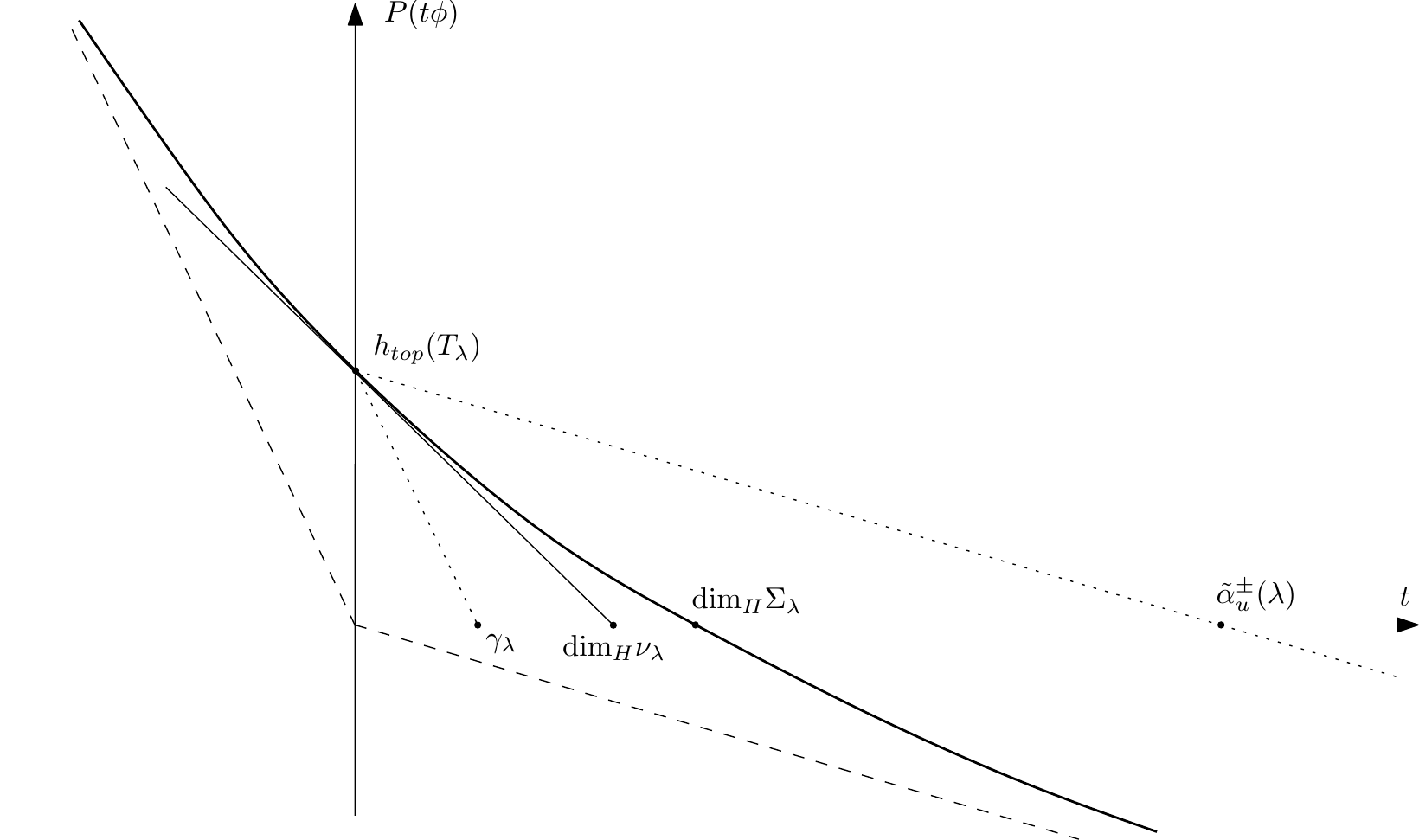}
\caption{Pressure function and spectral characteristics of the Fibonacci Hamiltonian.}
\label{fig:thermodyn}
\end{figure}

Now let us return to our case where $\sigma_A : \Sigma_A \to \Sigma_A$ is conjugate to $T_\lambda|_{\Omega_\lambda}$ and the potential is given by $\phi(x) = -\log \|DT_\lambda (x)|_{E^u}\|$ (suppressing the conjugacy). In \cite{DGY14} it was shown that this potential is not cohomological to a constant. For any $t \in \R$, consider the tangent line to the graph of $P(t)$ at the point $(t, P(t\phi))$. Since $P(t)$ is decreasing, there exists exactly one point of intersection of the tangent line with the $t$-axis, at the point $t_0 = -\frac{h_{\mu_t}}{\int \phi\, d\mu} = \frac{h_{\mu_t}}{Lyap^u\,\mu_t} = \mathrm{dim}_H\mu_t$. The last equality here is due to \cite{Man}. In particular, $\mathrm{dim}_H \mu_\mathrm{max} = \mathrm{dim}_H \nu_\lambda$ is given by the point of intersection of the tangent line to the graph of $P(t)$ at the point $(0, h_{top}(T_\lambda))$ with the $t$-axis. Also, due to Theorem~\ref{t.identities} the line $h_\mathrm{top} (T_\lambda) + t \cdot \inf_{\mu \in \frak{M}} \int \phi \, d\mu$ intersects the $t$-axis at the point $\gamma_\lambda$, and the line $h_\mathrm{top} (T_\lambda) + t \cdot \sup_{\mu \in \frak{M}} \int \phi \, d\mu$ intersects the $t$-axis at the point $\tilde \alpha^\pm_u(\lambda)$. Finally, due to \cite{MM}, the graph of $P(t)$ intersects the $t$-axis at the point $\mathrm{dim}_H\Sigma_\lambda$. These observations are illustrated in Figure~\ref{fig:thermodyn} and explain where the strict inequalities in Theorem~\ref{t.strictinequalities} come from once it is shown that $\phi$ is not cohomological to a constant.

\subsection{Square and Cubic Fibonacci Hamiltonians} \label{sec:2d3d}

Since spectral questions for Schr\"odinger operators in two (and higher) dimensions are hard to study, it is natural to consider a model where known one-dimensional results can be used. In particular, let us consider the Schr\"odinger operator
\begin{align}\label{e.sham}
[H^{(2)}_{\lambda_1, \lambda_2, \omega_1, \omega_2}& \psi] (m,n) =  \psi(m+1,n) + \psi(m-1,n) + \psi(m,n+1) + \psi(m,n-1) + \\  + &\left(\lambda_1 \chi_{[1-\alpha , 1)}(m\alpha +\omega_1 \!\!\! \mod 1) + \lambda_2\chi_{[1-\alpha , 1)}(n\alpha +\omega_2\!\!\! \mod 1) \right) \psi(m,n) \nonumber
\end{align}
in $\ell^2(\Z^2)$. The theory of tensor products of Hilbert spaces and operators then implies that $\sigma(H^{(2)}_{\lambda_1, \lambda_2, \omega_1, \omega_2}) = \Sigma_{\lambda_1} + \Sigma_{\lambda_2}$ for all $\omega_1, \omega_2$. This operator and its spectrum have been studied numerically and heuristically by Even-Dar Mandel and Lifshitz in a series of papers \cite{EL06, EL07, EL08} (a similar model was studied by Sire in \cite{Si89}). Their study suggested that at small coupling, $\Sigma_{\lambda_1} + \Sigma_{\lambda_2}$ is not a Cantor set; quite on the contrary, it has no gaps at all.

It turns out that this is indeed the case \cite{DG2}:

\begin{Thm}\label{t.4}
For $\lambda_1, \lambda_2 > 0$ sufficiently small, $\sigma(H^{(2)}_{\lambda_1, \lambda_2, \omega_1, \omega_2}) = \Sigma_{\lambda_1} + \Sigma_{\lambda_2}$ is an interval.
\end{Thm}

This result follows from the estimates for the thickness of $\Sigma_\lambda$ from Theorem~\ref{t.2} and Newhouse's Gap Lemma (Theorem~\ref{t.sumofcantorsets}).

Theorem~\ref{t.4} should be contrasted with the following result, which is an immediate consequence of Corollary~\ref{c.dyndefcantorset} and Theorem~\ref{t.DEGT}.

\begin{Thm}\label{t.5}
For $\lambda_1, \lambda_2 > 0$ sufficiently large, $\sigma(H^{(2)}_{\lambda_1, \lambda_2, \omega_1, \omega_2}) = \Sigma_{\lambda_1} + \Sigma_{\lambda_2}$ is a Cantor set.
\end{Thm}

The same statements hold for the cubic Fibonacci Hamiltonian (i.e., the analogously defined Schr\"odinger operator in $\ell^2(\Z^3)$ with spectrum $\Sigma_{\lambda_1} + \Sigma_{\lambda_2}+ \Sigma_{\lambda_3}$). Section~\ref{sec:2d3dcomp} shows numerical illustrations of the finite approximations $\Sigma_{k,\lambda} + \Sigma_{k,\lambda}$ and $\Sigma_{k,\lambda}+\Sigma_{k,\lambda}+\Sigma_{k,\lambda}$, along with an exploration of the number of disjoint intervals that make up these sets.

Moreover, the density of states measure of the family $\{ H^{(2)}_{\lambda_1, \lambda_2, \omega_1, \omega_2} \}_{ \lambda_j \in \R, \omega_j \in \T }$ can be expressed as the convolution of the density of states measures associated with the families $\{ H_{\lambda_1, \omega_1} \}_{\omega_1 \in \T}$ and $\{ H_{\lambda_2, \omega_2} \}_{\omega_2 \in \T}$, that is,
\begin{equation}\label{e.2ddos}
\nu^{(2)}_{\lambda_1, \lambda_2} = \nu_{\lambda_1} \ast \nu_{\lambda_2}.
\end{equation}
See the appendix in \cite{DGS} for further background on separable potentials and operators.
The following result was obtained by Damanik, Gorodetski and Solomyak in \cite{DGS}.

\begin{Thm}\label{cor.main}
Let $\nu^{(2)}_{{\lambda_1}, \lambda_2}$ be the density of states measure for the Square Fibonacci Hamiltonian \eqref{e.sham} with coupling constants ${\lambda_1}, \lambda_2$. There is $\lambda^*>0$ such that 
for almost every pair $(\lambda_1, \lambda_2) \in [0, \lambda^*)\times [0,\lambda^*)$, the measure $\nu^{(2)}_{{\lambda_1}, \lambda_2}$ is absolutely continuous with respect to Lebesgue measure.
\end{Thm}

In fact, it follows from the proof that {\rm (}with a uniform smallness condition{\rm )} for every  $\lambda_1\in [0, \lambda^*)$, the measure $\nu^{(2)}_{{\lambda_1}, \lambda_2}$ is absolutely continuous with respect to the Lebesgue measure for almost every $\lambda_2\in [0, \lambda^*)$.

\section{Sturmian Potentials}\label{s.6}

The Fibonacci potential is a special case of a Sturmian potential. The latter are obtained if $\alpha$ in the definition of the potential, $V(n) = \lambda \chi_{[1-\alpha,1)}(n \alpha + \omega \!\! \mod 1)$, is a general irrational number in $(0,1)$. The Fibonacci case corresponds to the choice $\alpha = \frac{\sqrt{5}-1}{2}$.

Given an irrational $\alpha \in (0,1)$, consider its continued fraction expansion
$$
\alpha = \cfrac{1}{a_1+ \cfrac{1}{a_2+ \cfrac{1}{a_3 + \cdots}}}
$$
with uniquely determined $a_k \in \Z_+$. Truncating the continued fraction expansion of $\alpha$ after $k$ steps yields the rational number $p_k/q_k$, which is the best rational approximant of $\alpha$ with denominator bounded by $q_{k+1} - 1$. The following recursions hold:
\begin{align*}
p_{k+1} & = a_{k+1} p_k + p_{k-1} , \quad p_0 = 0 , \; p_1 = 1, \\
q_{k+1} & = a_{k+1} q_k + q_{k-1} , \quad q_0 = 1 , \; q_1 = a_1.
\end{align*}
(In the Fibonacci case $\alpha = \frac{\sqrt{5}-1}{2}$, we have $a_k \equiv 1$ and $p_k/q_k = F_{k-1}/F_k$.) A number of the results for the Fibonacci Hamiltonian described in the previous section have been generalized to the Sturmian case under suitable assumptions on the continued fraction coefficients $\{ a_k \}$. In this section, we explain what these results are, and how the proofs had to be modified.

\subsection{Extension of the Trace Map Formalism}

Let us the denote the discrete Schr\"odinger operator on $\ell^2(\Z)$ with potential $V(n) = \lambda \chi_{[1-\alpha,1)}(n \alpha + \omega \!\! \mod 1)$ by $H_{\lambda,\alpha,\omega}$. Strong approximation again shows that the spectrum of $H_{\lambda,\alpha,\omega}$ does not depend on $\omega$, and may therefore be denoted by $\Sigma_{\lambda,\alpha}$. The one-step transfer matrices associated with the difference equation $H_{\lambda,\alpha,\omega} u = E u$ are given by
$$
T_{\lambda,\alpha,\omega}(m,E) = \begin{pmatrix} E - \lambda \chi_{[1-\alpha,1)}(m \alpha + \omega \!\!\!\! \mod 1) & -1 \\ 1 & 0 \end{pmatrix}.
$$

The matrices
$$
M_{-1}(E) = \begin{pmatrix} 1 & -\lambda \\ 0 & 1 \end{pmatrix} , \quad M_0(E) = \begin{pmatrix} E & -1 \\ 1 & 0 \end{pmatrix},
$$
and
$$
M_k(E) = T_{\lambda,\alpha,0}(q_k,E) \times \cdots \times T_{\lambda,\alpha,0}(1,E) \quad \text{ for } k \ge 1
$$
obey the recurrence relations
$$
M_{k+1}(E) = M_{k-1}(E) M_k(E)^{a_{k+1}}
$$
for $k \ge 0$; see \cite[Proposition~1]{BIST89}. Passing to the variables
$$
x_k(E) = \frac12 \mathrm{Tr} M_k(E),
$$
this in turn implies via the Cayley-Hamilton theorem that $x_{k+1}(E)$ can be expressed as an explicit function of (suitable Chebyshev polynomials applied to) $x_k(E), x_{k-1}(E), x_{k-2}(E)$ for $k \ge 1$; see \cite[Proposition~2]{BIST89}. These recursion relations exhibit the same conserved quantity as before; namely, with
$$
\tilde x_{k+1}(E) =  \frac12 \mathrm{Tr} (M_k(E) M_{k-1}(E)),
$$
we have
$$
\tilde x_{k+1}(E)^2 + x_k(E)^2 + x_{k-1}(E)^2 - 2 \tilde x_{k+1}(E) x_k(E) x_{k-1}(E) - 1 = \frac{\lambda^2}{4}
$$
for every $k \ge 0$; see \cite[Proposition~3]{BIST89}.

\subsection{Results Obtained via an Analysis of the Trace Recursions}

Notice that the key difference with the Fibonacci case is that, in general, the sequence of traces may not be obtained by iterating a single map. In this sense, there is in general no direct analog of the trace map. However, as we have just seen, the underlying structure of recurrence relations extends nicely. The substitute for the dynamical analysis of the Fibonacci trace map will have to lie in studying the dynamics of an initial point under the successive application of a sequence of maps, the elements of which are dictated by the continued fraction expansion of $\alpha$. These developments are still in their early stages. In the following we will concentrate on the known results that can be established by simply exploiting the recurrence relations, without employing sophisticated tools from dynamical systems theory.

\bigskip

The first result that establishes a clean analogy with the Fibonacci case is the following analog of Theorem~\ref{spectrum}, which was established in \cite{BIST89}.

\bthm\label{bistspectrum}
Fix $\lambda > 0$ and $\alpha \in (0,1)$ irrational. An energy $E$ belongs to the spectrum $\Sigma_{\lambda,\alpha}$ if and only if the sequence $\{ x_k(E) \}$ is bounded.
\ethm

The proof of Theorem~\ref{bistspectrum} follows the same line of reasoning as the proof of Theorem~\ref{spectrum}, which was outlined in the previous section. In particular, one obtains canonical covers of the spectrum, which are useful in the estimation of its dimension. Let us make this explicit. As before, define the sets
$$
\sigma_{\lambda,\alpha,k} = \{ E \in \R : |x_k(E)| \le 1 \}
$$
and
$$
\Sigma_{\lambda,\alpha,k} = \sigma_k \cup \sigma_{k+1}.
$$
The same reasoning shows that the sets $\Sigma_{\lambda,\alpha,k}$ are decreasing in $k$ and the spectrum is the limiting set, that is,
$$
\Sigma_{\lambda,\alpha} = \bigcap_{k \ge 1} \Sigma_{\lambda,\alpha,k};
$$
see \cite[Proposition~4]{BIST89}.

\bigskip

A refinement of this description of the spectrum in the Sturmian case due to Raymond \cite{Ra} allowed Liu and Wen to obtain the following estimates for the Hausdorff dimension of the spectrum in the large coupling regime \cite{LW}.

\begin{Thm}\label{t.qinghua1}
Suppose $\lambda > 20$ and $\alpha \in (0,1)$ is irrational with continued fraction coefficients $\{ a_k \}$. Denote
$$
M_* = \liminf_{k \to \infty} (a_1 \cdots a_k)^{1/k} \in [1,\infty].
$$

{\rm (a)} If $M_* = \infty$, then $\dim_H \Sigma_{\lambda,\alpha} = 1$.

{\rm (b)} If $M_* < \infty$, then $\dim_H \Sigma_{\lambda,\alpha}$ belongs to the open interval $(0,1)$ and obeys the estimates
$$
\dim_H \Sigma_{\lambda,\alpha} \le \frac{2 \log M_* + \log 3}{2 \log M_* - \log \frac{3}{\lambda - 8}}
$$
and
$$
\dim_H \Sigma_{\lambda,\alpha} \ge \max \left\{ \frac{\log 2}{10 \log 2 - 3 \log \frac{1}{4(\lambda - 8)}} , \frac{\log M_* - \log 3}{\log M_* - \log \frac{1}{12(\lambda - 8)}} \right\}.
$$
\end{Thm}

A study of the box counting dimension of $\Sigma_{\lambda,\alpha}$ in the case of bounded $\{ a_k \}$ was carried out in the follow-up paper \cite{FLW} by Fan, Liu, and Wen. Among other things, they showed that for $\lambda > 20$, the Hausdorff dimension and the box counting dimension of $\Sigma_{\lambda,\alpha}$ coincide whenever the sequence $\{ a_k \}$ is eventually periodic. The analysis of the case of unbounded $\{ a_k \}$ was carried out by Liu, Qu, and Wen in \cite{LQW}. On the one hand, these papers establish the following companion result to Theorem~\ref{t.qinghua1}.

\begin{Thm}\label{t.qinghua2}
Suppose $\lambda \ge 24$ and $\alpha \in (0,1)$ is irrational with continued fraction coefficients $\{ a_k \}$. Denote
$$
M^* = \limsup_{k \to \infty} (a_1 \cdots a_k)^{1/k} \in [1,\infty].
$$

{\rm (a)} If $M^* = \infty$, then $\dim_B^+ \Sigma_{\lambda,\alpha} = 1$.

{\rm (b)} If $M^* < \infty$, then $\dim_B^+ \Sigma_{\lambda,\alpha}$ belongs to the open interval $(0,1)$.
\end{Thm}

Here $\dim_B^+ S$ denotes the upper box counting dimension of the set $S$. Note that Theorems~\ref{t.qinghua1} and \ref{t.qinghua2} imply in particular that for suitable choices of $\alpha$ and $\lambda$, we may have $\dim_H \Sigma_{\lambda,\alpha} < 1$ and $\dim_B^+ \Sigma_{\lambda,\alpha} = 1$.

On the other hand, Liu, Qu, and Wen also study in \cite{LQW} the large coupling asymptotics of these dimensions. Namely they show that the limits $\lim_{\lambda \to \infty} \dim_H \Sigma_{\lambda,\alpha} \cdot \log \lambda$ and $\lim_{\lambda \to \infty} \dim_B^+ \Sigma_{\lambda,\alpha} \cdot \log \lambda$ exist, and provide a description of these limits.

\bigskip

The transport exponents in the Sturmian case were studied in the papers \cite{D98, DKL, DST, DT05, M10}. The following result from \cite{DT05} gives dynamical lower bounds for all values of $\lambda$ and $p$, provided $\alpha$ has bounded continued fraction coefficients.

\begin{Thm}
Suppose $\lambda > 0$ and $\alpha \in (0,1)$ is irrational with $a_k \le C$. With
$$
\gamma = D \, \log (2 + \sqrt{8 + \lambda^2}) \cdot \limsup_{n \to \infty} \frac{1}{n} \sum_{k = 1}^n a_k
$$
{\rm (}where $D$ is some universal constant{\rm )} and
$$
\kappa = \frac{\log (\sqrt{17} / 4)}{(C+1)^5},
$$
the transport exponents associated with the operator $H_{\lambda,\alpha,0}$ and the initial state $\psi(0) = \delta_0$ obey
$$
\tilde \beta^-(p) \ge \begin{cases} \frac{p+2\kappa}{(p+1)(\gamma + \kappa + 1/2)}, & p \le 2 \alpha+1; \\ \frac{1}{\gamma+1}, & p > 2 \alpha +1. \end{cases}
$$
\end{Thm}

The following result from \cite{M10} gives dynamical upper bounds in the large coupling regime.

\begin{Thm}
Suppose $\lambda > 20$ and $\alpha \in (0,1)$ is irrational with continued fraction coefficients $\{ a_k \}$ and corresponding rational approximants $\{ p_k/q_k \}$. Denote
$$
D = \limsup_{k \to \infty} \frac1k \log q_k.
$$
Then, the transport exponents associated with the operator $H_{\lambda,\alpha,0}$ and the initial state $\psi(0) = \delta_0$ obey
$$
\tilde \alpha_u^\pm \le \frac{2D}{\log \frac{\lambda - 8}{3}}.
$$
Moreover, if $a_k \ge 2$ for all $k$, then
$$
\tilde \alpha_u^\pm \le \frac{D}{\log \frac{\lambda - 8}{3}}.
$$
\end{Thm}

\section{Numerical Results and Computational Issues}\label{s.7}

In this section, we provide numerical illustrations of a number of
the results described in this survey.  These calculations focus on
the Fibonacci Hamiltonian, though many could readily be adapted to
the Sturmian potentials described in the last section.
We begin by studying approximations to the spectrum for the
Fibonacci model in one dimension, then investigate estimates
of the integrated density of states based on
spectra of finite sections of the operator.
Finally, we address upper bounds on the spectrum in
two and three dimensions.
In all cases, we set the phase $\omega$ to zero.

\subsection{Spectral Approximations for the Fibonacci Hamiltonian}

We begin by calculating the spectrum $\sigma_k$ for the $k$th periodic
approximations to the Fibonacci potential.  The analysis described in
Section~\ref{s.5} suggests several ways to compute $\sigma_k$, which
turn out to have varying degrees of utility.

Given a candidate energy $E$, one can test if $E\in \sigma_k$ by
iterating the trace recurrence~(\ref{e.tracerec}) and testing if
$|x_k(E)|\le 1$.  In principle, this simple approach enables
investigation for arbitrarily large values of $k$.
However, two key obstacles restrict the utility of this method:
(i) it does not readily yield the entire set $\sigma_k$;
(ii) as $k$ increases, the intervals that comprise $\sigma_k$
become exponentially narrow, beyond the resolution of the
standard floating point number system in which such calculations
are typically performed.
However, this approach can yield some useful results, particularly
in the small coupling regime where the decay of the interval widths
is most gradual, or when one is only interested in some narrow set
of energy values. (This method of calculation was used to produce
illustrations in~\cite{DG2}.)

To obtain the entire set $\sigma_k$, one might instead use the
recurrence~(\ref{e.tracerec}) to construct the degree-$F_k$
polynomial $x_k(E)$, then determine the regions where
$|x_k(E)|\le 1$ by finding the zeros of the polynomials
$x_k(E)+1$ and $x_k(E)-1$ using a standard root-finding
algorithm.  For all but the smallest $k$ this approach is
untenable.  Coefficients of $x_k(E)$ grow exponentially
in $k$; e.g., for $\lambda=4$,
\begin{eqnarray*}
   x_6(E) \!\!\!&=&\!\!\! {\textstyle{1\over 2}} E^{13} - 16 E^{12} + \textstyle{435\over 2} E^{11} - 1616 E^{10} + {13905\over2} E^9
            - 16272 E^8 + 13330 E^7 \\
    && \hspace*{-1em}{}\textstyle + 20160 E^6 - 37133 E^5 - 17056 E^4 + {61013 \over 2} E^3 + 25104 E^2 + {13021\over2} E + 560.
\end{eqnarray*}
The magnitude of these coefficients,
compounded by the proximity of the roots for larger values of $\lambda$ and $k$,
leads to inaccurate root calculations,
a phenomenon well studied by numerical analysts; see, e.g. \cite{Mos86,Wil84}.
Indeed, it is not uncommon for the computed roots to be so inaccurate as to
have significant spurious imaginary parts.

There is a more robust approach to computing the approximate
Fibonacci spectrum $\sigma_k$.
One can view $\sigma_k$ as the exact spectrum of a related
Schr\"odinger operator with a potential having period $F_k$.
The spectrum of this operator is the union of $F_k$ non-degenerate
intervals whose endpoints are given by the eigenvalues of the
two $F_k$-dimensional matrices $J_{k+}$ and $J_{k-}$:
\[ J_{k\pm} = \left(\begin{array}{ccccc}
                   v_{1,k} & 1 & & & \pm 1 \\
                    1 & v_{2,k} & \ddots \\
                      & \ddots & \ddots & \ddots \\
                      & & \ddots & v_{F_k-1,k} & 1 \\
                   \pm1   & & & 1 & v_{F_k,k}
               \end{array}\right),\]
with unspecified entries beyond the tridiagonal section equal to zero;
see, e.g., \cite[Ch.~7]{Tes00}.
Here the potential values $v_{n,k}$ are given by
\begin{equation} \label{eq:fibpot}
    v_{n,k} = \lambda \chi_{[1-F_{k-1}/F_k,1)}(n F_{k-1}/F_k\ {\rm mod}\ 1).
\end{equation}
This approach, which we use for the computations described below,
has also been employed in the context of Fibonacci computations
by Even-Dar Mandel and Lifshitz~\cite{EL06},
and for the almost Mathieu operator by Lamoureux~\cite{Lam97}.

The standard procedure for computing all the eigenvalues of a symmetric
matrix begins by applying a unitary similarity transformation to reduce
the matrix to symmetric tridiagonal form.%
\footnote{Methods such as the Lanczos algorithm excel at computing
\emph{a few} eigenvalues of large symmetric matrices~\cite[Ch.~13]{Par98}.
These methods are not feasible here, for \emph{all} eigenvalues
of $J_{k\pm}$ are required.  However, if one is only interested in
a narrow band of energies, these methods can be highly effective.}
Floating-point arithmetic introduces errors into this process, resulting
in the exact tridiagonal reduction of a matrix that differs
from the intended matrix by a factor that scales with the precision of the
floating point arithmetic system, the coupling constant $\lambda$,
and the dimension $F_k$.
The eigenvalues of this tridiagonal matrix are then approximated
to high accuracy via a procedure known as QR iteration~\cite{Par98}.
Remarkably, this iteration does not introduce significant errors beyond
those incurred by the reduction to tridiagonal form; for a discussion
of this accuracy, see \cite{And99,WR71}. Overall, this process requires
$O(F_k^3)$ floating point arithmetic operations and the storage of
$O(F_k^2)$ floating point numbers.  (The conventional procedure for
reducing the matrix to tridiagonal form destroys the zero structure
present in $J_{k\pm}$.)
Of course, the upper estimate
$\Sigma_{k,\lambda} = \sigma_{k,\lambda} \cup \sigma_{k+1,\lambda}$
then requires computation of all eigenvalues of four matrices.

\begin{figure}
\begin{center}
   \includegraphics[scale=0.3]{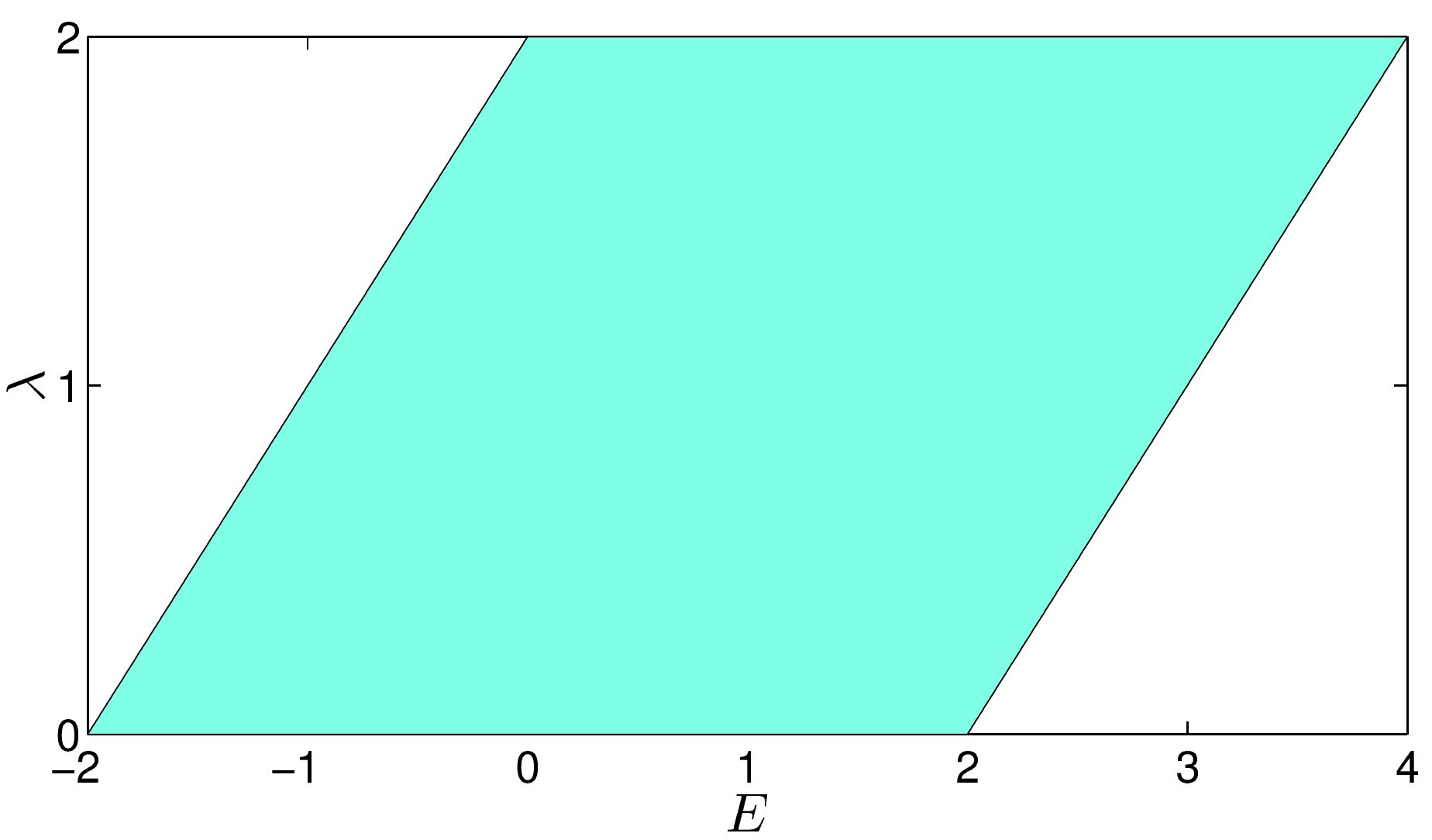}\quad
   \includegraphics[scale=0.3]{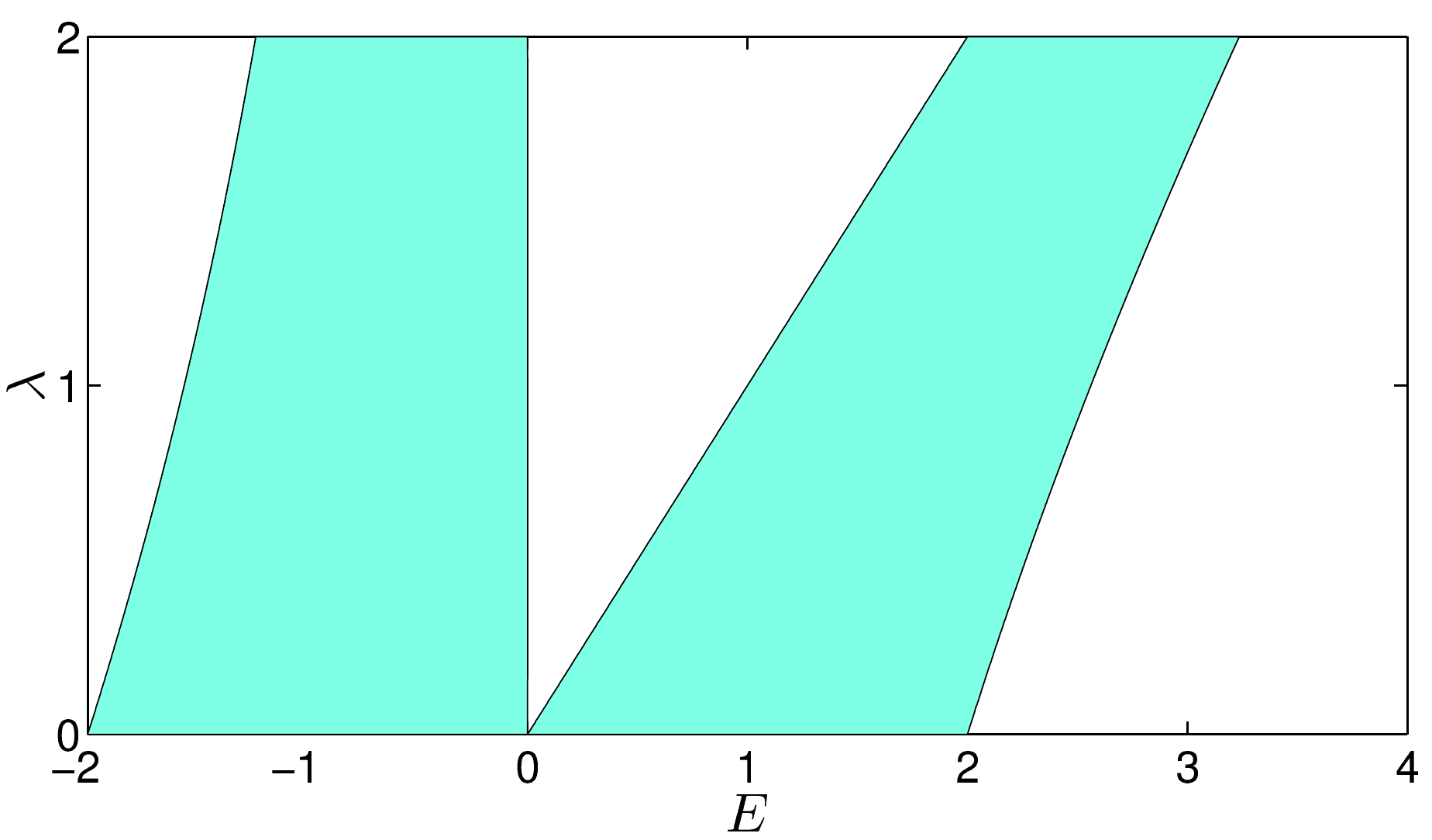}

   \includegraphics[scale=0.3]{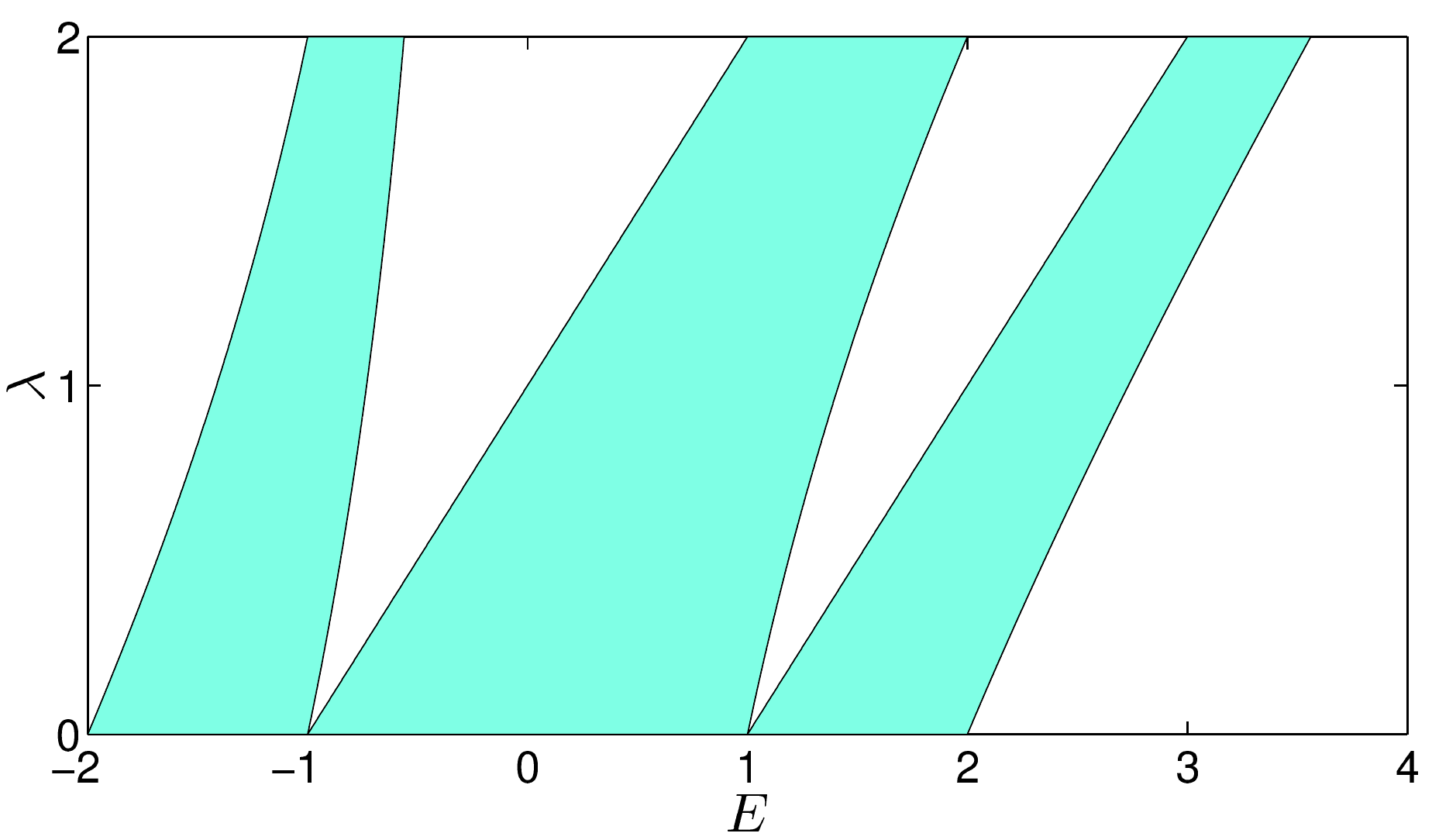}\quad
   \includegraphics[scale=0.3]{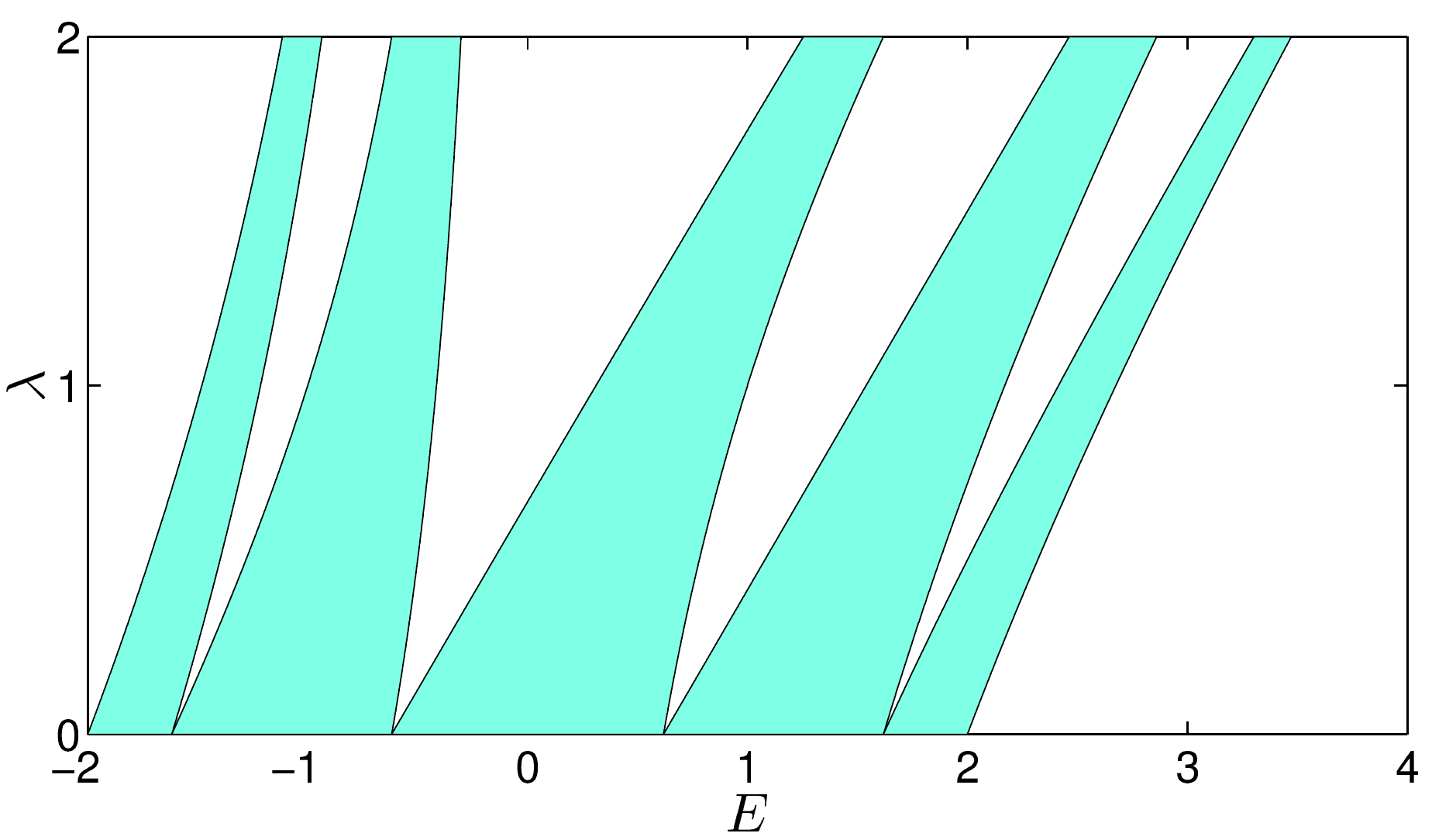}

   \includegraphics[scale=0.3]{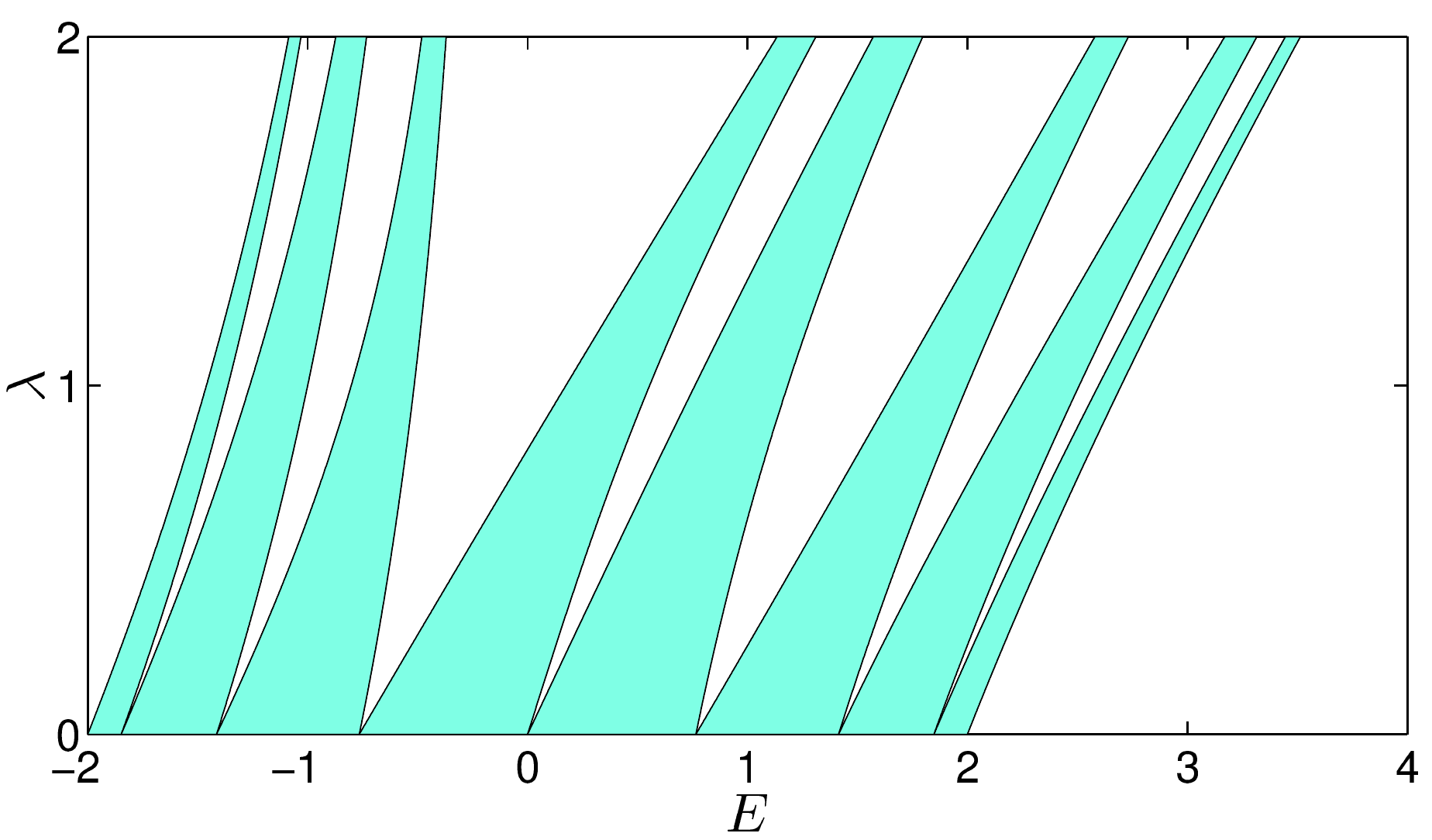}\quad
   \includegraphics[scale=0.3]{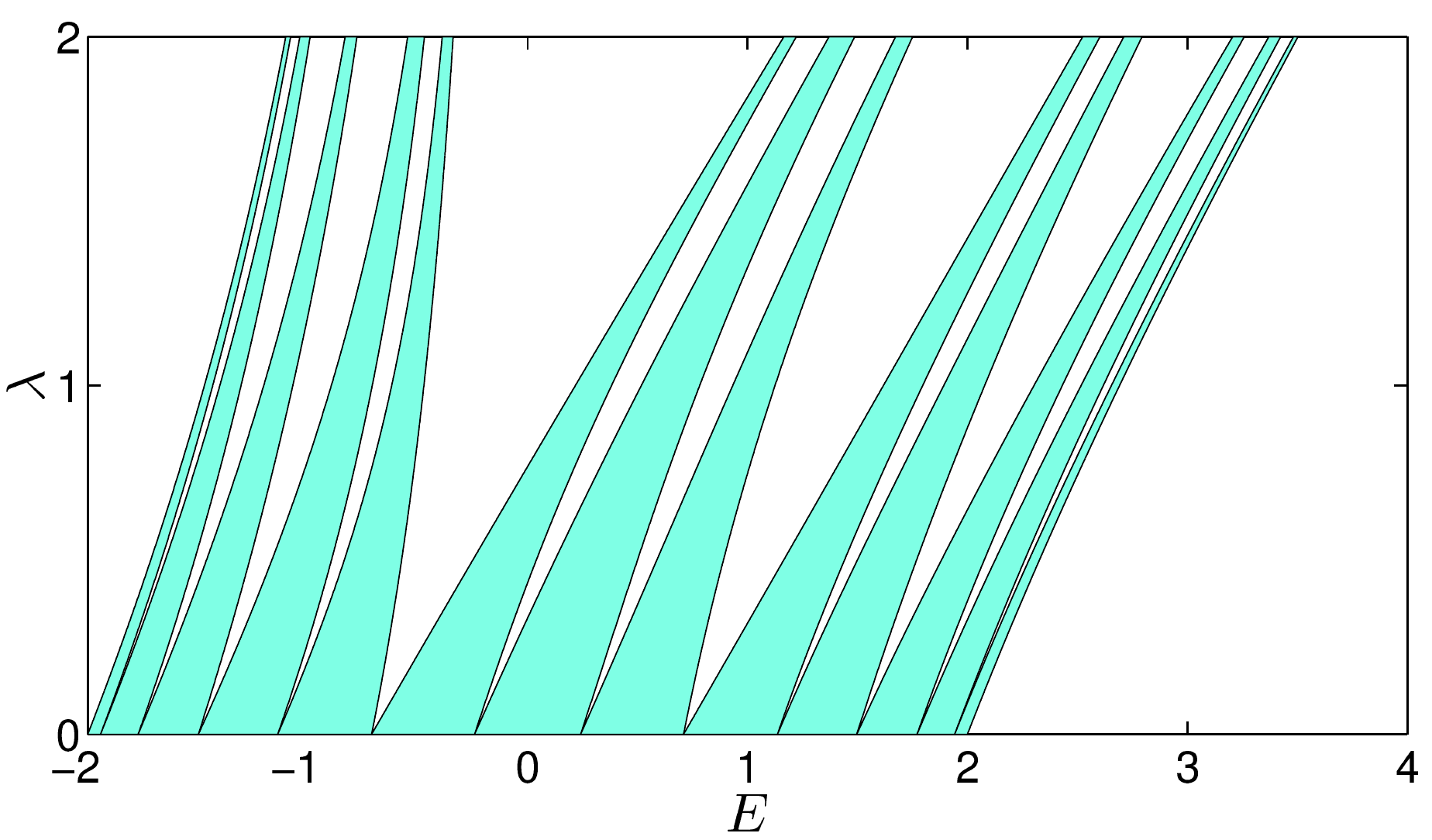}

   \includegraphics[scale=0.3]{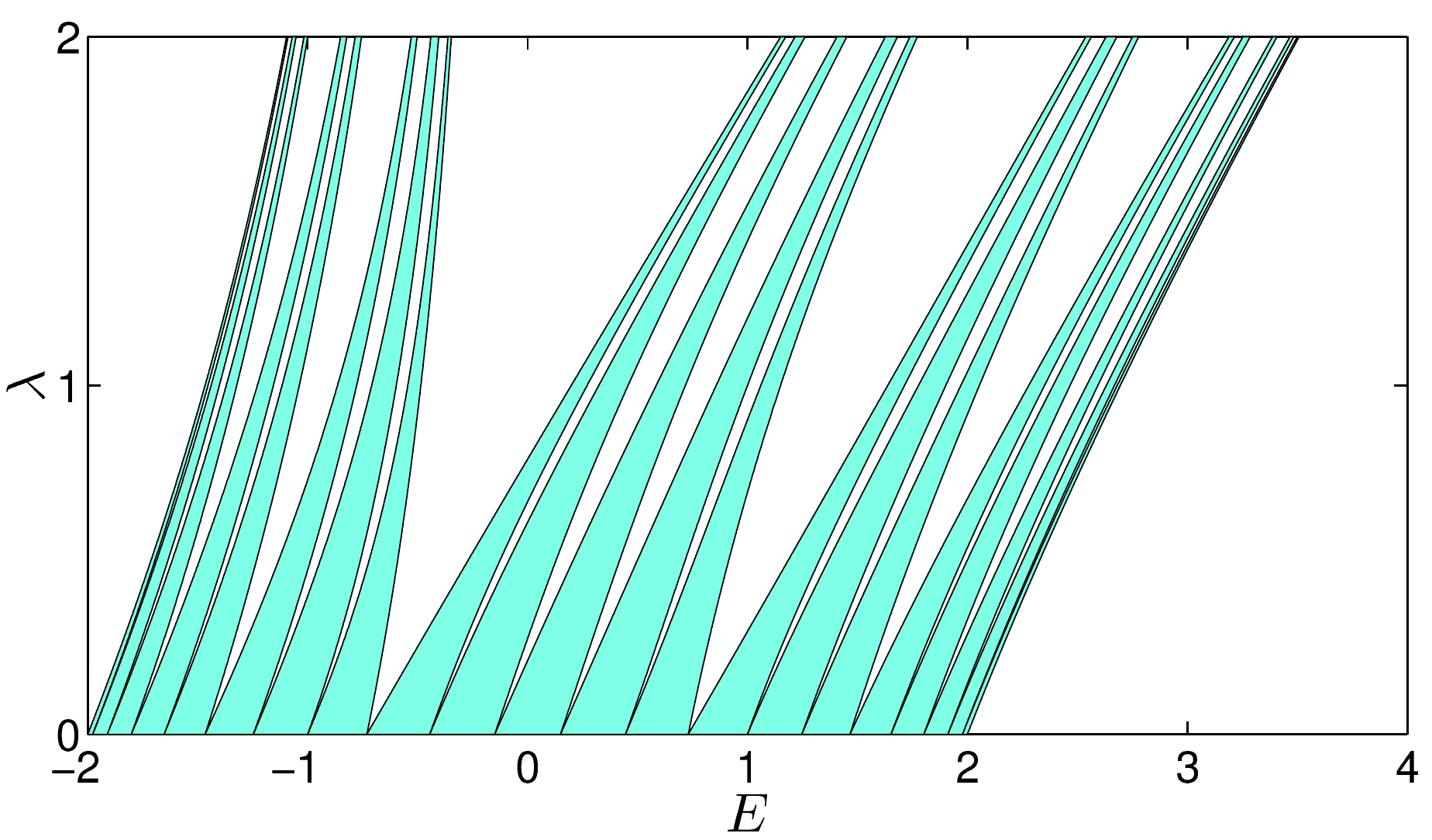}\quad
   \includegraphics[scale=0.3]{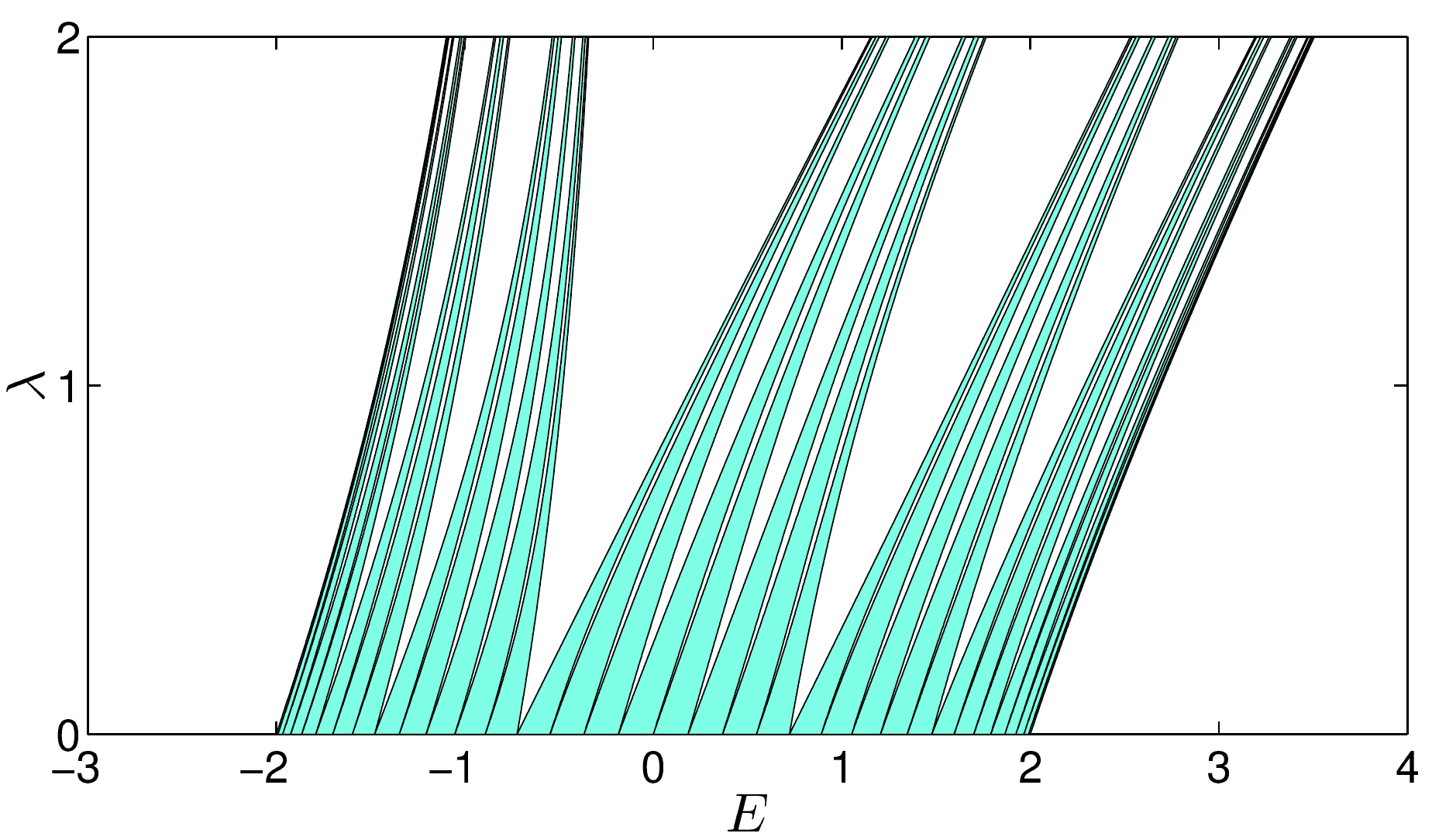}

\begin{picture}(0,0)
\put(-35,28){$k=7$} \put(135,28){$k=8$}
\put(-35,123){$k=5$} \put(135,123){$k=6$}
\put(-35,218){$k=3$} \put(135,218){$k=4$}
\put(-35,312){$k=1$} \put(135,312){$k=2$}
\end{picture}
\end{center}

\vspace*{-1.5em}
\caption{\label{fig:sig}
Spectra of the periodic approximations $\sigma_{k,\lambda}$ for the Fibonacci Hamiltonian,
as a function of $\lambda\in[0,2]$.
For $k=8$ and all $\lambda>0$, $\sigma_{k,\lambda}$ is the union of $F_8 = 34$ disjoint intervals.
}
\end{figure}
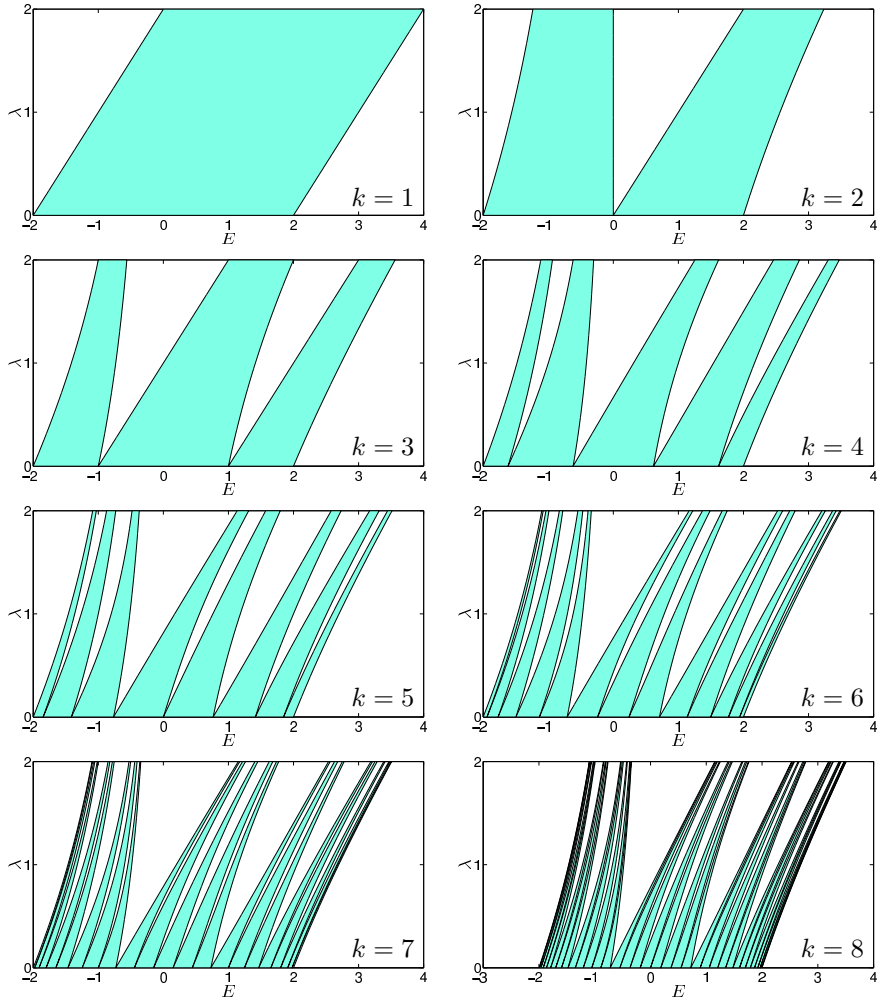

\begin{figure}
\begin{center}
   \includegraphics[scale=0.3]{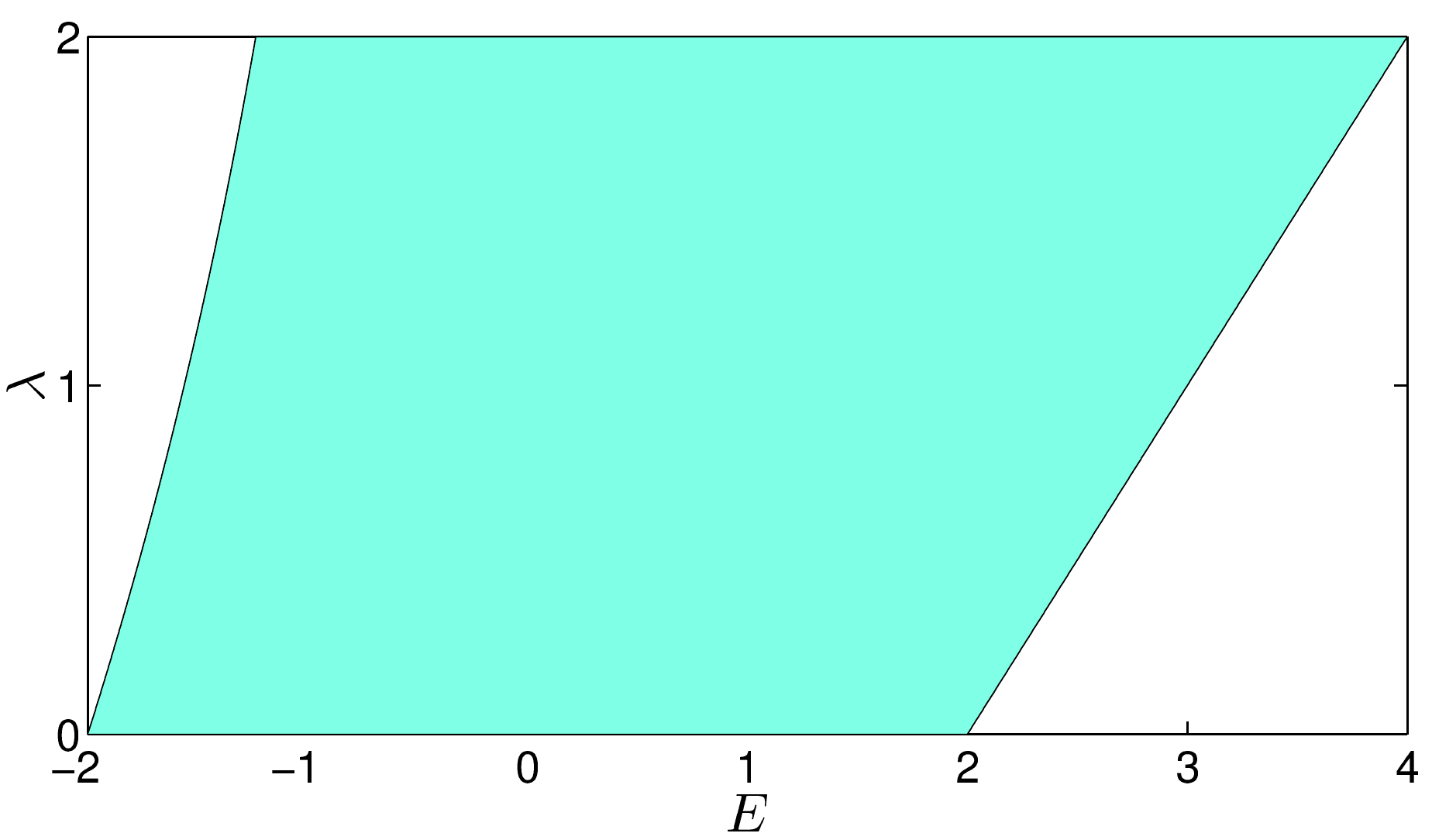}\quad
   \includegraphics[scale=0.3]{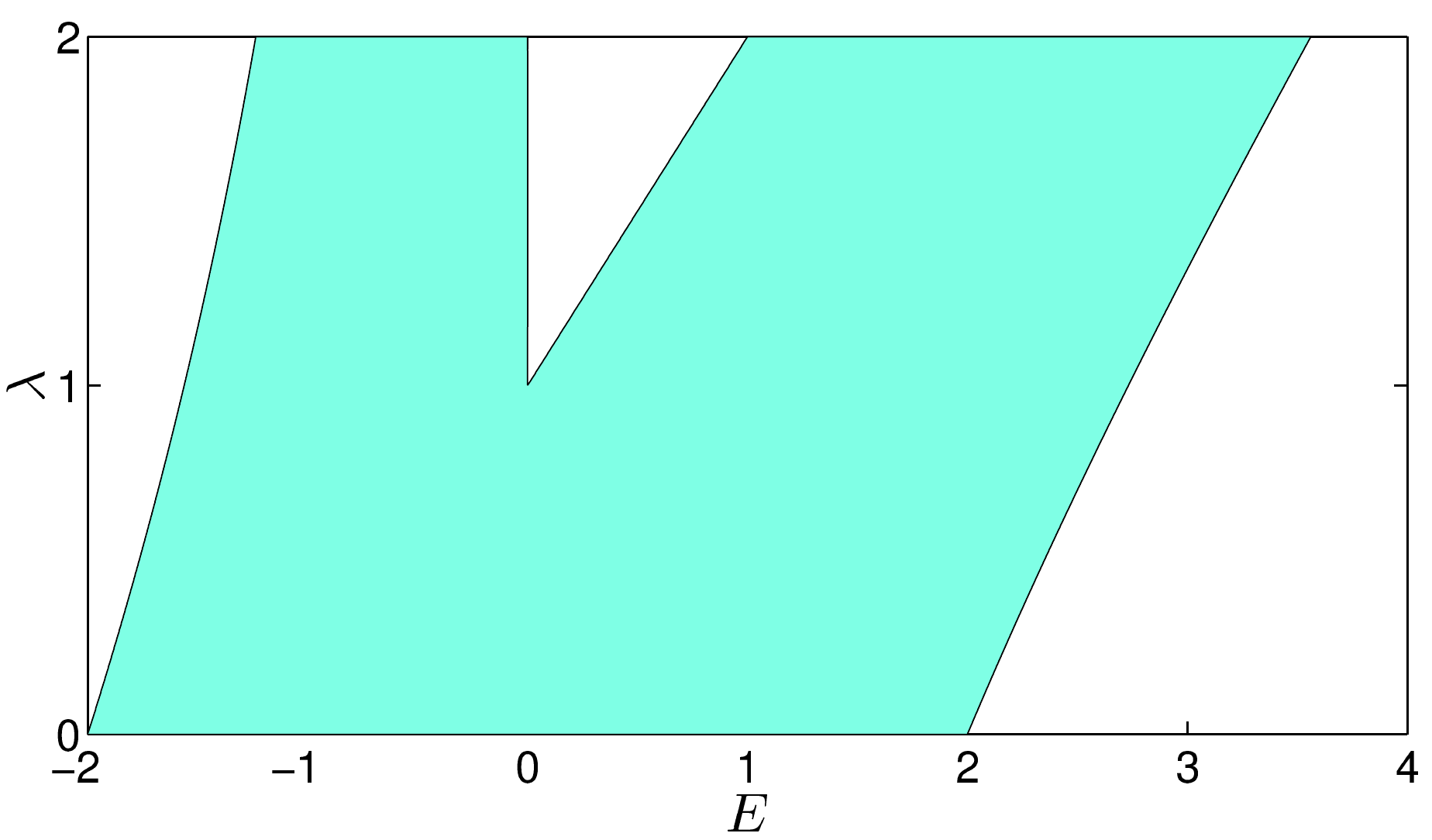}

   \includegraphics[scale=0.3]{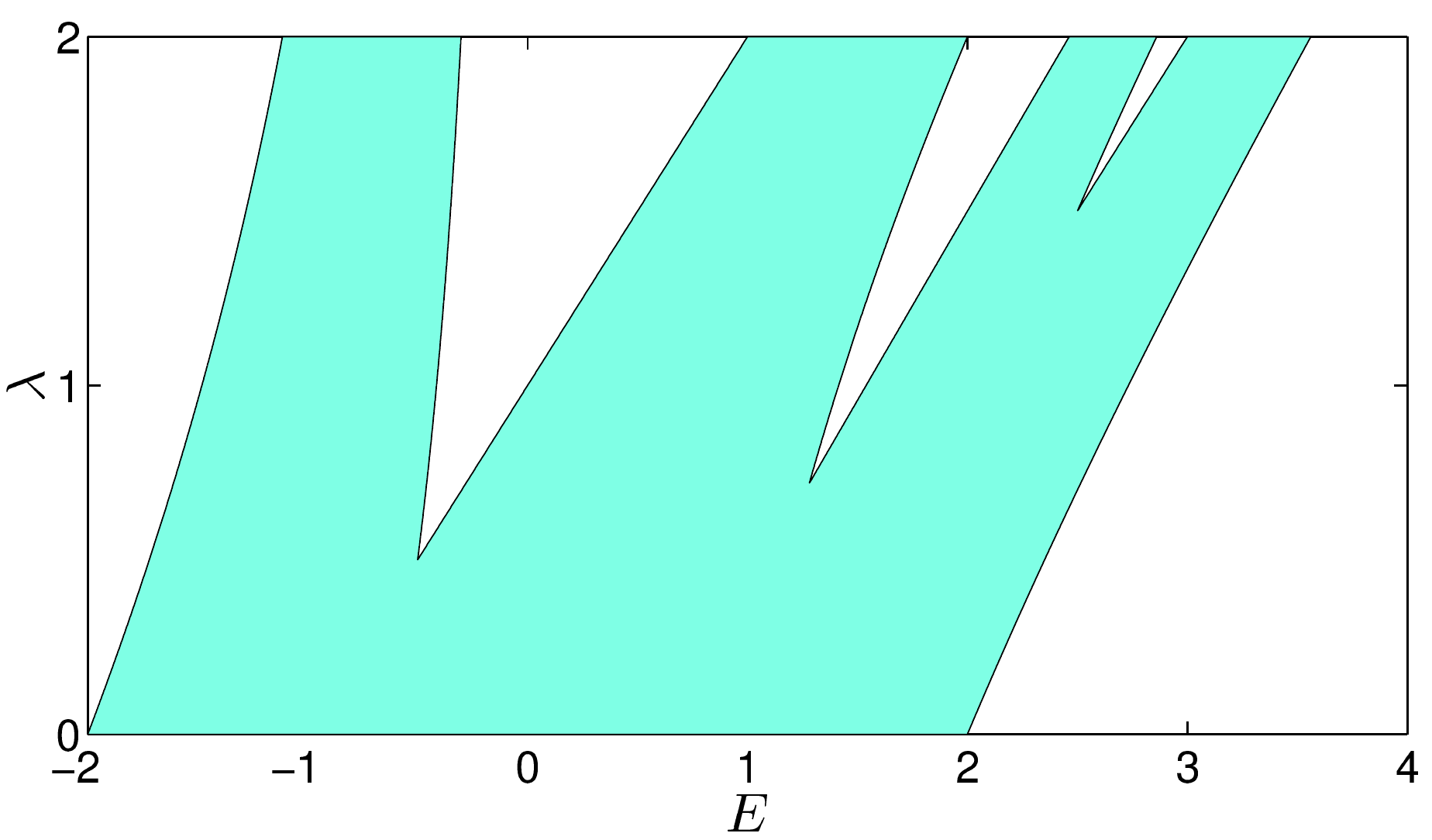}\quad
   \includegraphics[scale=0.3]{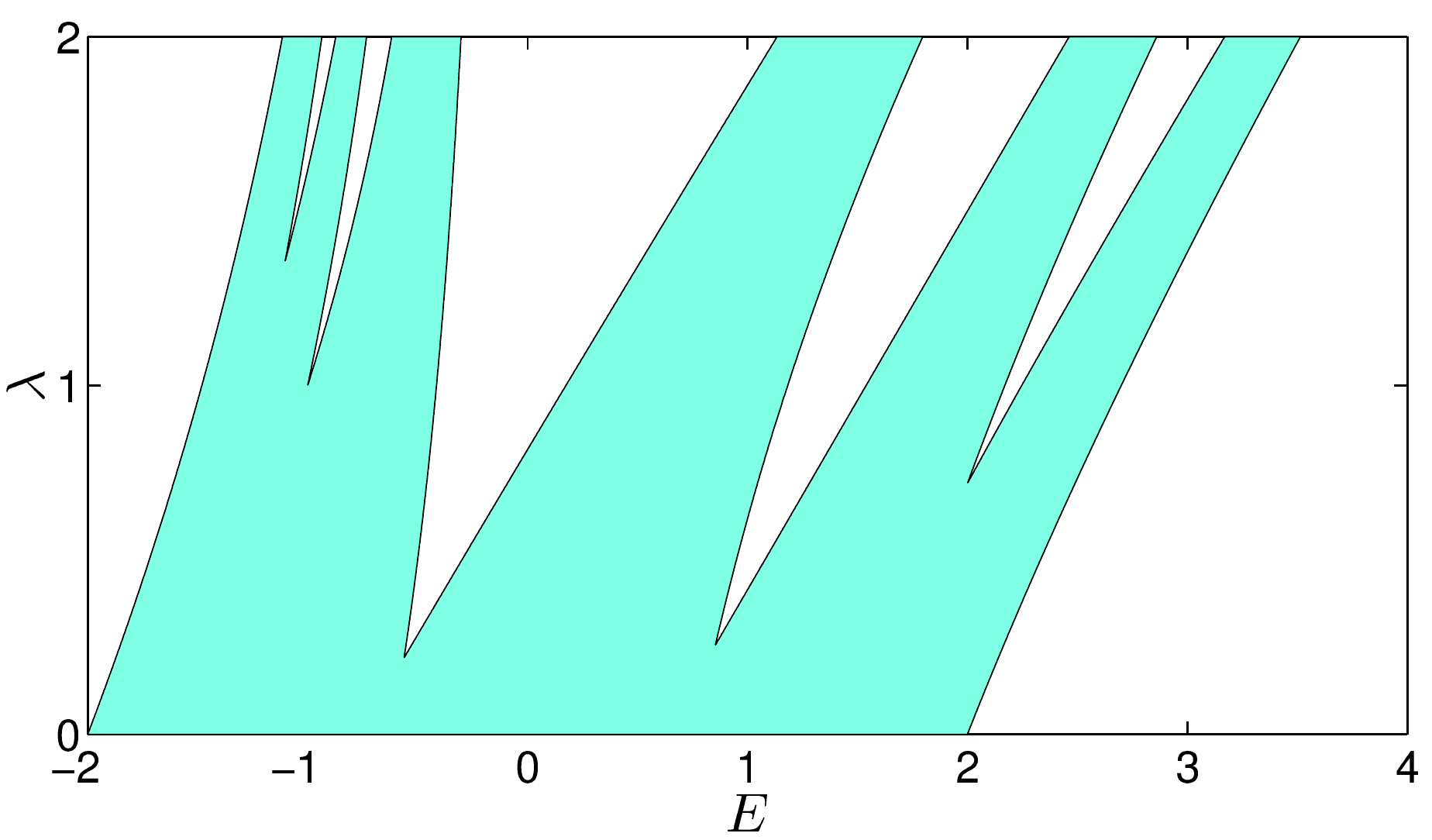}

   \includegraphics[scale=0.3]{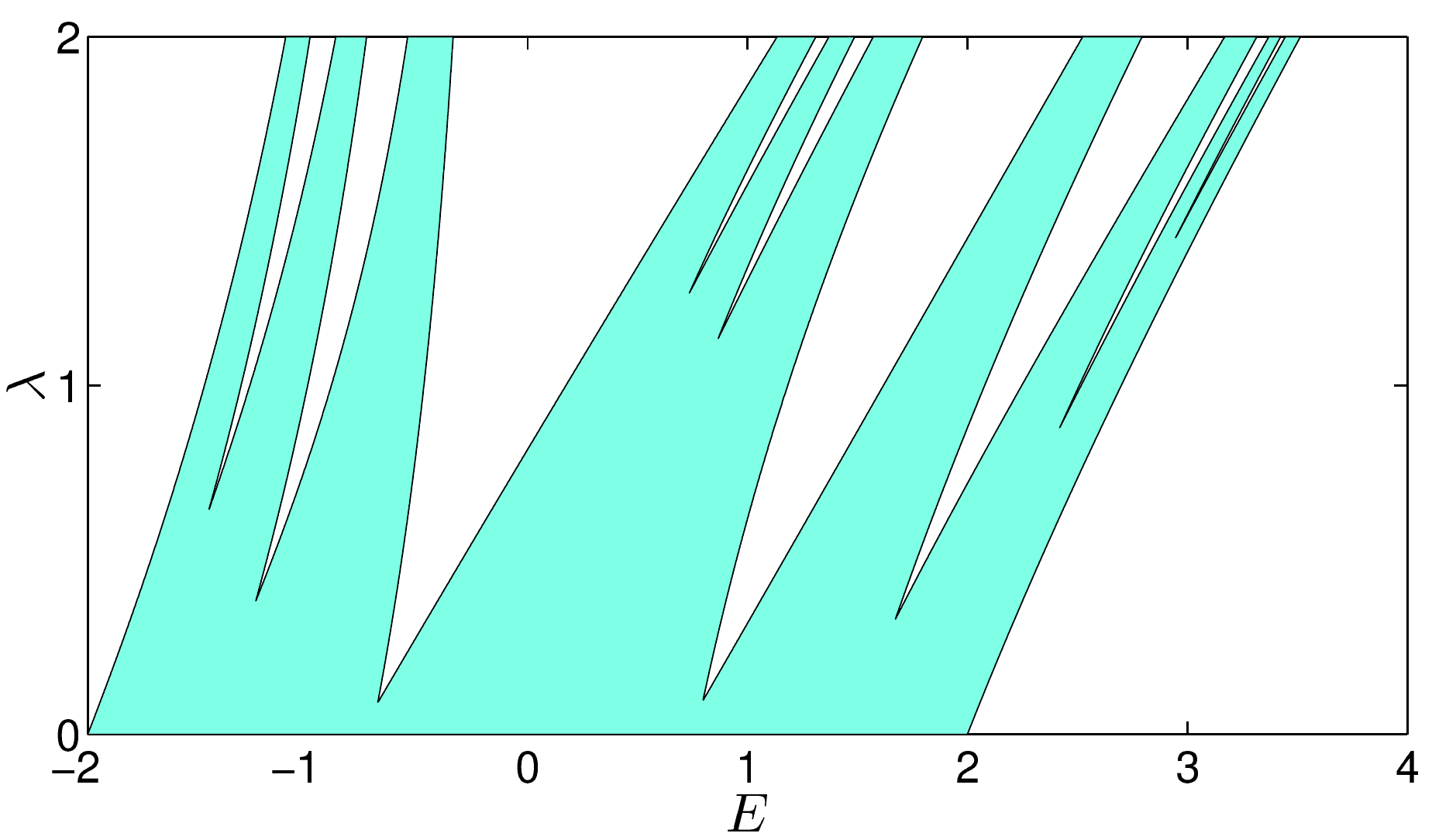}\quad
   \includegraphics[scale=0.3]{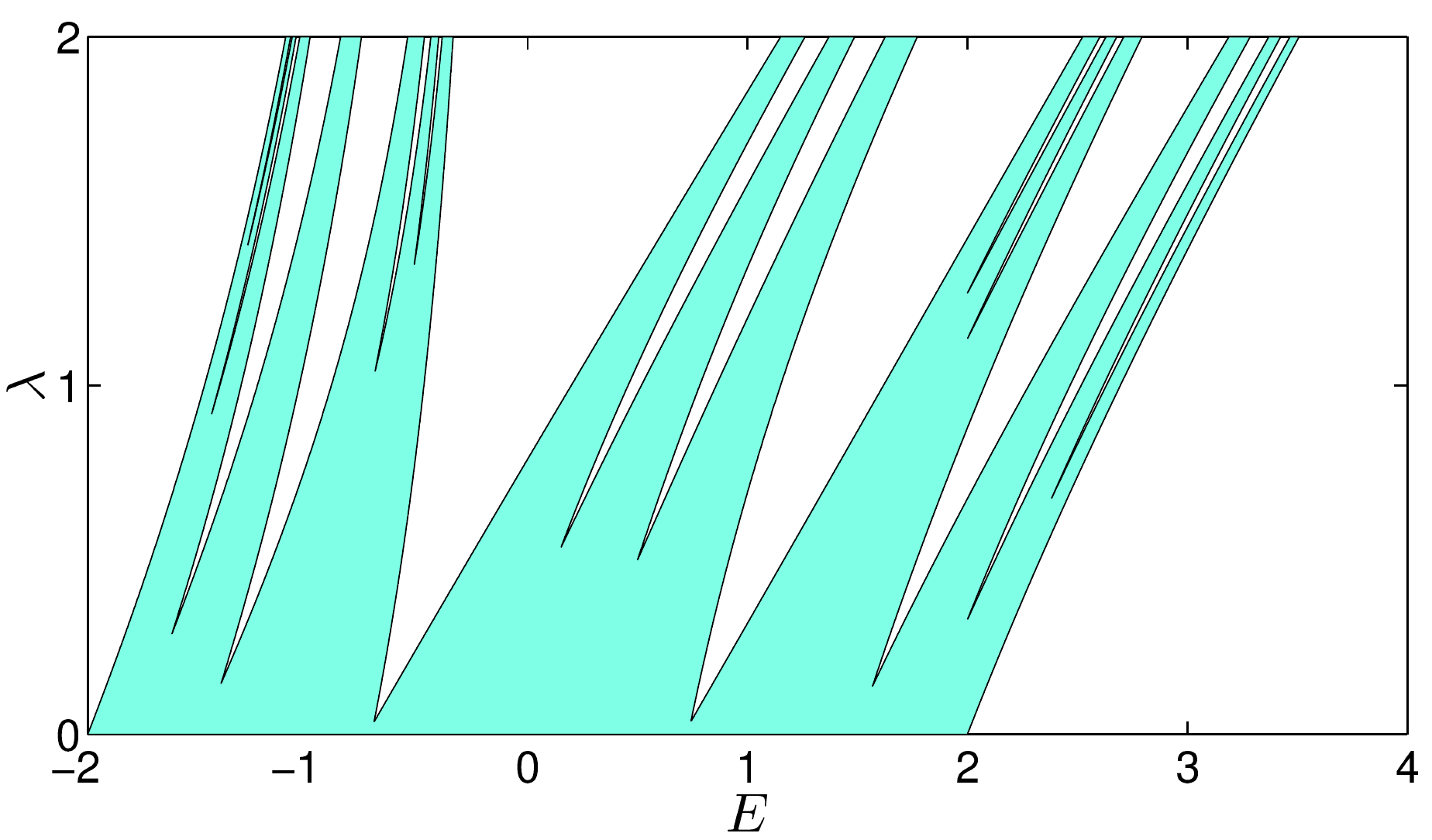}

   \includegraphics[scale=0.3]{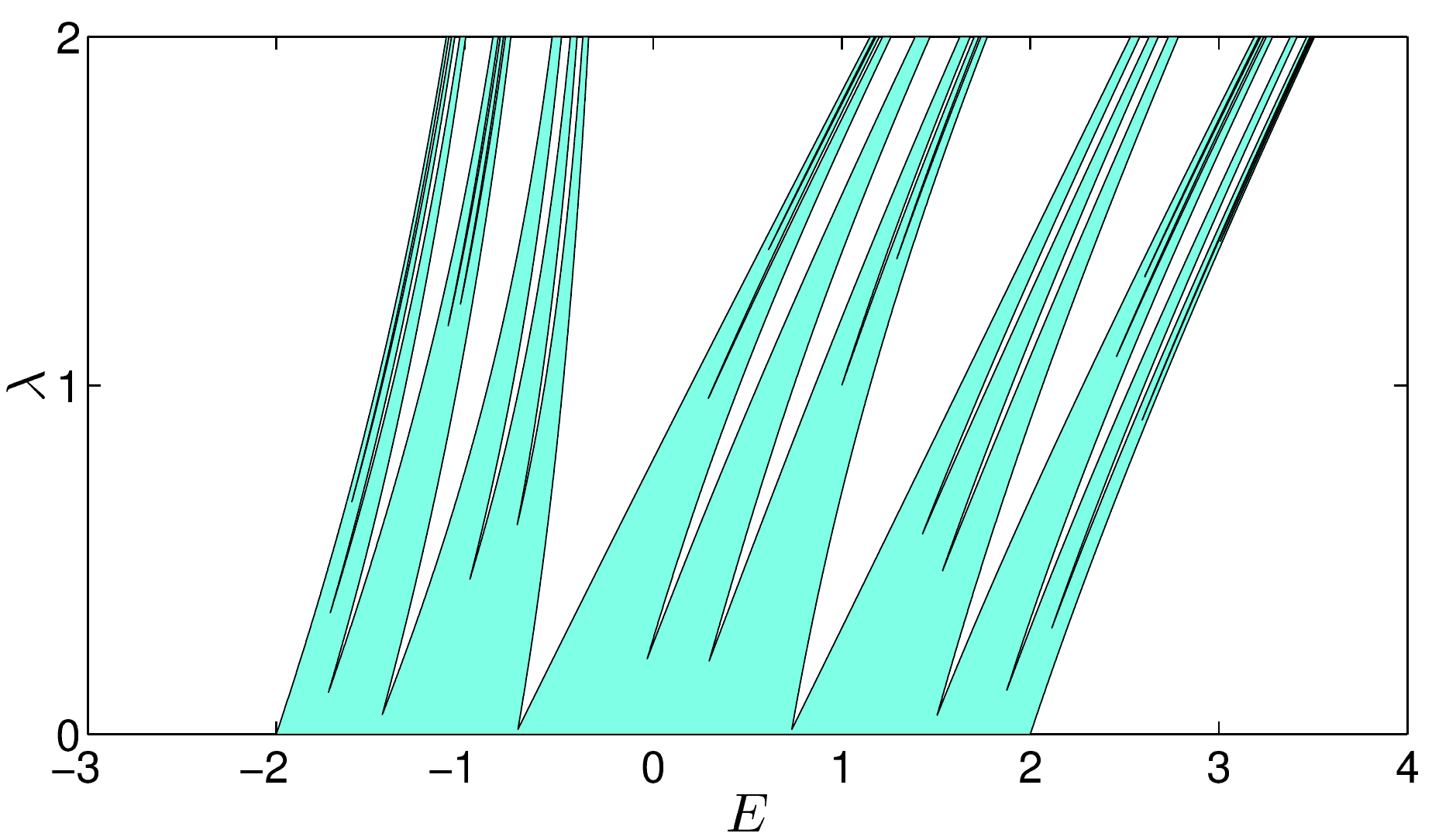}\quad
   \includegraphics[scale=0.3]{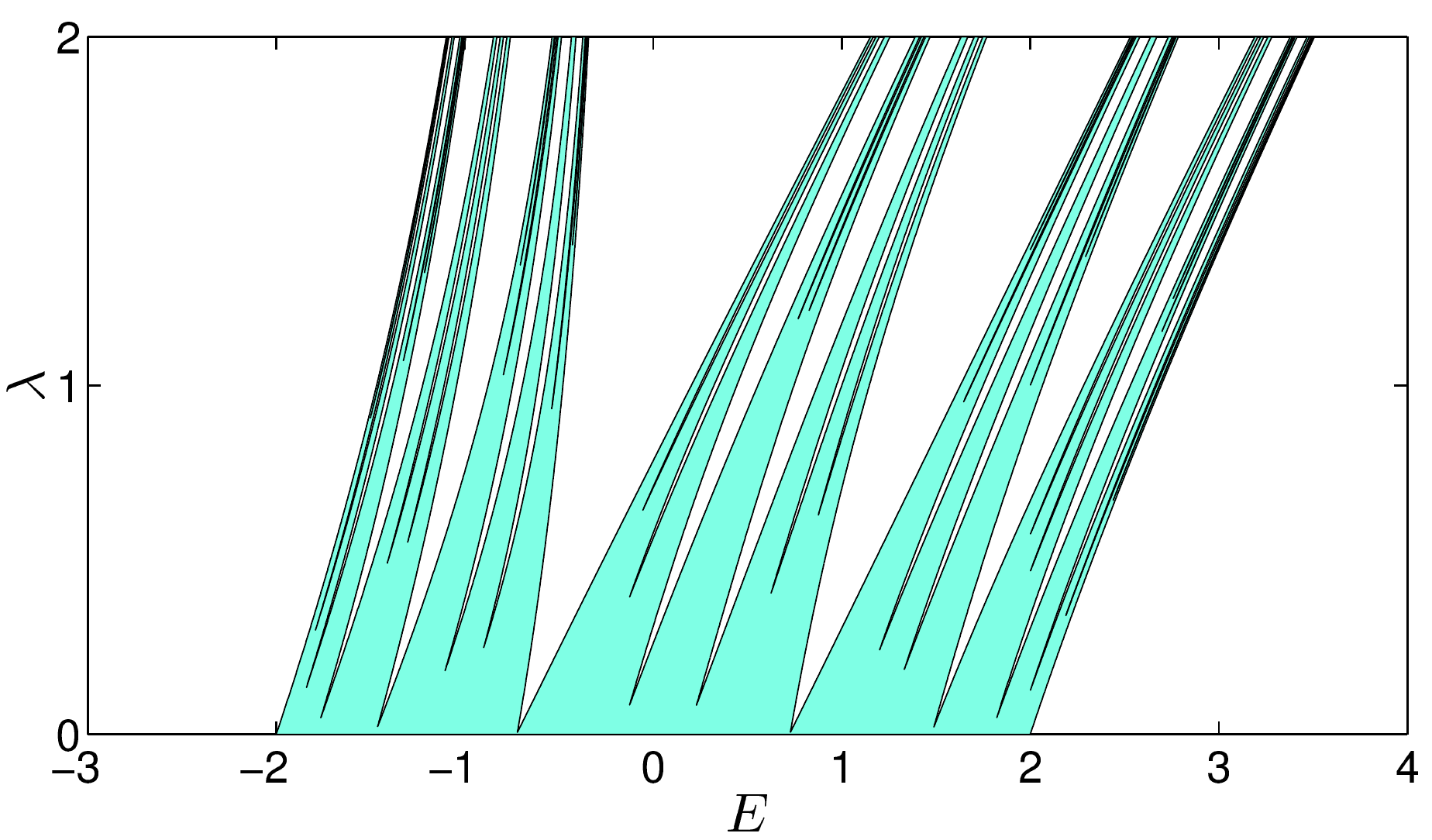}

\begin{picture}(0,0)
\put(-35,28){$k=7$} \put(135,28){$k=8$}
\put(-35,123){$k=5$} \put(135,123){$k=6$}
\put(-35,218){$k=3$} \put(135,218){$k=4$}
\put(-35,312){$k=1$} \put(135,312){$k=2$}
\end{picture}
\end{center}

\vspace*{-1.5em}
\caption{\label{fig:1d}
The upper bound $\Sigma_{k,\lambda} = \sigma_{k,\lambda} \cup \sigma_{k+1,\lambda}$
on the Fibonacci spectrum, as a function of $\lambda\in[0,2]$.
For $k=8$ and $\lambda=2$, $\Sigma_{k,\lambda}$ is the union of 42 disjoint intervals.
}
\end{figure}

Significant insight can be gleaned from numerical calculations
involving small to moderate values of $k$.
For example, Figure~\ref{fig:sig} shows $\sigma_{k,\lambda}$ for
$\lambda\in[0,2]$ and $k=1,\ldots, 8$, while
Figure~\ref{fig:1d} shows the upper bounds $\Sigma_{k,\lambda}$
for the same range of $\lambda$ and $k$.
Since $\Sigma_{8,\lambda} = \sigma_{8,\lambda} \cup \sigma_{9,\lambda}$,
for $\lambda>0$ the spectrum is the union of 34 and 55 intervals.

\begin{figure}
\begin{center}
\includegraphics[scale=0.4]{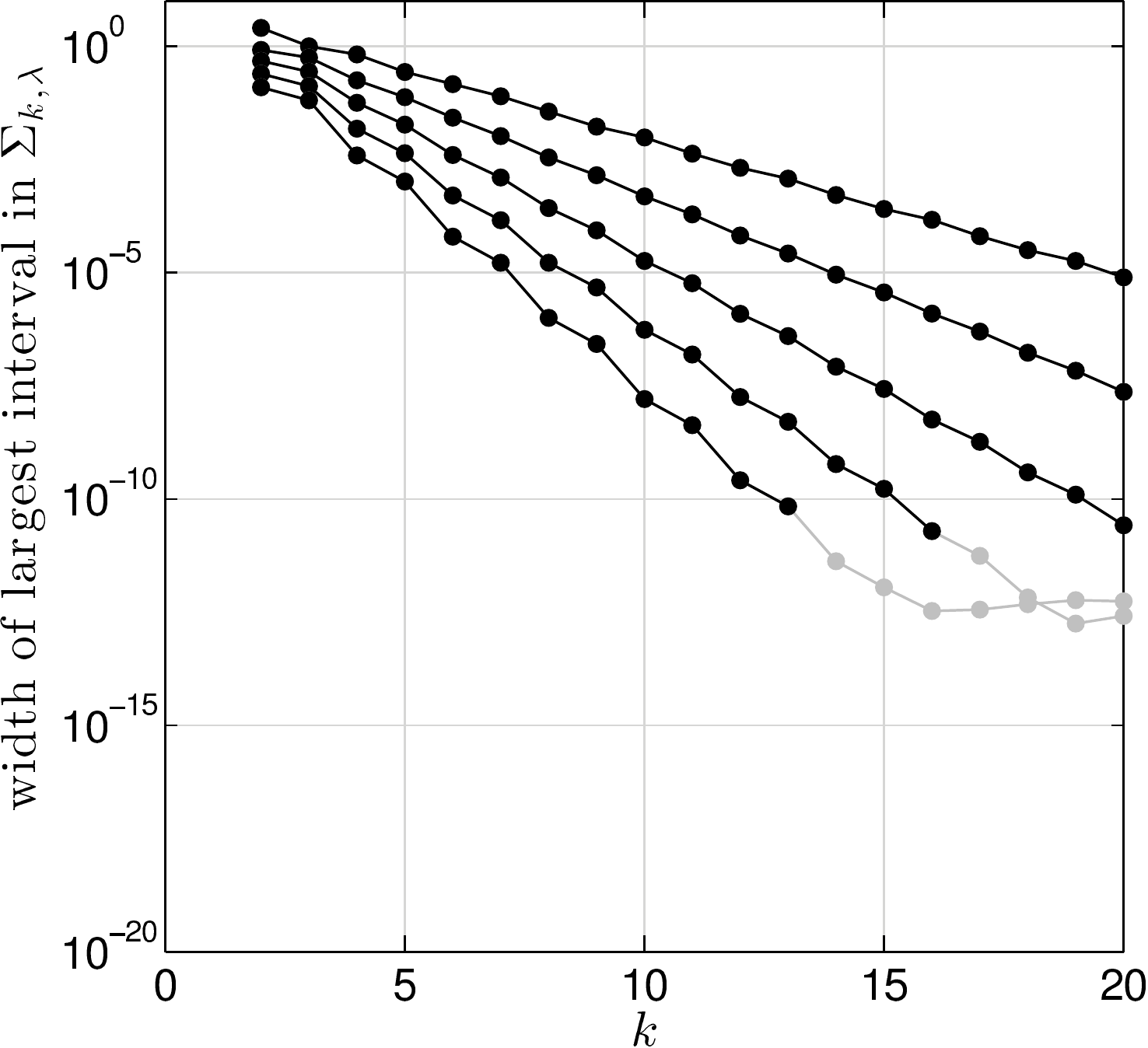}\quad
\includegraphics[scale=0.4]{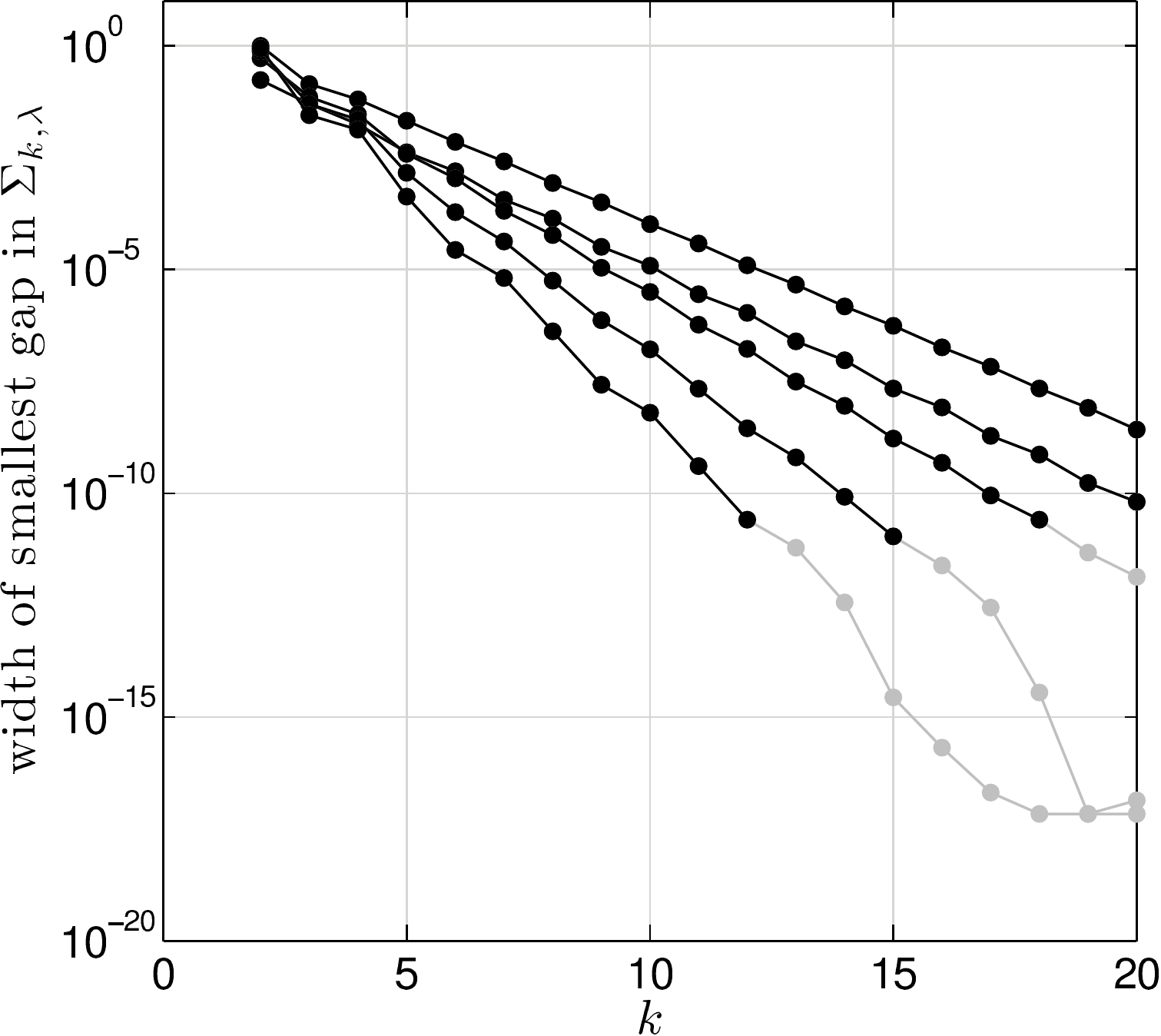}
\begin{picture}(0,0)
\put(-255,137){\rotatebox{-17}{\footnotesize $\lambda=2$}}
\put(-290,108){\rotatebox{-37}{\footnotesize $\lambda=32$}}
\put(-75,125){\rotatebox{-25}{\footnotesize $\lambda=2$}}
\put(-103,100){\rotatebox{-37}{\footnotesize $\lambda=32$}}
\end{picture}
\end{center}

\vspace*{-1em}
\caption{\label{fig:gap_int}
Exponential decay of the largest intervals and smallest gaps
in the approximations $\Sigma_{k,\lambda}$,
for coupling constants $\lambda = 2, 4, 8, 16, 32$,
as computed in MATLAB's double-precision floating point arithmetic.
The data points that are plotted in gray
are likely dominated by computational errors.
}
\end{figure}

To develop conjectures (e.g., regarding ${\rm dim}\ \Sigma_\lambda$), one would like to use approximations to $\Sigma_\lambda$ for larger $k$. Two fundamental challenges arise: (i) the $O(F_k^3)$ work and $O(F_k^2)$ storage becomes prohibitive; (ii) while non-degenerate, the intervals in $\sigma_{k,\lambda}$
become exponentially small and exponentially close together. This phenomenon is illustrated in Figure~\ref{fig:gap_int}. The utility of the numerical results degrade when the size of these bands and gaps approaches the order of the error in the numerical computation.%
\footnote{ More subtly, the formula~(\ref{eq:fibpot}) incurs significant rounding errors for large $n$ and $k$, resulting in errors on the diagonal of $J_{k\pm}$ of size $\lambda$. For greater accuracy, one should use the equivalent formulation $v_{n,k} = \lambda \chi_{[F_k-F_{k-1},F_k)}(n F_{k-1}\ {\rm mod}\ F_k)$,
which is more robust.
}
On contemporary commodity computers, computations of $\Sigma_{k,\lambda}$ up to roughly $k=20$ (requiring all eigenvalues of matrices of dimension $F_{20} = \mbox{10,946}$ and $F_{21}=\mbox{17,711}$) is feasible, provided $\lambda$ is sufficiently small for the results to be accurate. Recently Puelz has proposed an improved approach that ameliorates challenge (i) above by reducing the required work to $O(F_k^2)$ and storage to $O(F_k)$, and challenge~(ii) by using extended precision arithmetic~\cite{Pue}.

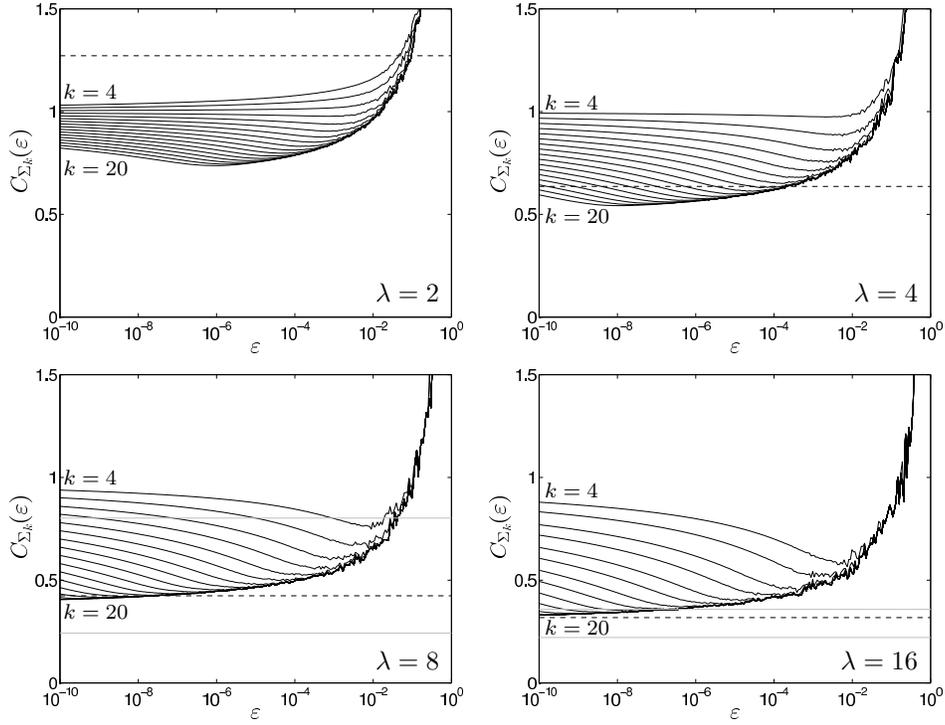
\begin{figure}
\begin{center}
\includegraphics[scale=0.33]{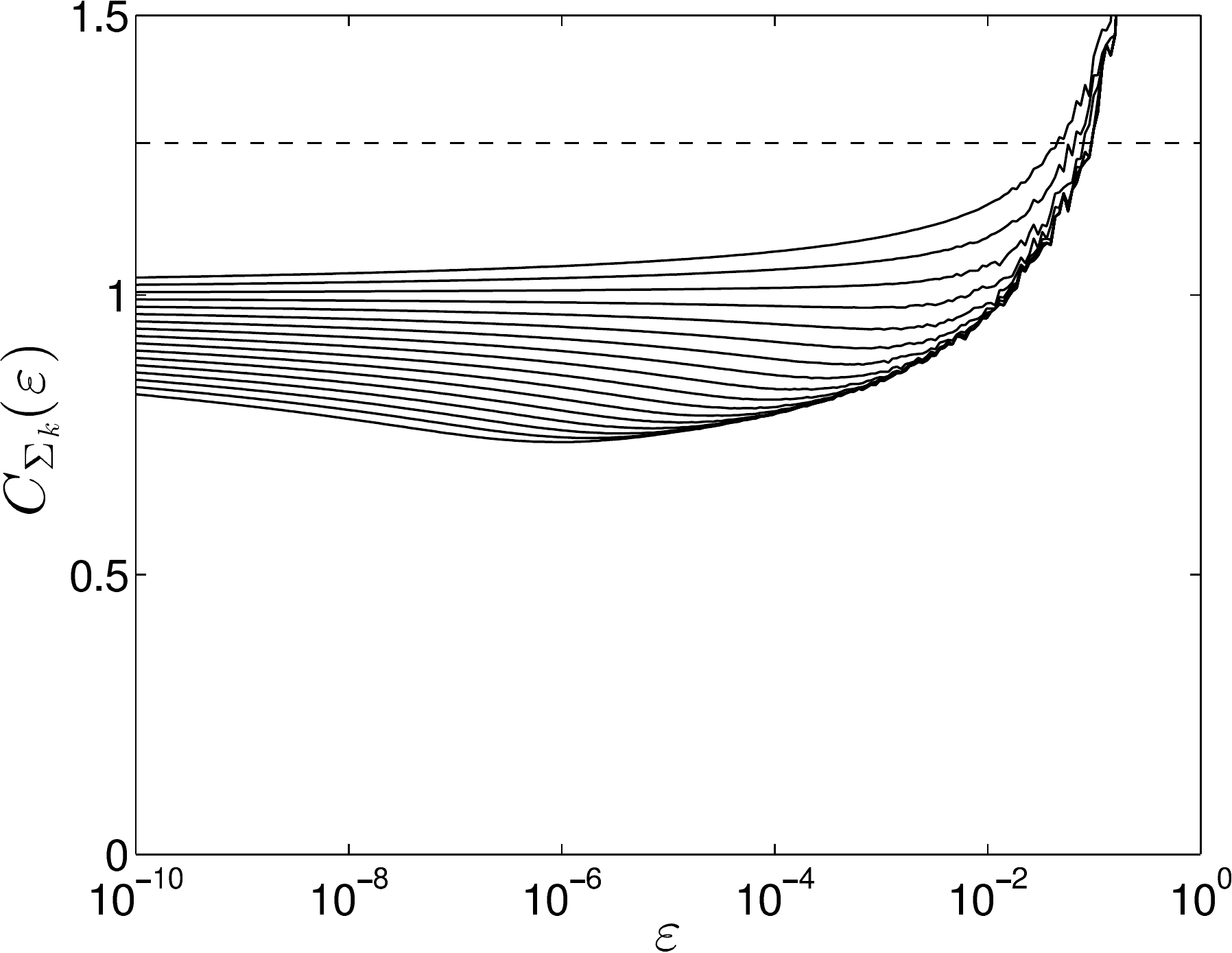}\quad
\includegraphics[scale=0.33]{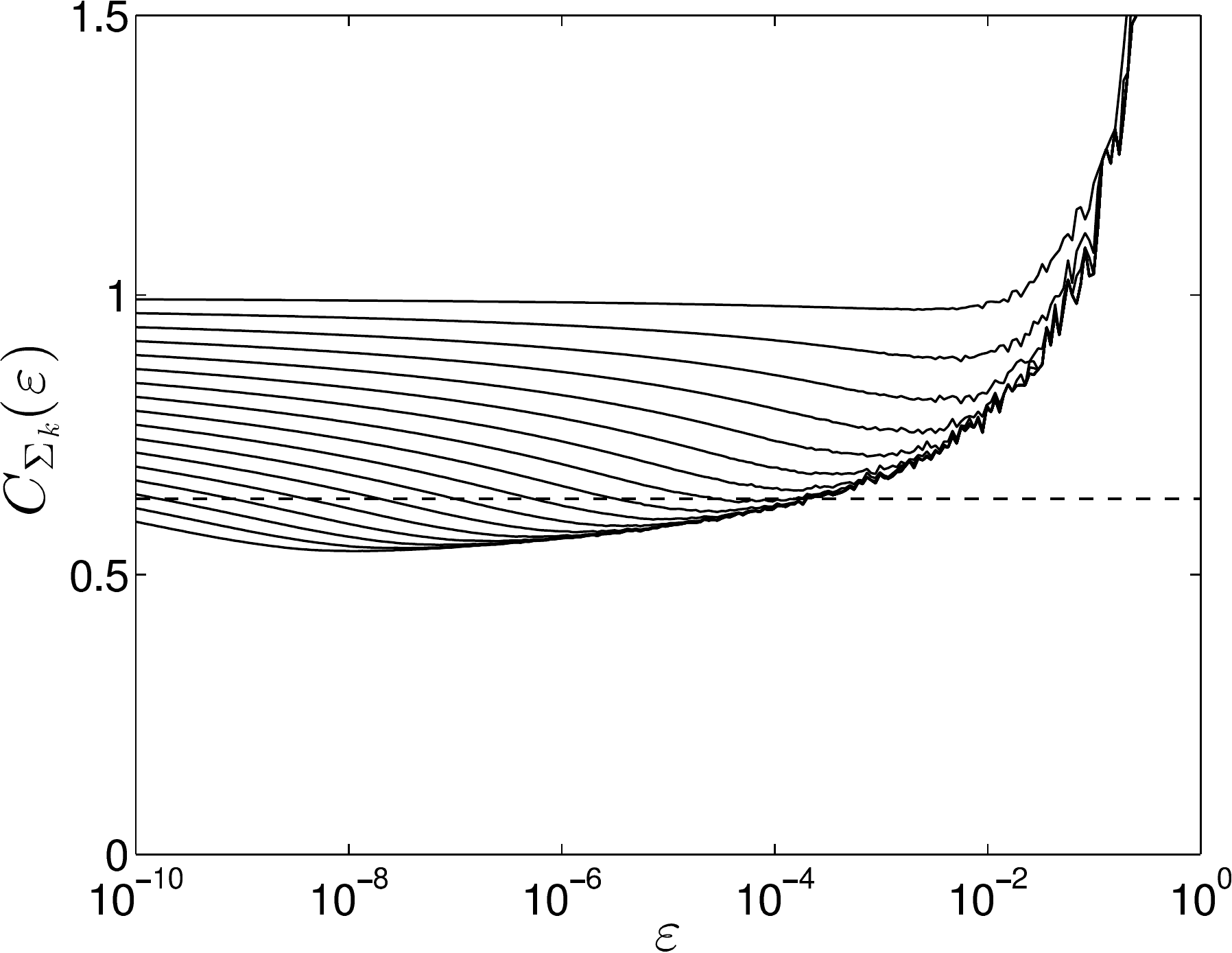}

\vspace*{0.5em}

\includegraphics[scale=0.33]{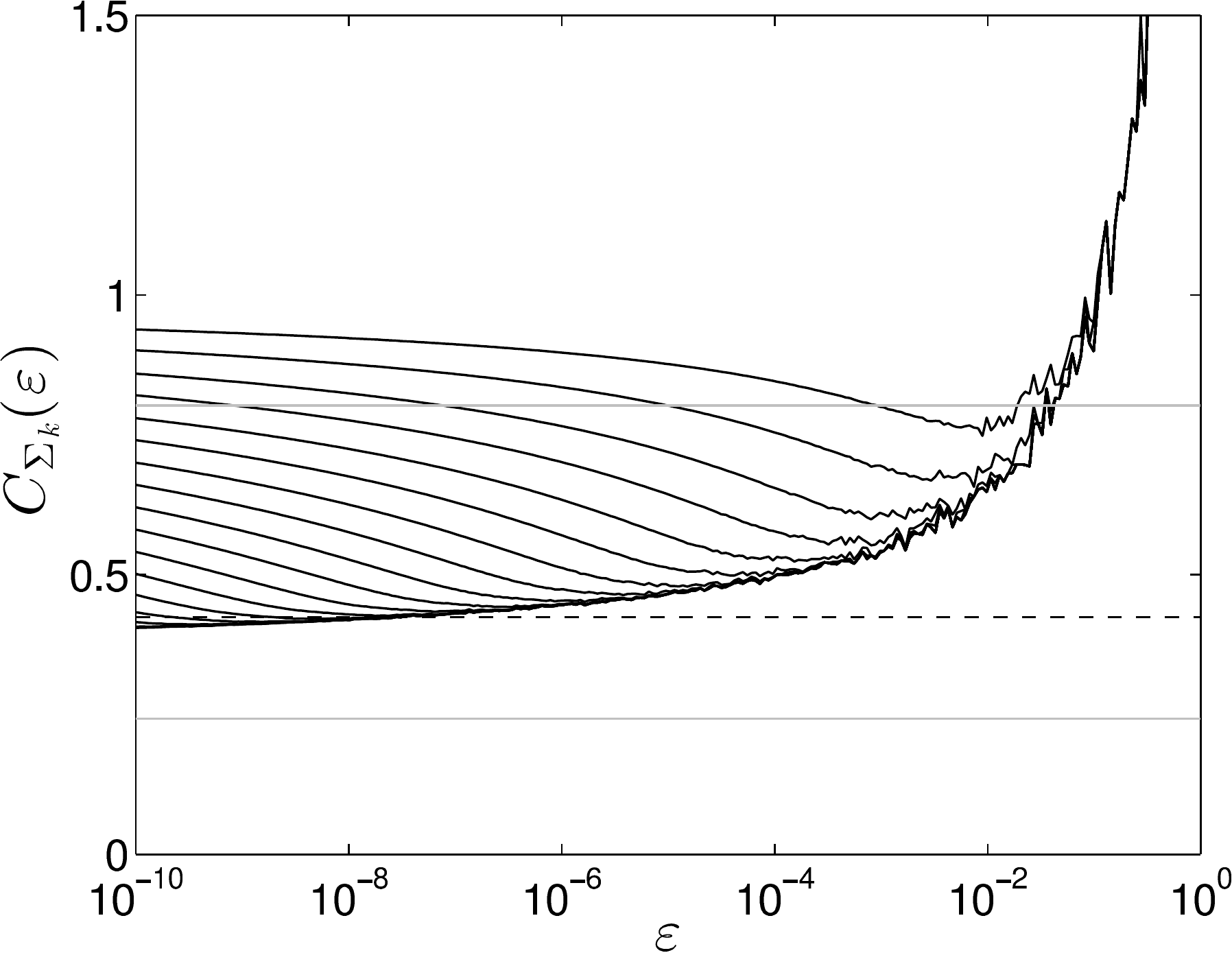}\quad
\includegraphics[scale=0.33]{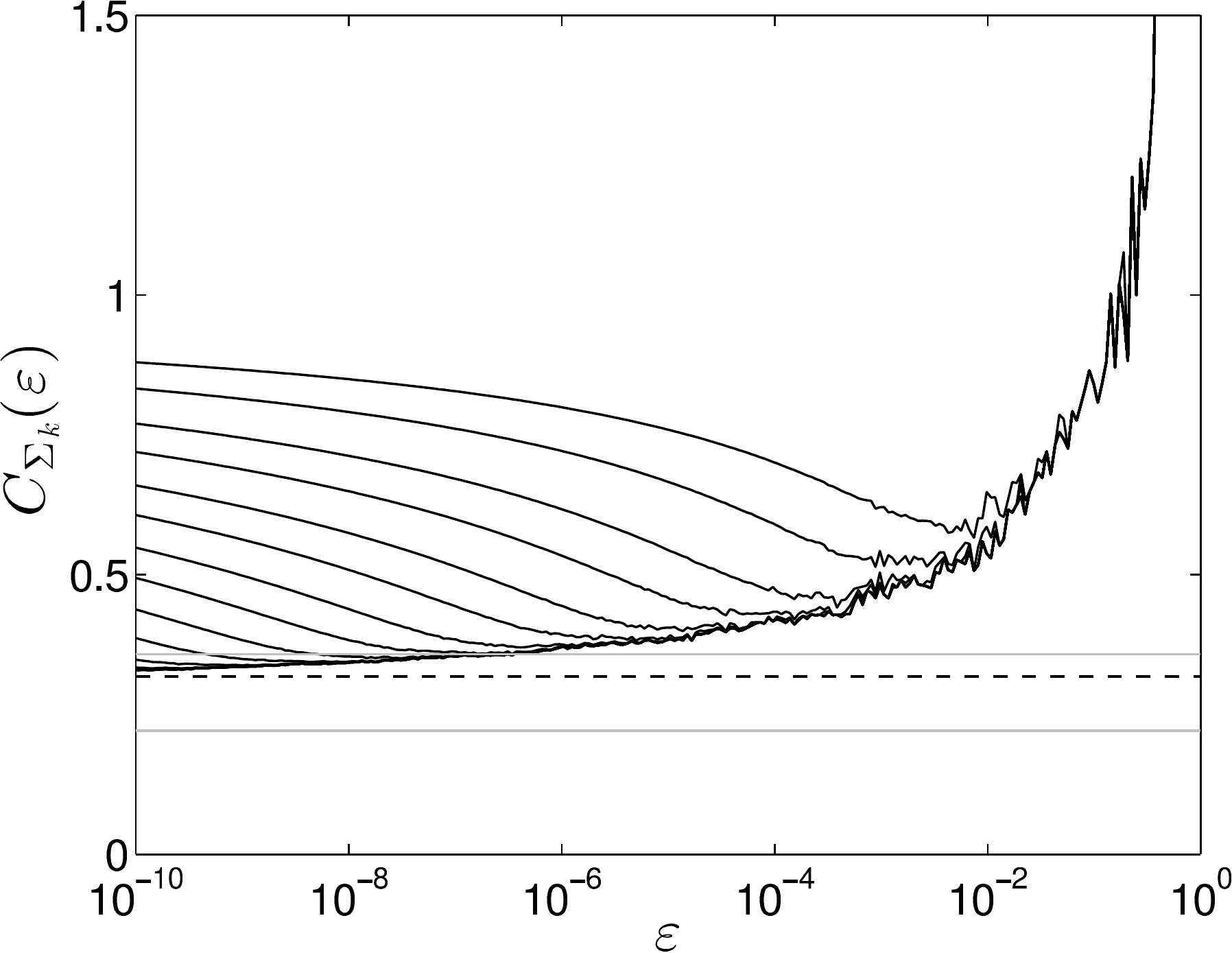}
\end{center}

\begin{picture}(0,0)(-176,-14)
\put(-150,231){{\footnotesize $k=4$}}
\put(-150,188){{\footnotesize $k=20$}}
\put(-332,235){{\footnotesize $k=4$}}
\put(-332,206){{\footnotesize $k=20$}}
\put(-332,89){{\footnotesize $k=4$}}
\put(-332,38){{\footnotesize $k=20$}}
\put(-150,84){{\footnotesize $k=4$}}
\put(-150,32){{\footnotesize $k=20$}}
\put(-214,158){{$\lambda=2$}}
\put(-214,19){{$\lambda=8$}}
\put(-33,158){{$\lambda=4$}}
\put(-38,19){{$\lambda=16$}}
\end{picture}

\vspace*{-1.5em}
\caption{\label{fig:plot_bcd}
Estimates of ${\rm dim}_B(\Sigma_\lambda)$ for various values of $\lambda$, based on the upper bounds $\Sigma_{k,\lambda}$ for various $k$. The dashed horizontal line denotes $\log(1+\sqrt{2})/\log(\lambda)$, to which ${\rm dim}_B(\Sigma_\lambda)$ tends as $\lambda\to\infty$
(Theorem~\ref{t.DEGT}). The gray horizontal lines in the bottom plots show the upper and lower bounds~(\ref{eq:bcdup})--(\ref{eq:bcdlow}).
}
\end{figure}

To estimate the box-counting dimension of $\Sigma_\lambda$ (assuming it exists), we use the definition
\[ {\rm dim}_B(S) = \lim_{\varepsilon\to0} {\log C_S(\varepsilon) \over \log 1/\varepsilon},\]
where $C_S(\varepsilon)$ counts the number of intervals of width $\varepsilon$ that intersect $S$,
\[ C_S(\varepsilon) := \#\{ j\in \Z: [j\varepsilon, (j+1)\varepsilon) \cap S \ne \emptyset\}.\]
Note that ${\rm dim}_B(\Sigma_{k,\lambda})=1$ for all $k$, since $\Sigma_{k,\lambda}$ is the union of finitely many closed intervals. Still, one gains insight into ${\rm dim}_B(\Sigma_\lambda)$ from $\log(C_{\Sigma_{k,\lambda}}(\varepsilon))/\log(1/\varepsilon)$ for finite values of $\varepsilon$ and various $k$, as can be seen in Figure~\ref{fig:plot_bcd}. For fixed $\lambda$, the resulting estimates of ${\rm dim}_B(\Sigma_\lambda)$ (taken, e.g., as $\inf_{\varepsilon\in(0,1)} \log(C_{\Sigma_{k,\lambda}}(\varepsilon))/\log(1/\varepsilon)$) apparently improve as $k$ increases; lower values of $k$ are suitable for larger values of $\lambda$. However, with this approach it is difficult to accurately estimate the critical value at which ${\rm dim}_B(\Sigma_\lambda)= 1/2$.\footnote{We are interested in this critical value because as soon as ${\rm dim}_B(\Sigma_\lambda)$ falls below $1/2$, we can be sure that the sum set $\Sigma_\lambda + \Sigma_\lambda$ is a zero-measure Cantor set, and this is an issue of interest for reasons we will discuss in Subsection~\ref{sec:2d3dcomp}.}  (A rough estimate, suggested from Figure~\ref{fig:plot_bcd}, is $\lambda \approx 4$; see the discussion preceding Problem~\ref{prob:cantorval} below.)
More accurate approximations will require computations with larger values of $k$ than are feasible with the method described above.

\begin{figure}
\begin{center}
   \includegraphics[scale=0.5]{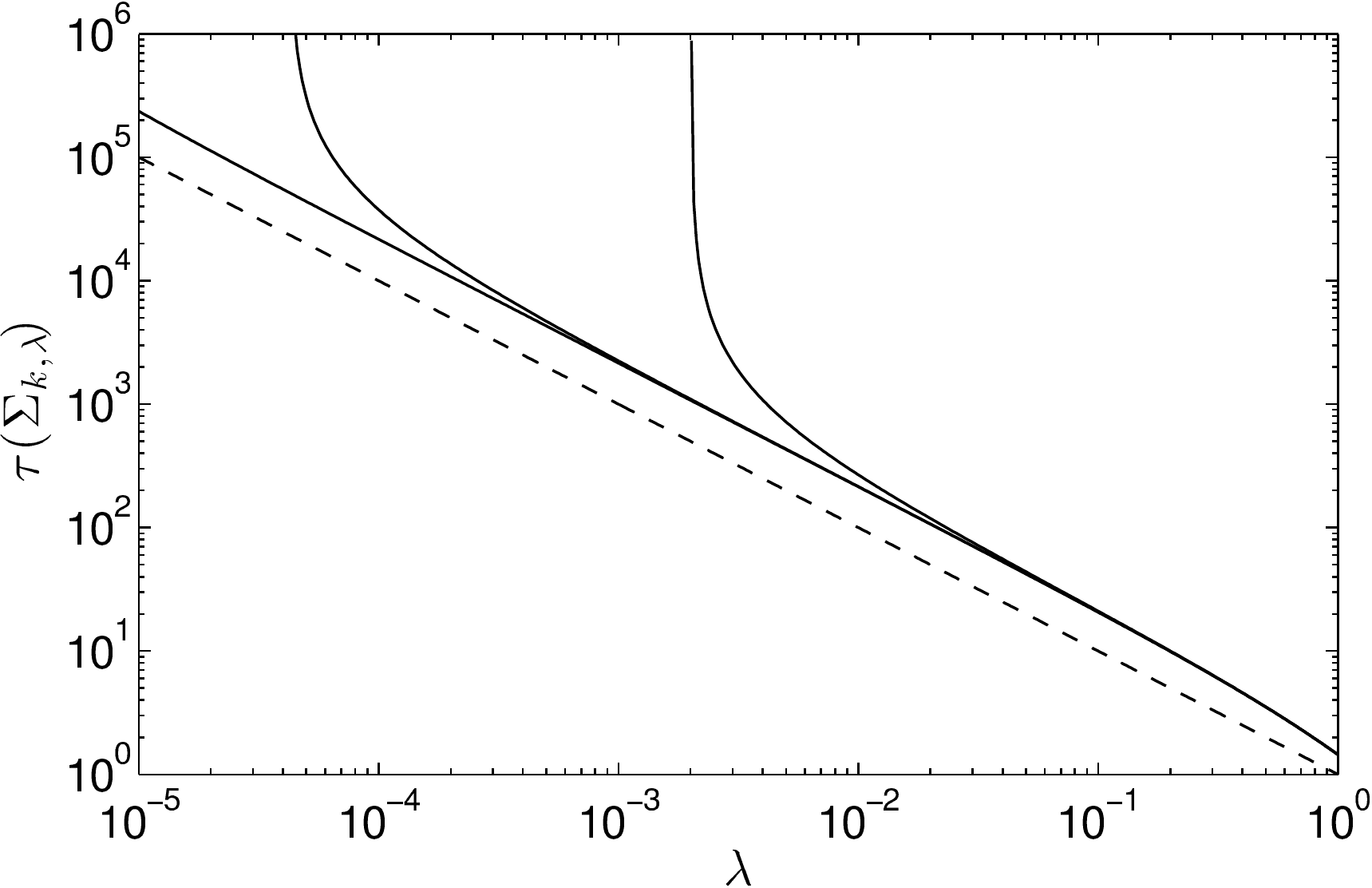}

\begin{picture}(0,0)
   \put(-97,153){\rotatebox{-27}{\footnotesize $k=17$}}
   \put(-43,105){\rotatebox{-27}{\footnotesize $\lambda^{-1}$}}
   \put(-59,140){\footnotesize $k=13$}
   \put(4,140){\footnotesize $k=9$}
\end{picture}
\end{center}

\vspace*{-2em}
\caption{\label{fig:thick}
Thickness of $\Sigma_{k,\lambda}$ as a function of $\lambda$ for three values of $k$,
consistent with Theorem~\ref{t.2}.
}
\end{figure}
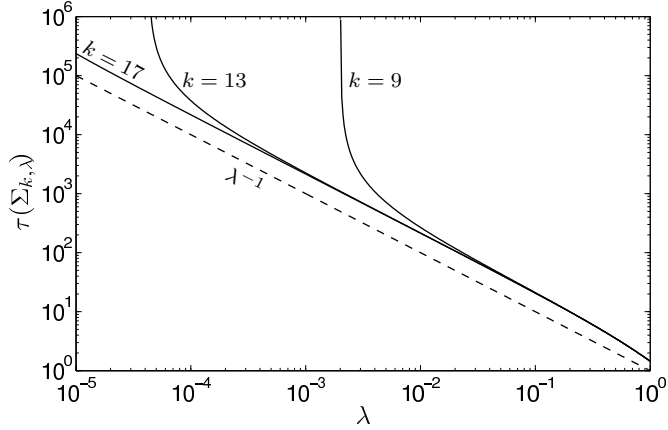

Finally, Figure~\ref{fig:thick} explores numerical computations of the thickness, defined in~(\ref{eq:thick}). As established in Theorem~\ref{t.2}, the thickness $\tau(\Sigma_\lambda)$ behaves like $1/\lambda$ as $\lambda \downarrow 0$.  As $\lambda$ decreases we see this behavior mirrored in the upper bounds $\Sigma_{k,\lambda}$, up to some point where $\tau(\Sigma_{k,\lambda})$ rapidly increases: $\Sigma_{k,\lambda}$ is the union of no more than $F_k+F_{k+1}$ intervals separated by gaps that diminish as $\lambda\downarrow 0$.

\subsection{Density of States for the Fibonacci Model}

We next turn to an investigation of the exponent of H\"older continuity of the
integrated density of states (IDS) for the Fibonacci model, discussed in
Section~\ref{sec:fibids}.
To estimate $N_\lambda(E)$ in equation~(\ref{eq:idsN}), one must compute all
the eigenvalues of $H_{\lambda,[1,n]}$, the restriction of $H_\lambda$ to
sites $[1,n]$ with Dirichlet boundary conditions.
This restriction is an $n\times n$ tridiagonal matrix; because this matrix
lacks the corner entries present in $J_{k\pm}$ in the last subsection,
its eigenvalues can be efficiently computed for large values of $n$
(say $n\le 10^6$ on contemporary desktop computers).
While computational complexity is no longer such a constraint,
accuracy still is: for large $n$ and $\lambda$, some eigenvalues of
$H_{\lambda, [1,n]}$ are closer than the precision of the floating point
arithmetic, rendering, for example, $|E_1-E_2|=0$ for theoretically
distinct eigenvalues $E_1$ and $E_2$ of $H_{\lambda,[1,n]}$.%
\footnote{By its structure, $H_{\lambda,[1,n]}$ must have $n$ distinct
eigenvalues.
Similar scenarios with exceptionally close distinct eigenvalues
are well-known in the numerical analysis community; see, e.g.,
Wilkinson's $W_{21}^+$ matrix~\cite[Sec.~7.7]{Par98}.}

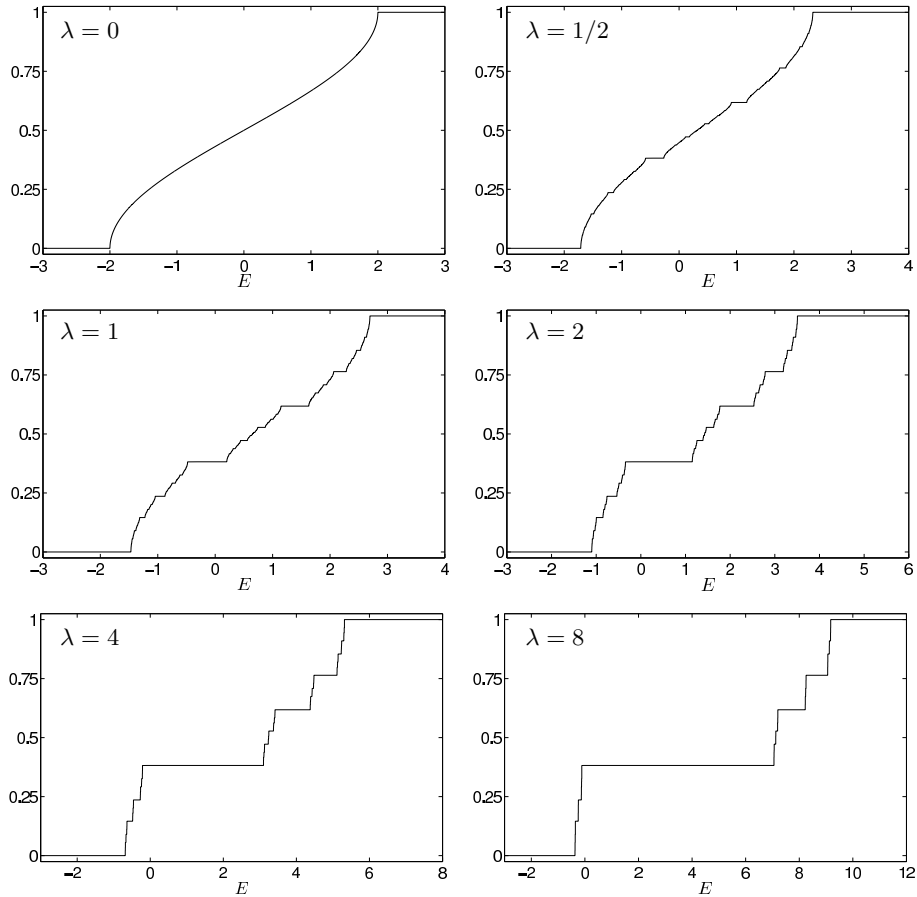
\begin{figure}
\begin{center}
\includegraphics[scale=0.35]{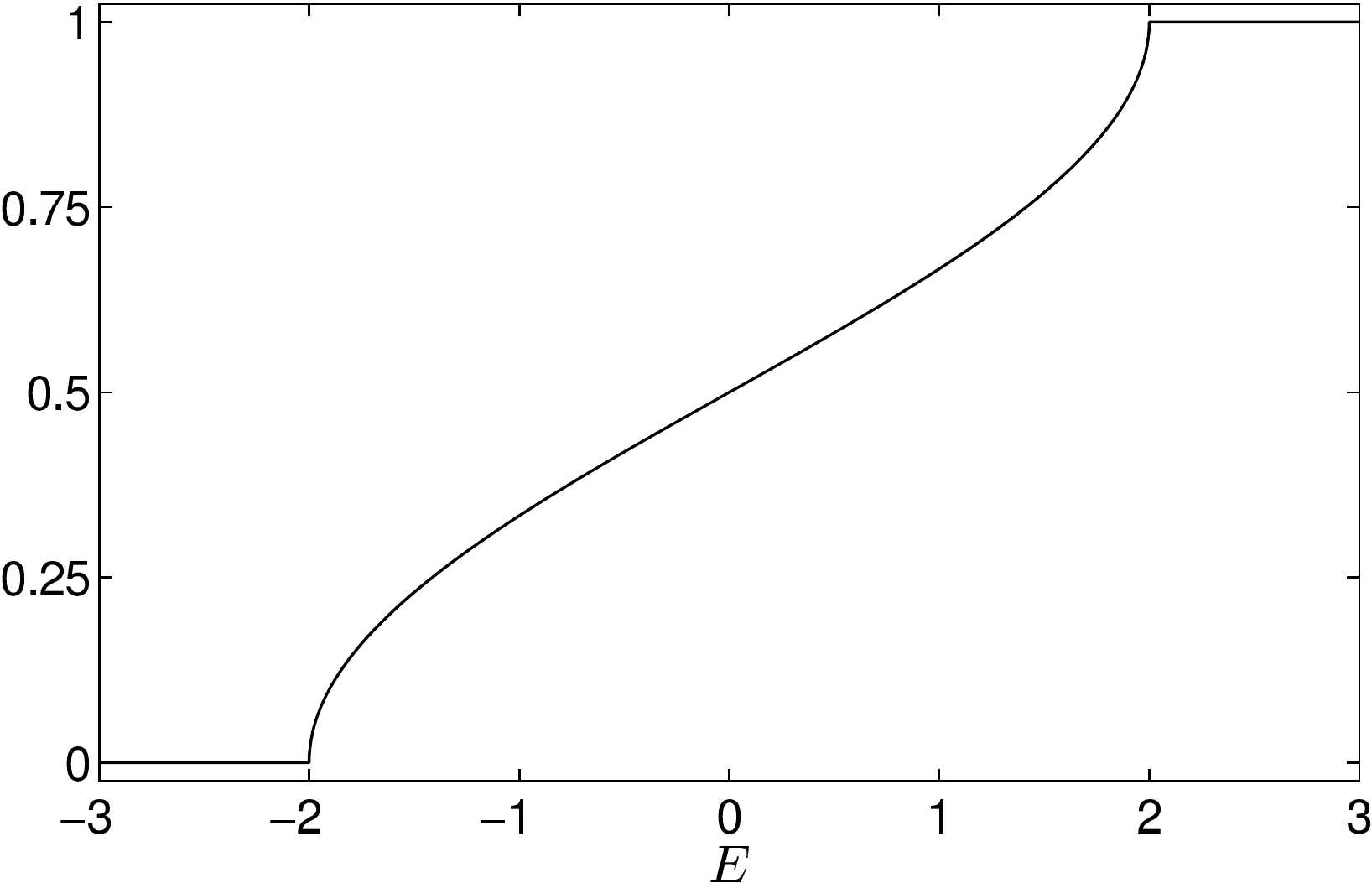}\quad
\includegraphics[scale=0.35]{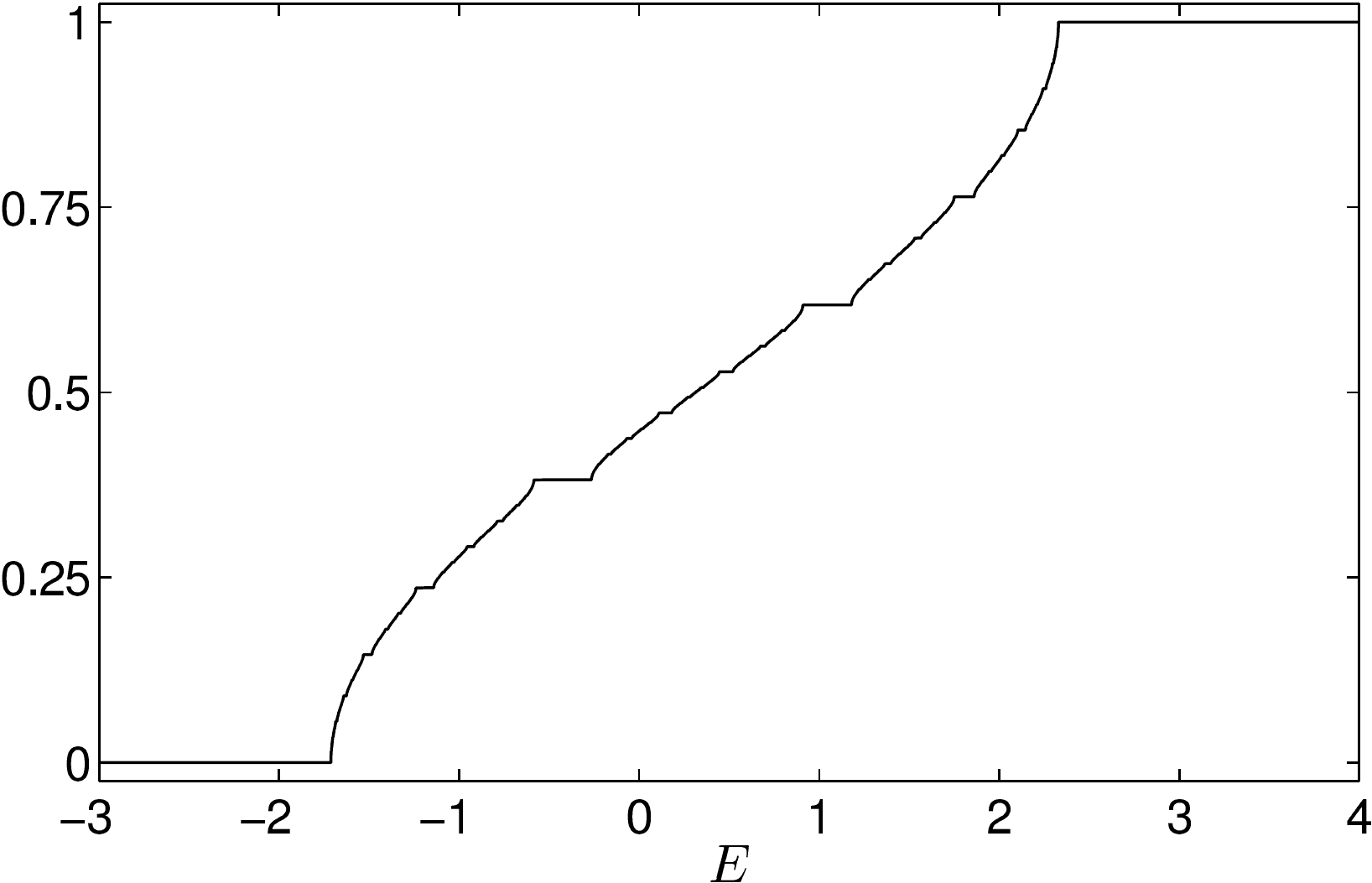}

\vspace*{.75em}
\includegraphics[scale=0.35]{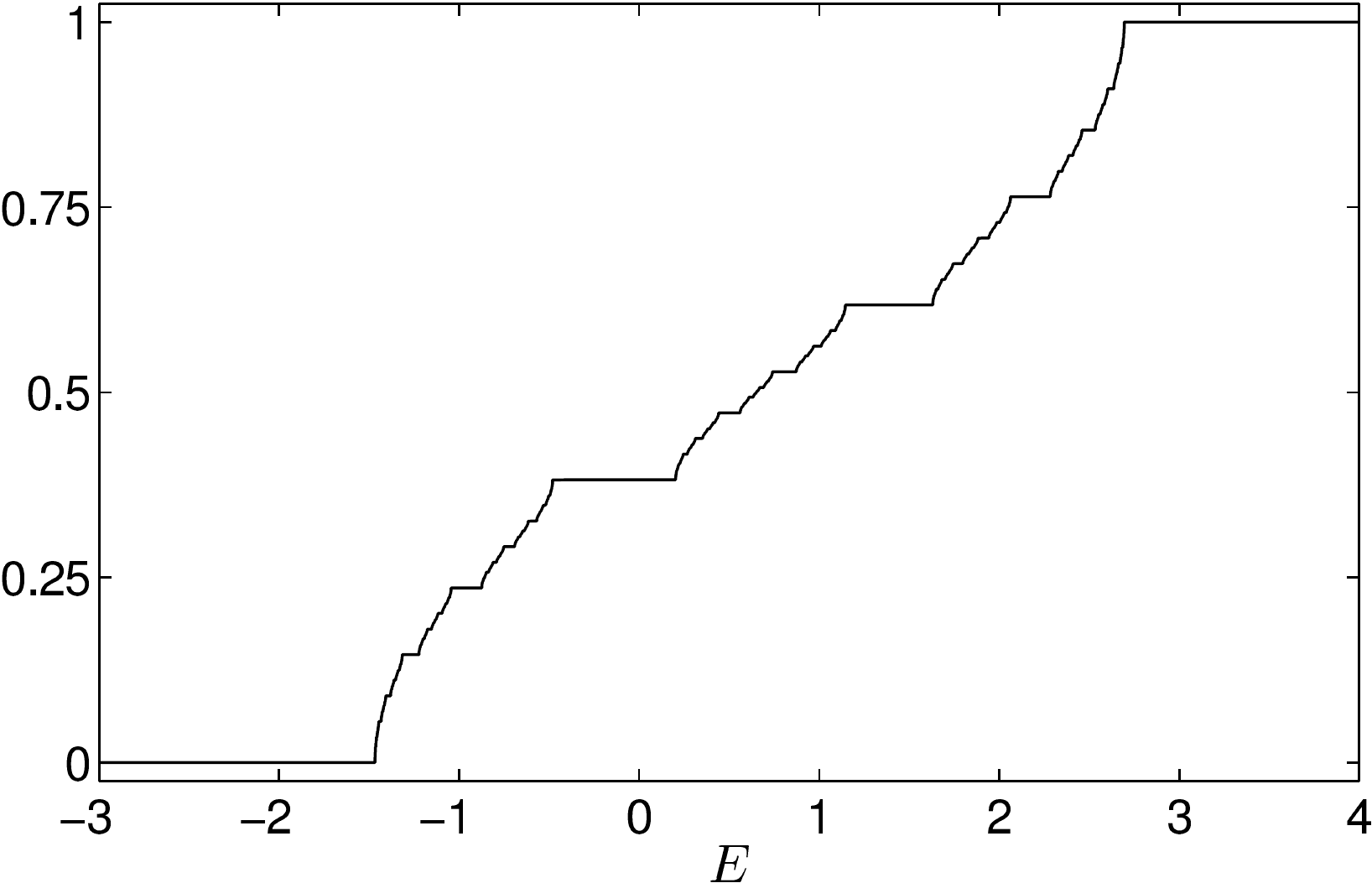}\quad
\includegraphics[scale=0.35]{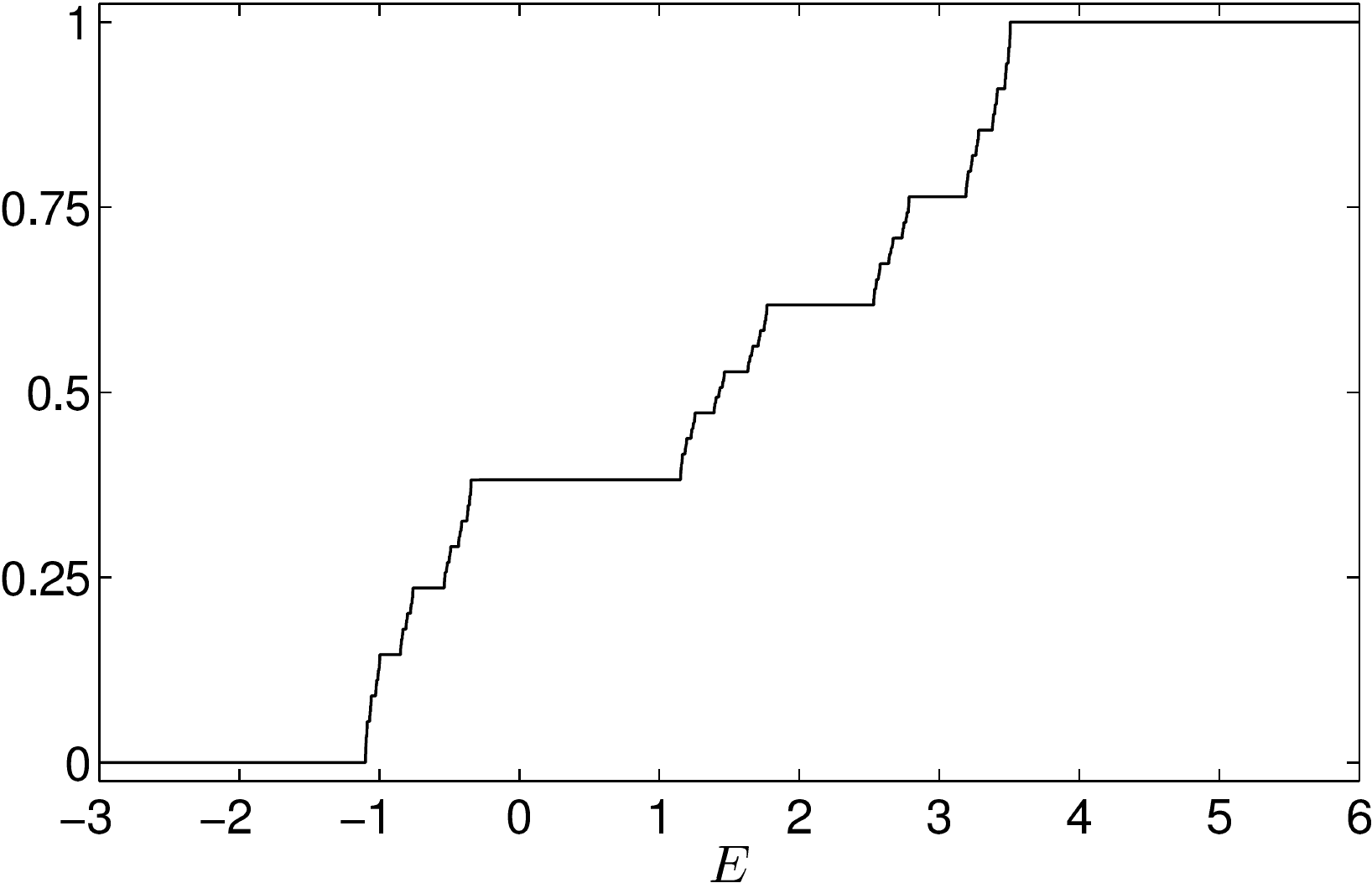}

\vspace*{.75em}
\includegraphics[scale=0.35]{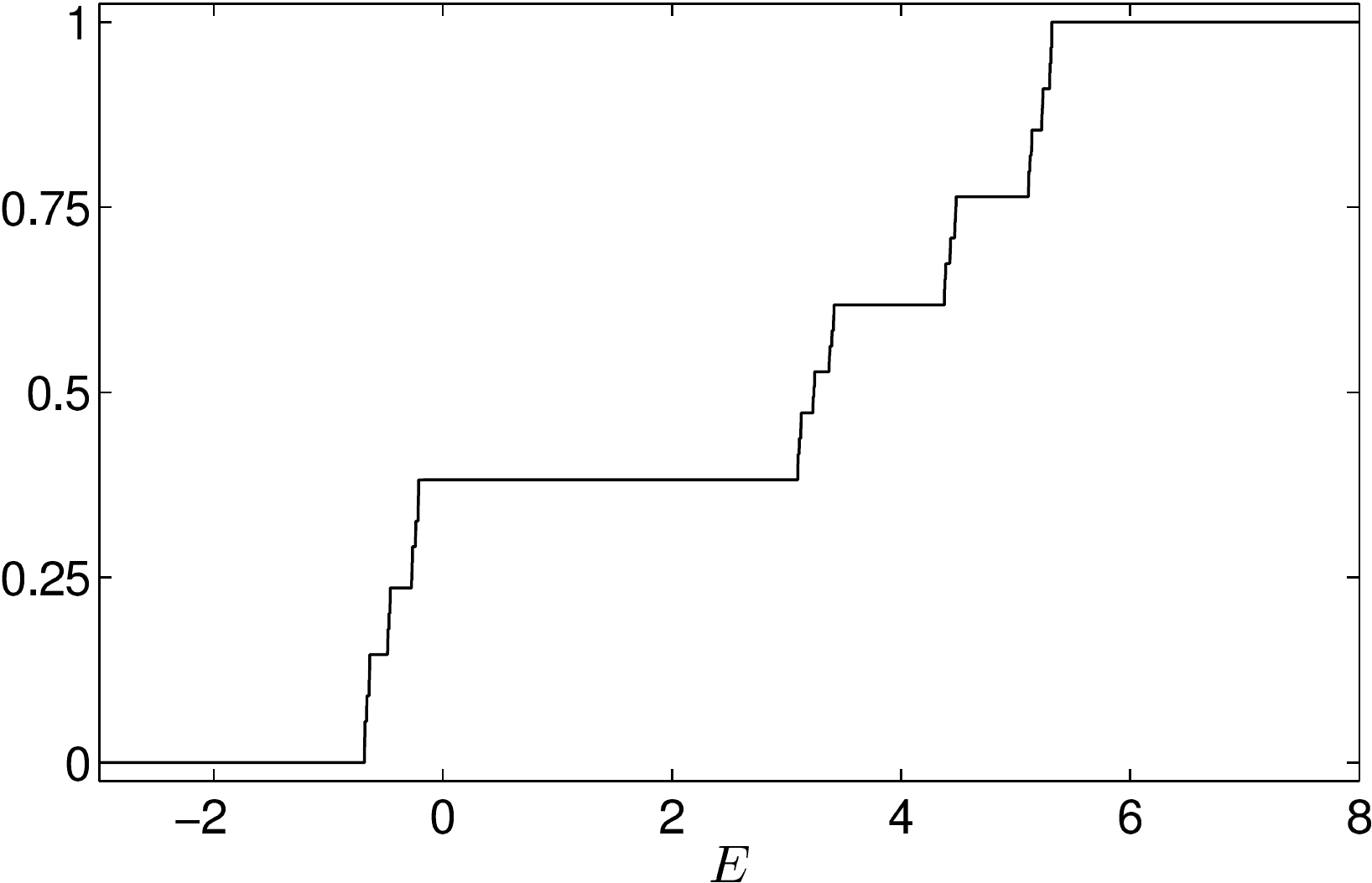}\quad
\includegraphics[scale=0.35]{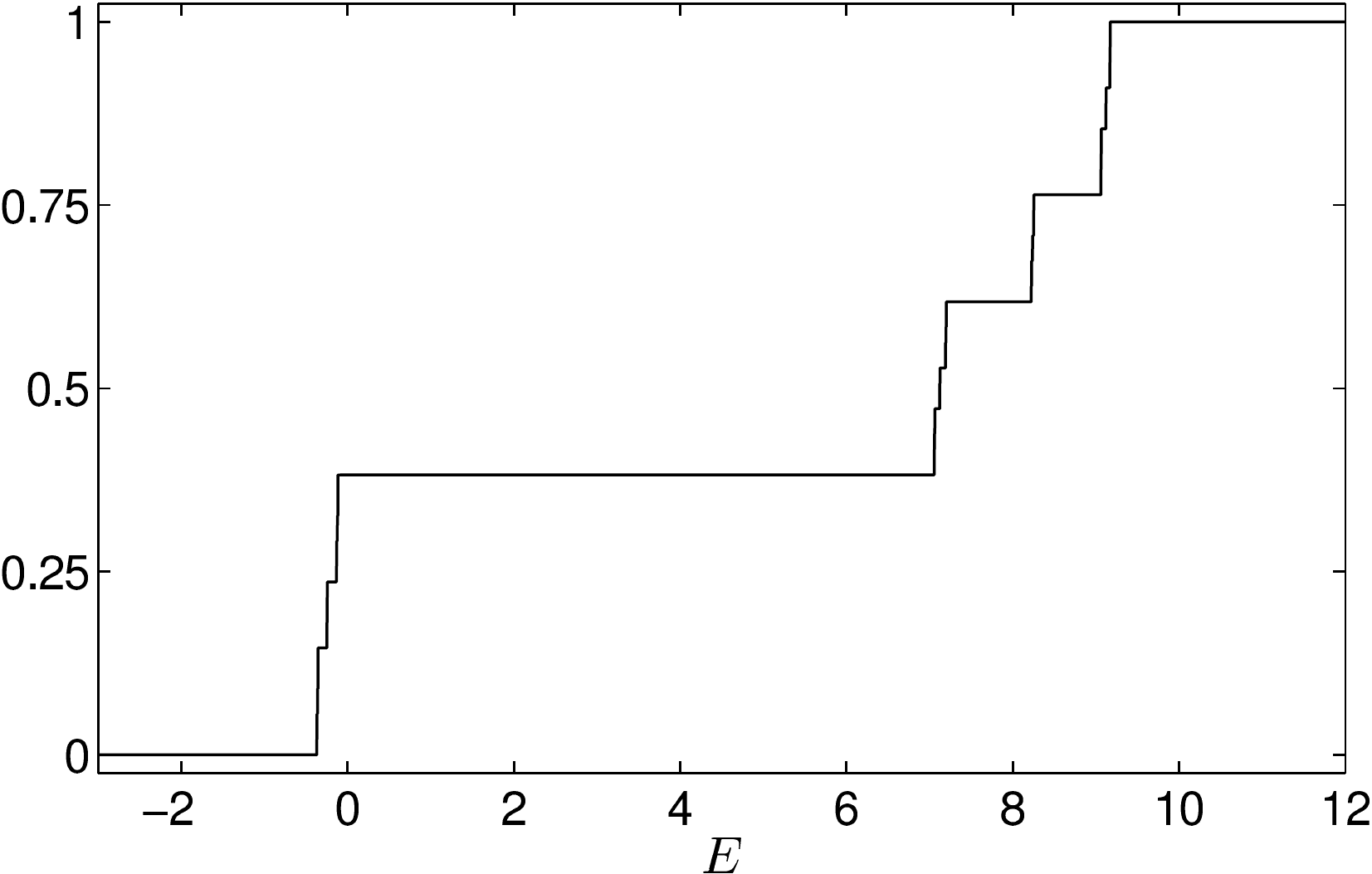}

\begin{picture}(0,0)
\put(-152,336){\small $\lambda=0$}
\put(24,336){\small $\lambda=1/2$}
\put(-152,222){\small $\lambda=1$}
\put(24,222){\small $\lambda=2$}
\put(-152,106){\small $\lambda=4$}
\put(24,106){\small $\lambda=8$}
\end{picture}
\end{center}

\vspace*{-1.75em}
\caption{\label{fig:ids_plot}
Approximations to the integrated density of states for the Fibonacci
model with various values of the coupling constant, $\lambda$,
based on $n=\mbox{10,000}$.
}
\end{figure}

\begin{figure}
\begin{center}
\includegraphics[scale=0.35]{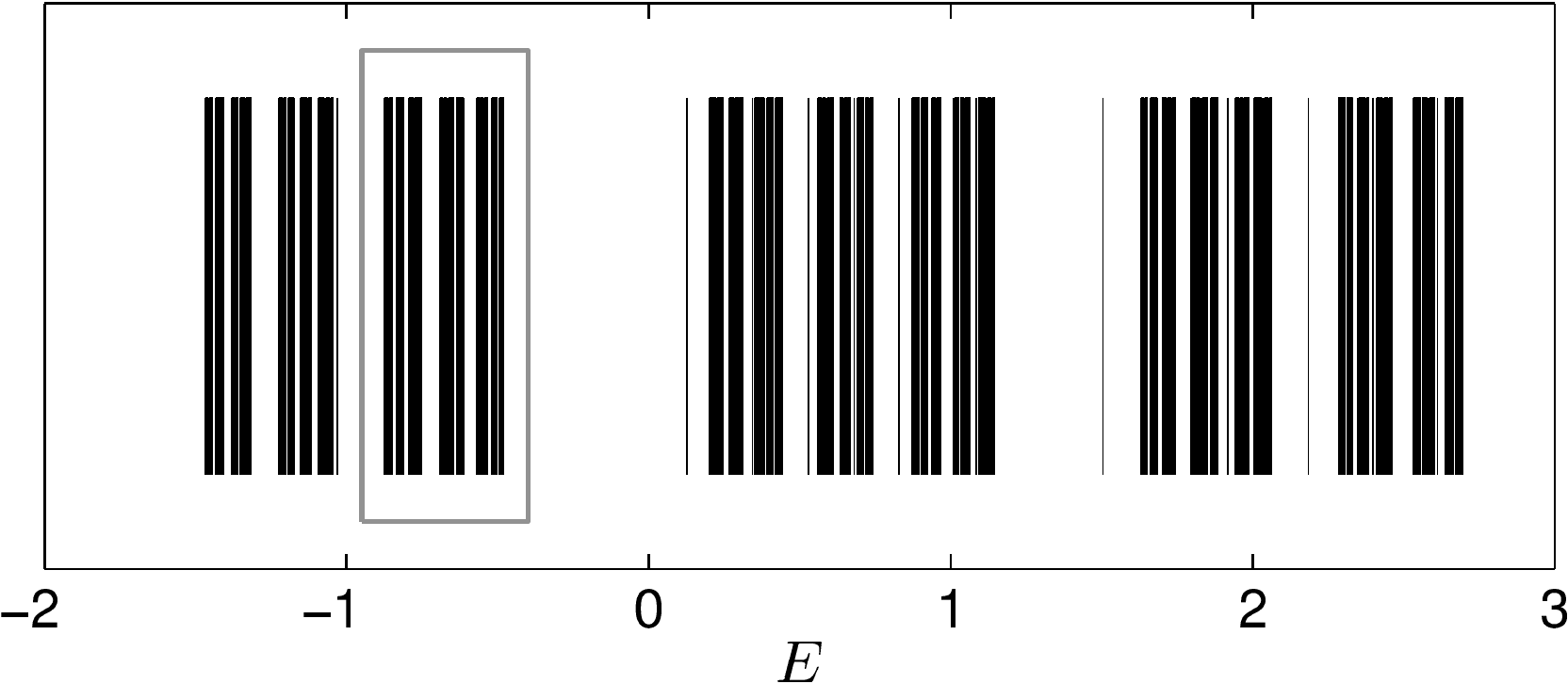}\quad
\includegraphics[scale=0.35]{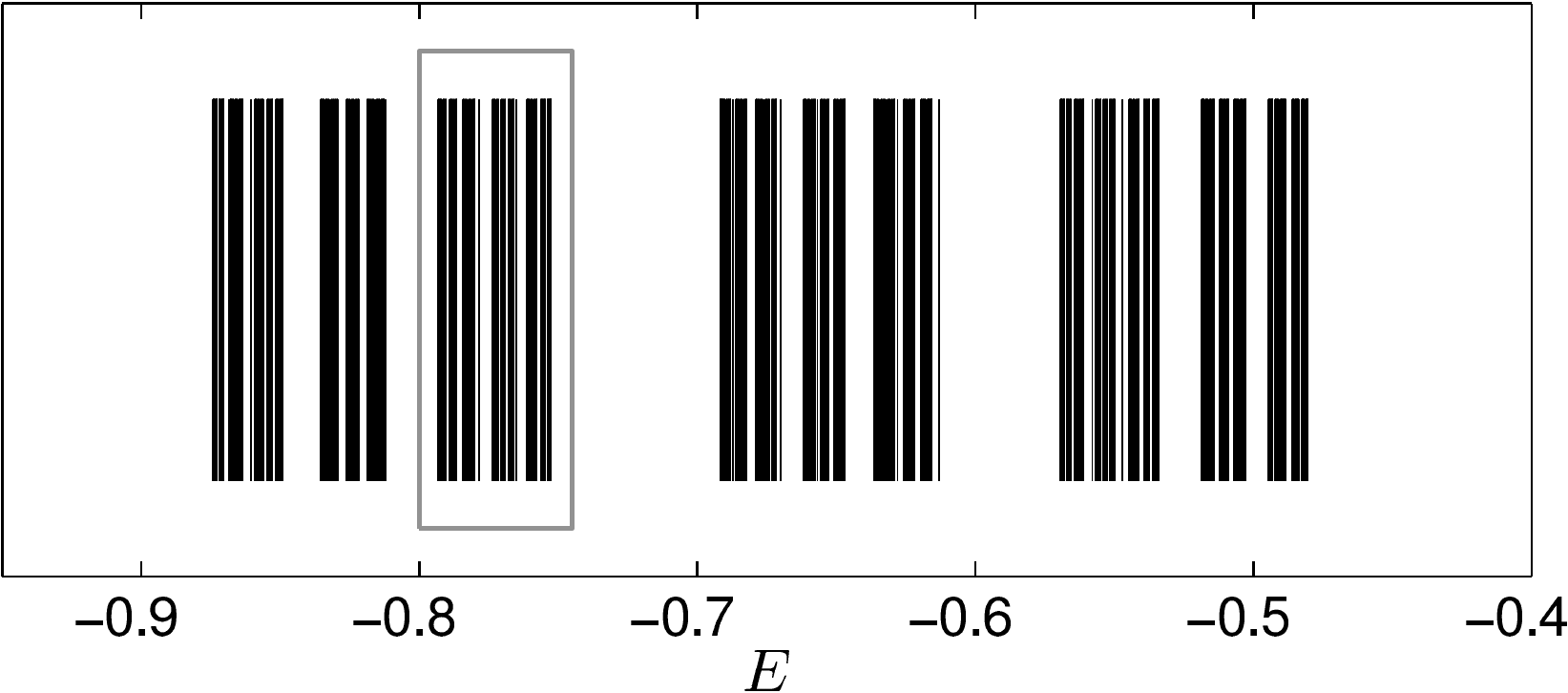}

\vspace*{.75em}
\hspace*{-6pt}
\includegraphics[scale=0.35]{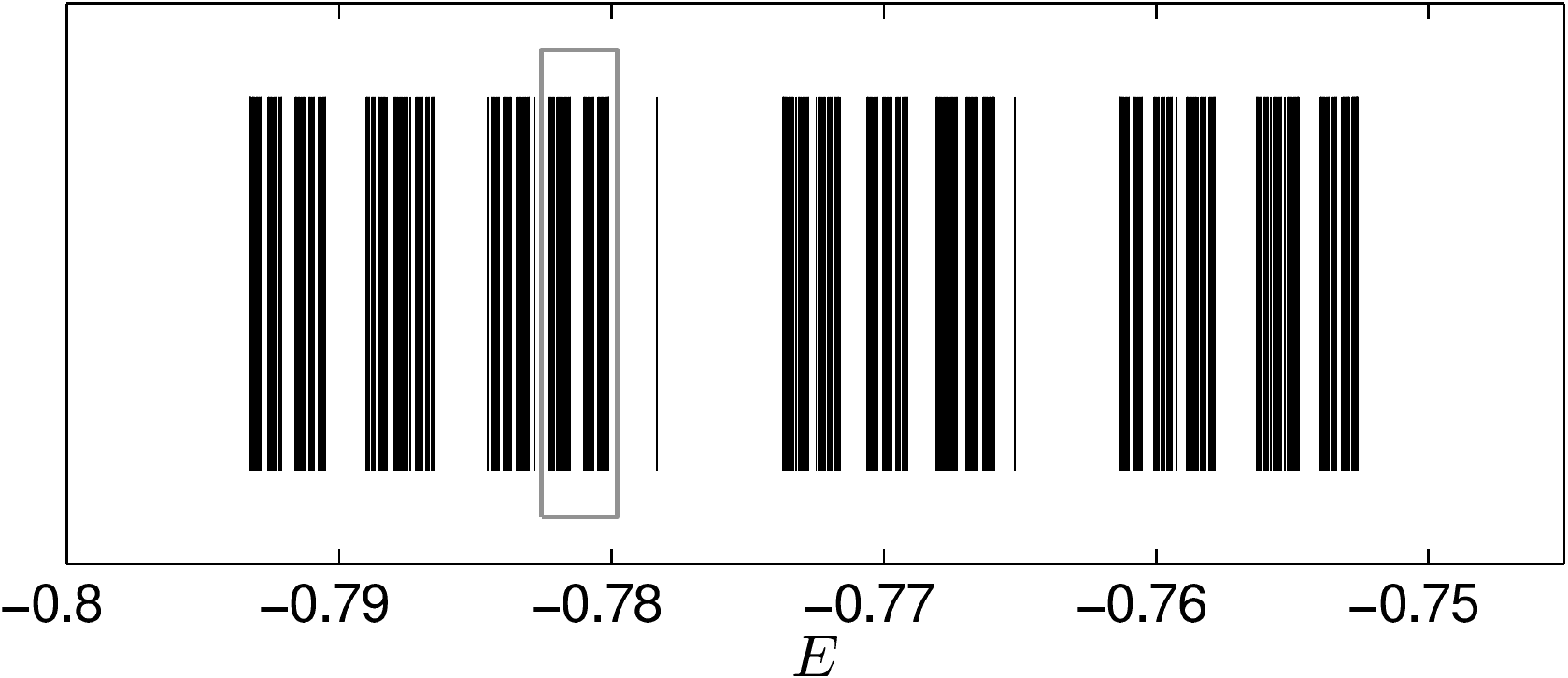}\kern3pt
\includegraphics[scale=0.35]{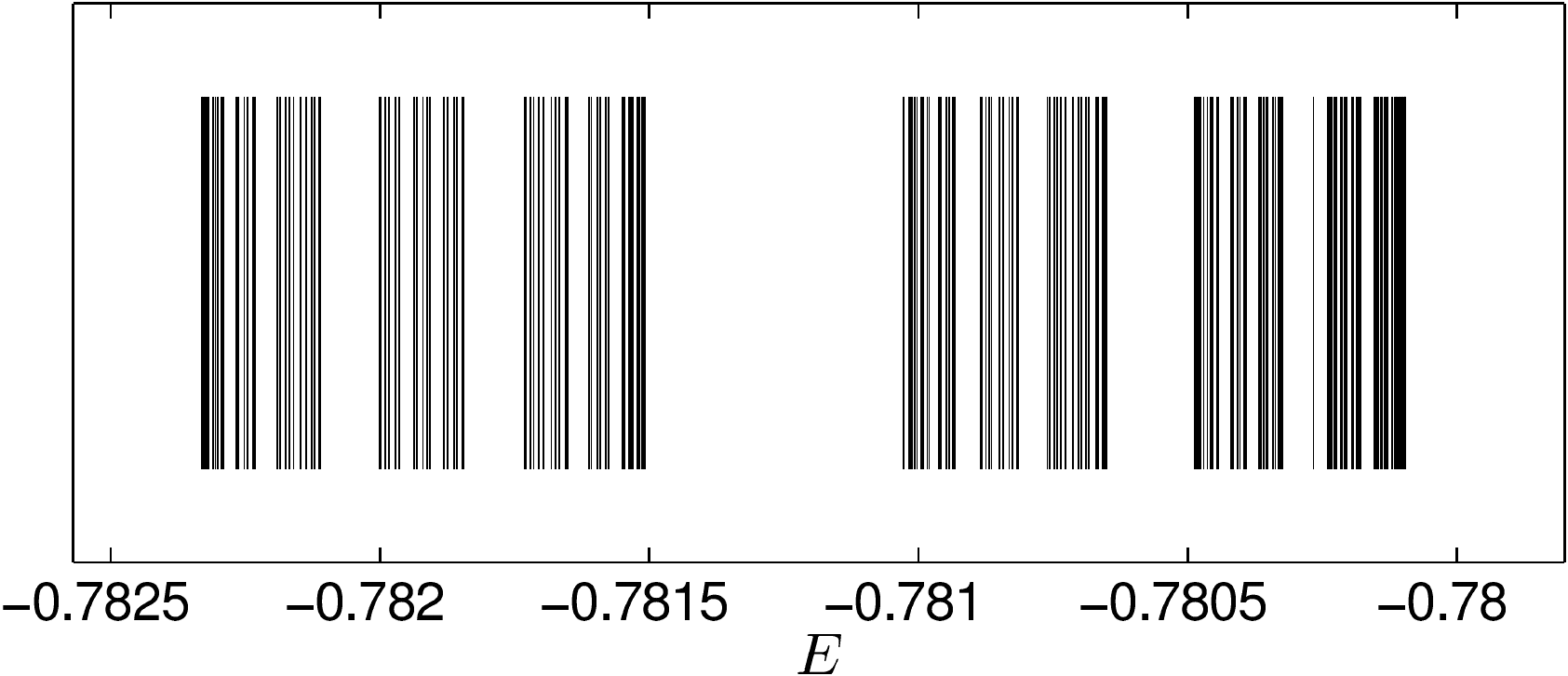}
\end{center}

\vspace*{-.25em}
\caption{\label{fig:zoomspec}
Eigenvalues of $H_{\lambda,[1,n]}$ for $\lambda=1$ and $n=\mbox{100,000}$,
drawn as vertical lines to aid visibility.  The first plot shows
the entire spectrum; the gray boxes denote the region on which the
next plot zooms.
}
\end{figure}

Figure~\ref{fig:ids_plot} shows estimates of the IDS based on computations
with $n=\mbox{10,000}$ for $\lambda$ values ranging from the trivial case
of no coupling ($\lambda=0$) to strong coupling ($\lambda=8$).
The fine structure of the spectrum is evident in Figure~\ref{fig:zoomspec},
which repeatedly zooms in upon subsets of the spectrum of the finite
section $H_{\lambda,[1,n]}$  for $\lambda=1$ and $n=\mbox{100,000}$.
(The numerical concerns discussed in the last paragraph do not
affect these figures.)

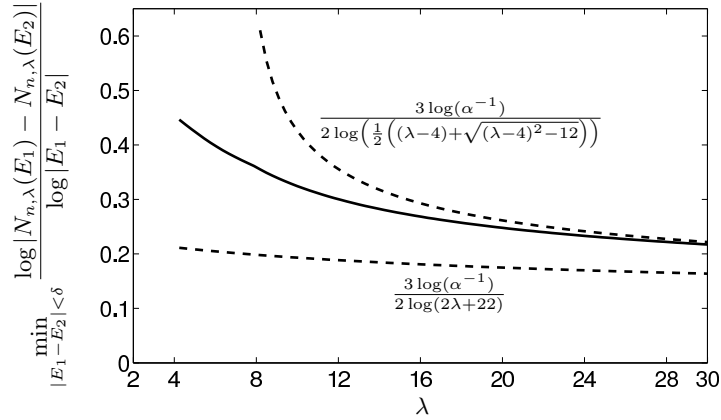
\begin{figure}
\begin{center}
\hspace*{3em}\includegraphics[scale=0.5]{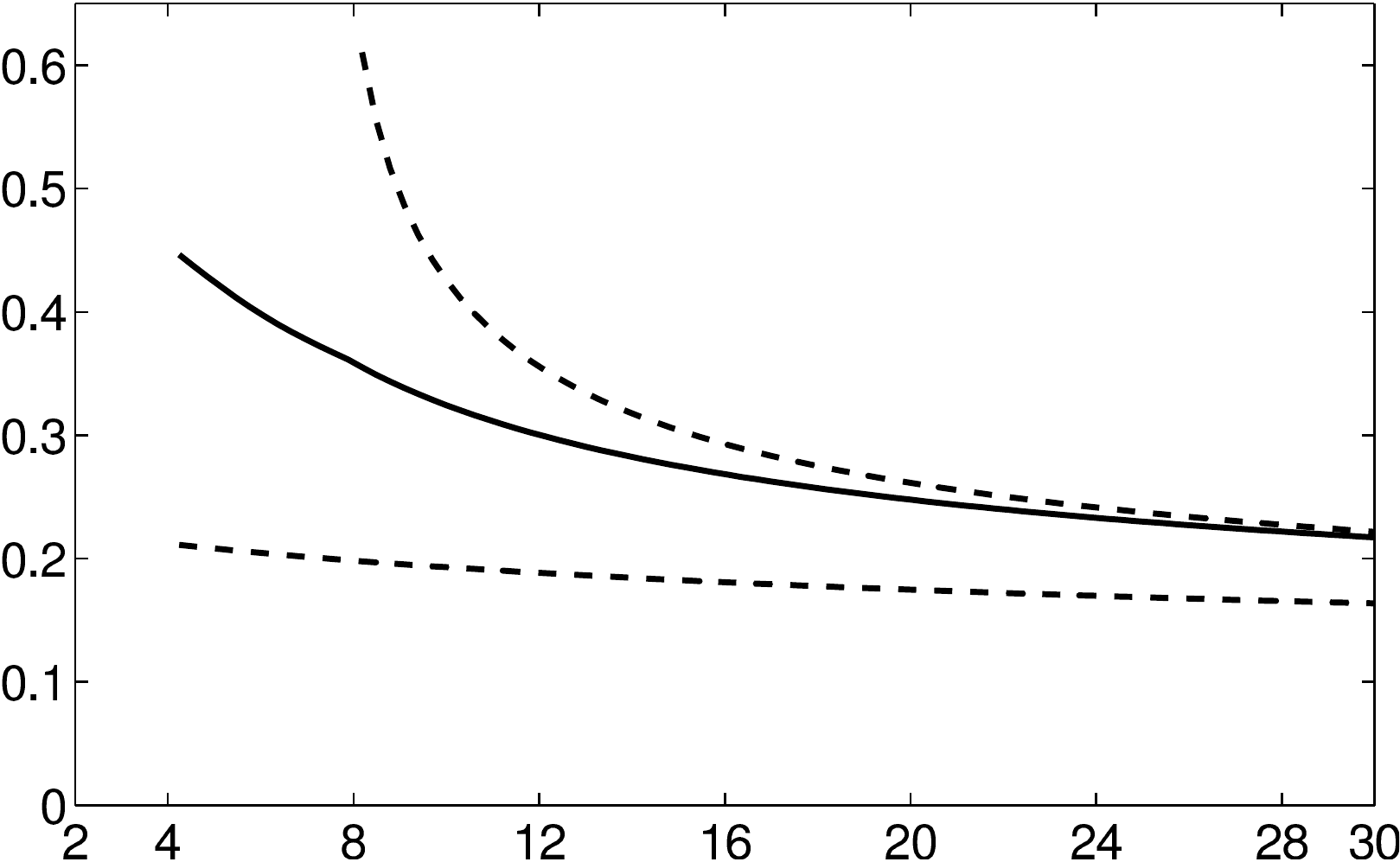}
\begin{picture}(0,0)
\put(-155,100){\footnotesize $\frac{3\log(\alpha^{-1})}{2\log \left(\frac{1}{2} \left( (\lambda - 4) + \sqrt{(\lambda - 4)^2 - 12} \right)\right)}$}
\put(-127,33){\small $\frac{3\log(\alpha^{-1})}{2\log(2\lambda + 22)}$}
\put(-118,-10){\small $\lambda$}
\put(-270,0){\small \rotatebox{90}{$\displaystyle{\min_{|E_1-E_2|<\delta} {\log|N_{n,\lambda}(E_1)-N_{n,\lambda}(E_2)| \over \log|E_1-E_2|}}$}}
\end{picture}
\end{center}

\vspace*{.25em}
\caption{\label{fig:ids_511}
Illustration of Theorem~\ref{main}, based on numerically-computed eigenvalues from
finite sections  $H_{\lambda,[1,n]}$ for $n=\mbox{10,000}$.  Here $\delta=0.025$ and
the minimization is over $E_1, E_2 \in \sigma(H_{\lambda,[1,n]})$.
}
\end{figure}

We now explore the H\"older continuity of the integrated density of states.
In consideration of~(\ref{eq:idsN}), define
\[ N_{n,\lambda}(E) = \lim_{n \to \infty} \frac{\#\{\text{eigenvalues of }
      H_{\lambda, [1,n]} \text{ that are} \le E \}}{n}.\]
Figure~\ref{fig:ids_511} investigates the large $\lambda$ behavior of
the H\"older exponent addressed in Theorem~\ref{main}, based on
computations with finite sections of dimension $n=\mbox{10,000}$.
Indeed, we see asymptotic behavior like ${3\log(\alpha^{-1})\over 2\log \lambda}$,
and moreover the figure suggests that the dimension of the measure is smooth
in this regime.

\subsection{Spectral Estimates for Square and Cubic Fibonacci Hamiltonians} \label{sec:2d3dcomp}

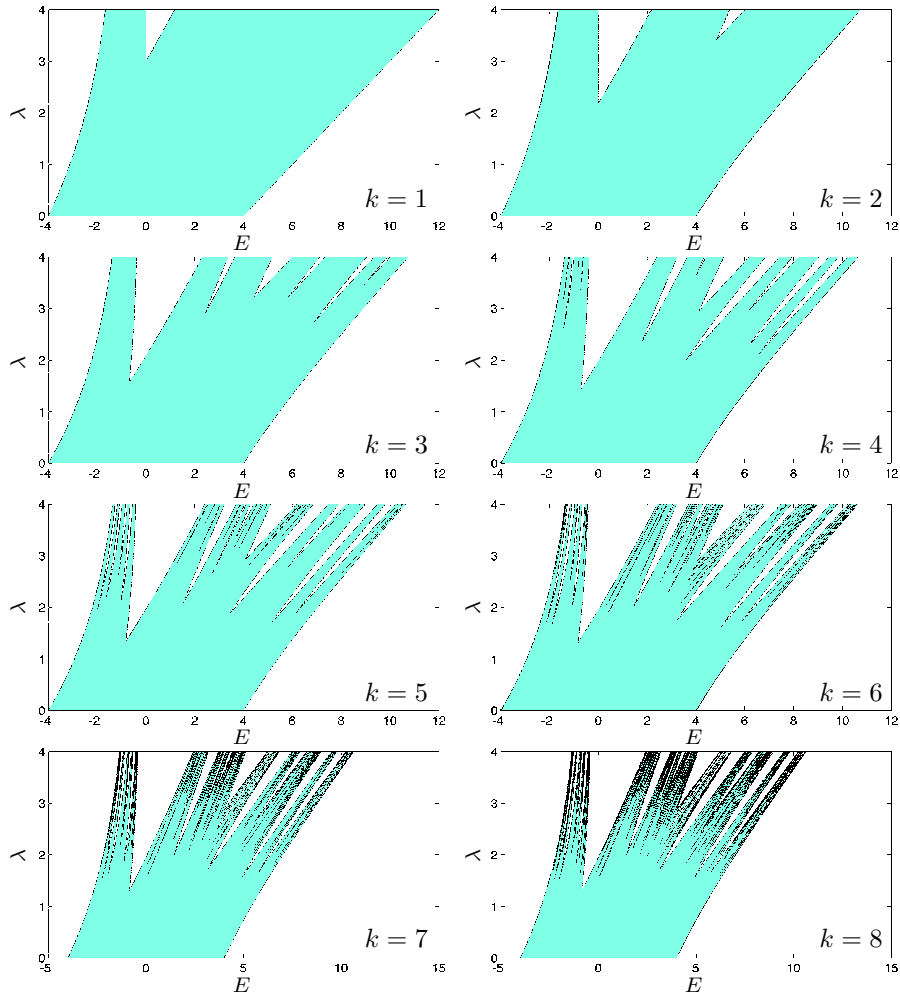
\begin{figure}
\begin{center}
   \includegraphics[scale=0.3]{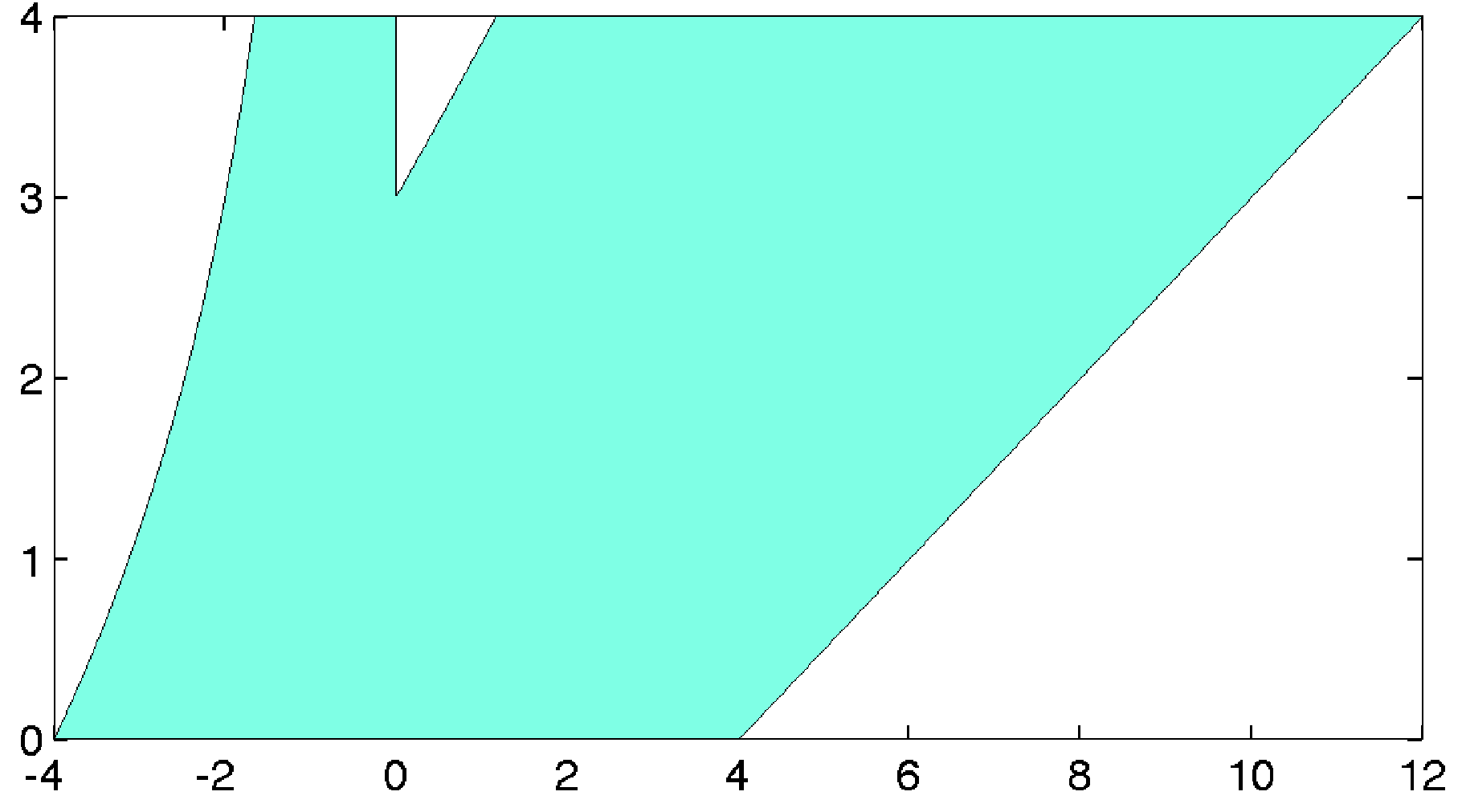}\hspace{10pt}
   \includegraphics[scale=0.3]{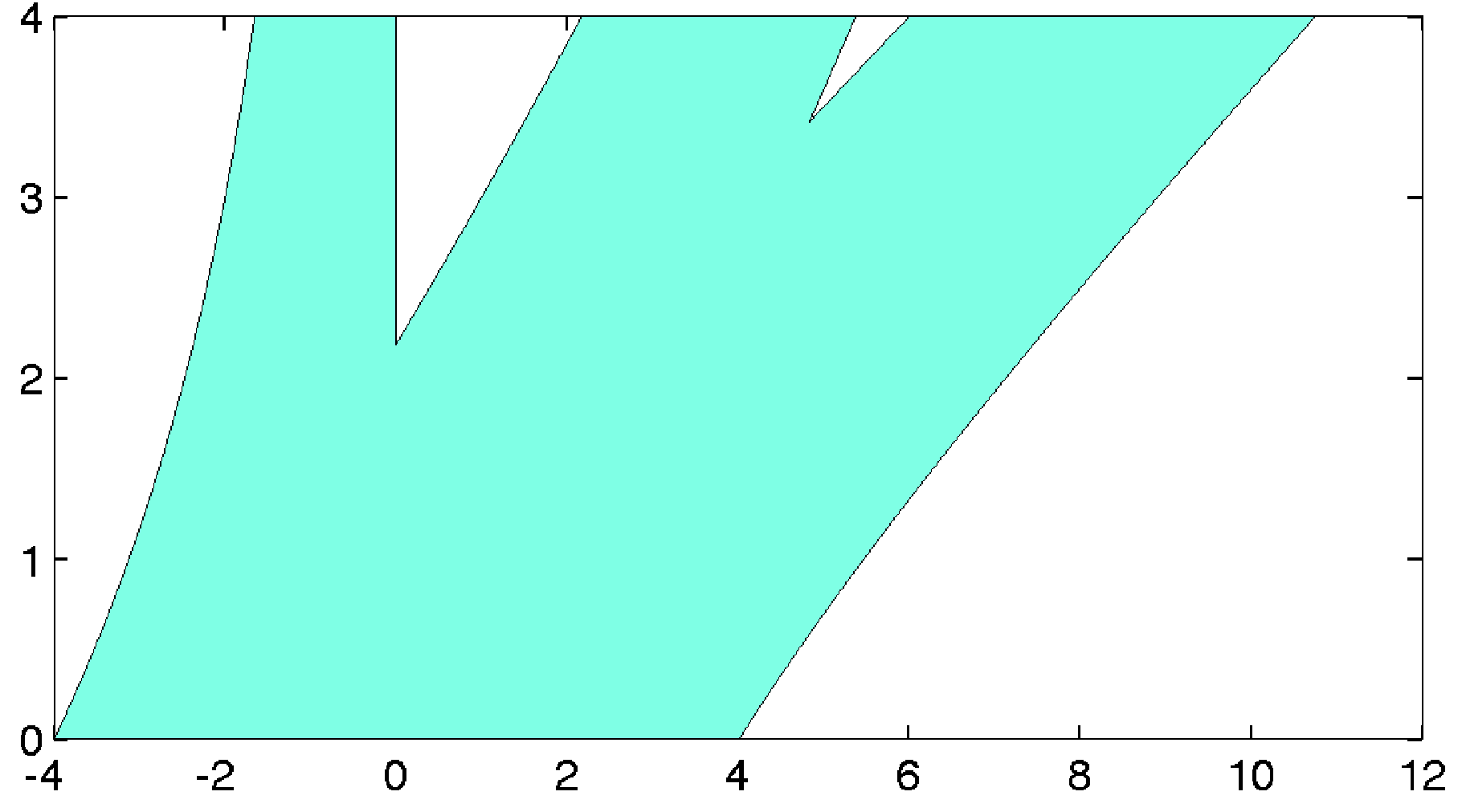}

\vspace*{.5em}
   \includegraphics[scale=0.3]{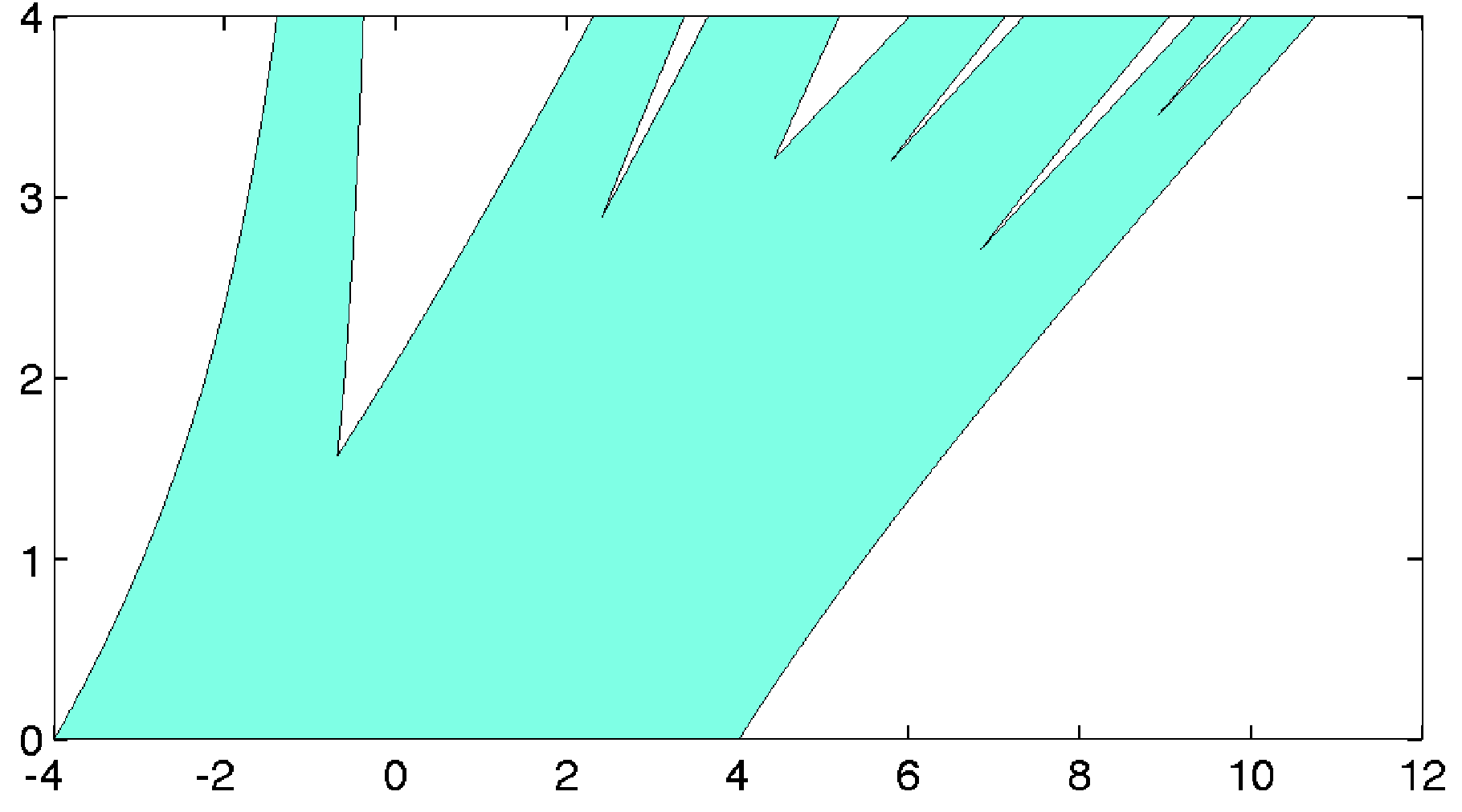}\hspace{10pt}
   \includegraphics[scale=0.3]{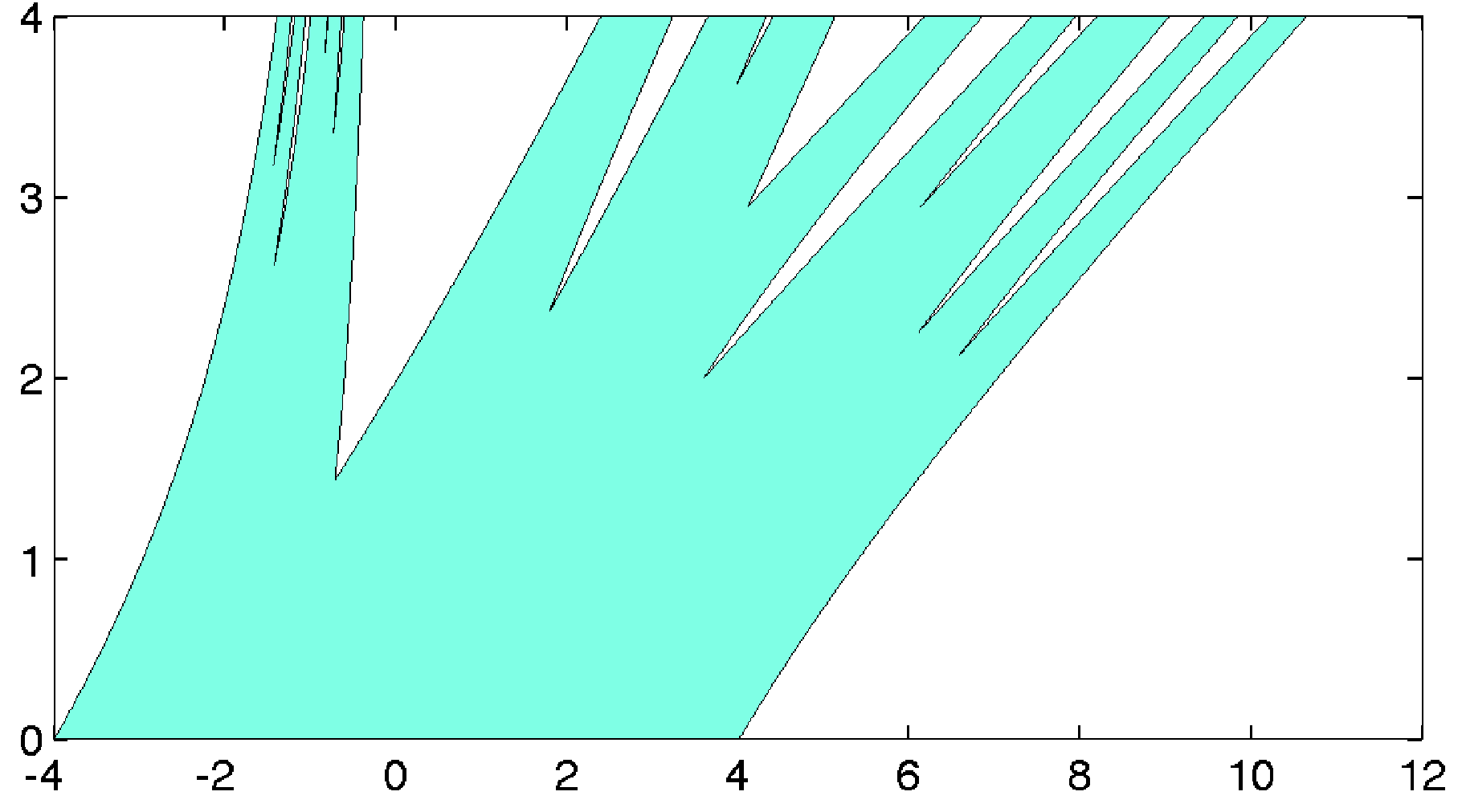}

\vspace*{.5em}
   \includegraphics[scale=0.3]{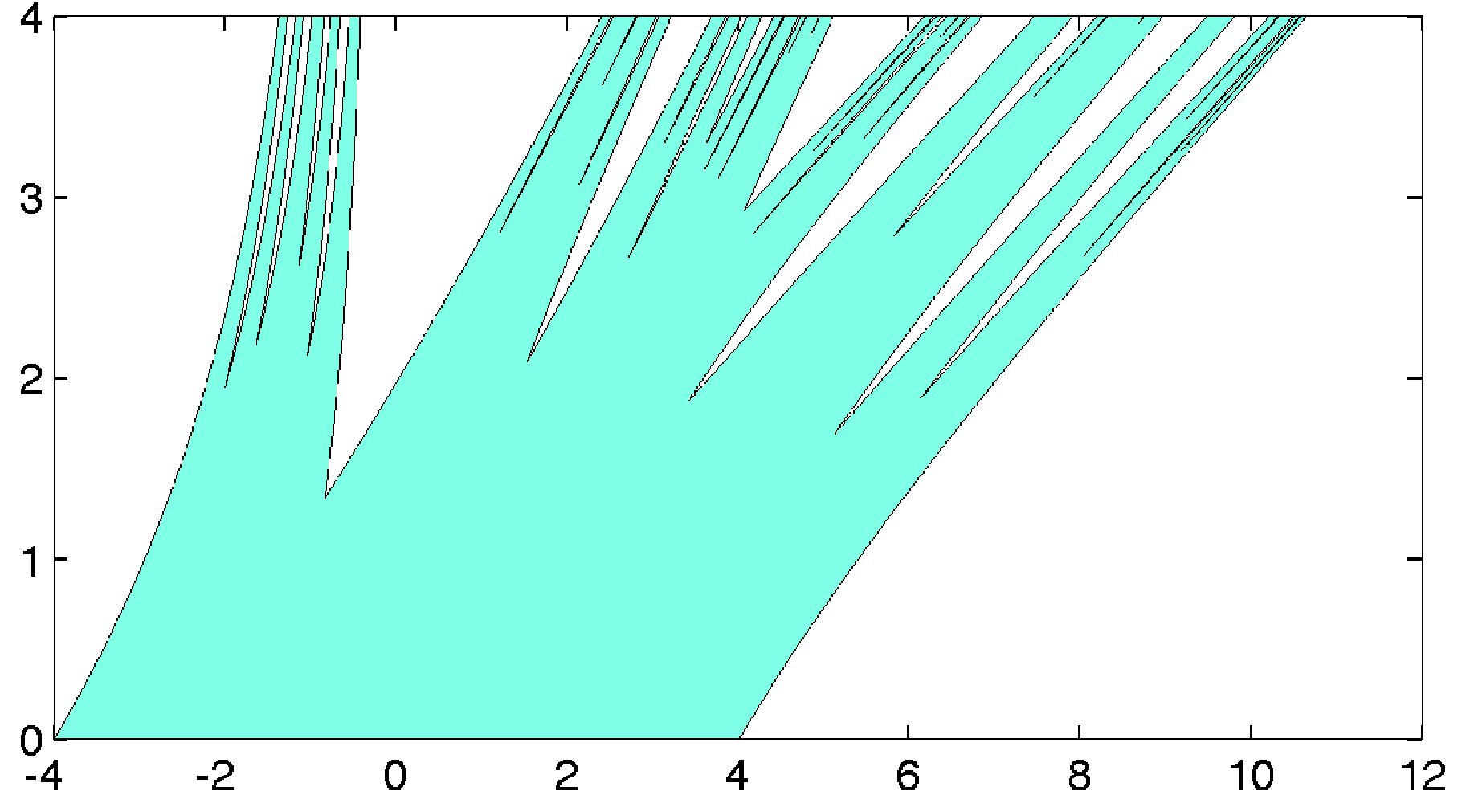}\hspace{10pt}
   \includegraphics[scale=0.3]{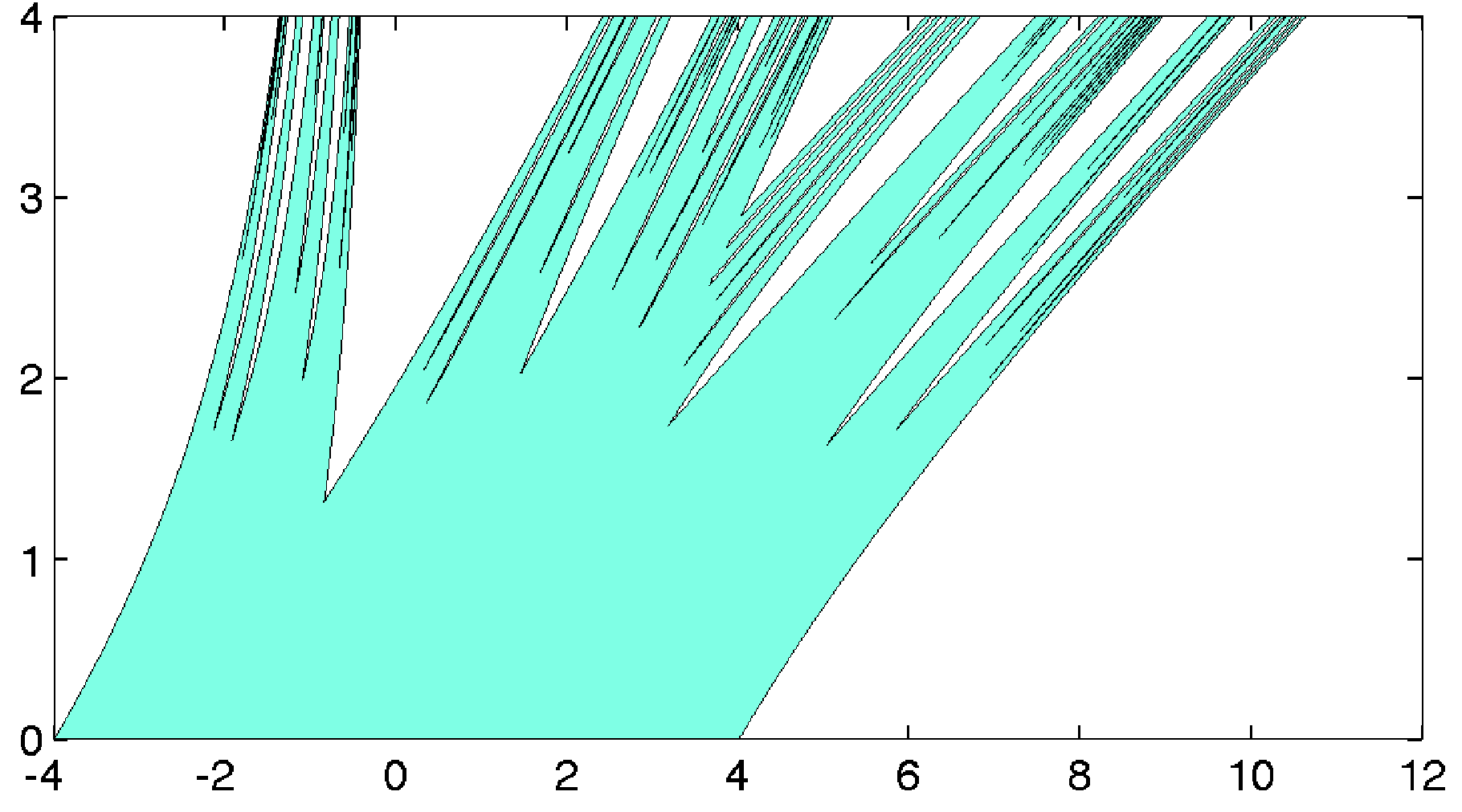}

\vspace*{.5em}
   \includegraphics[scale=0.3]{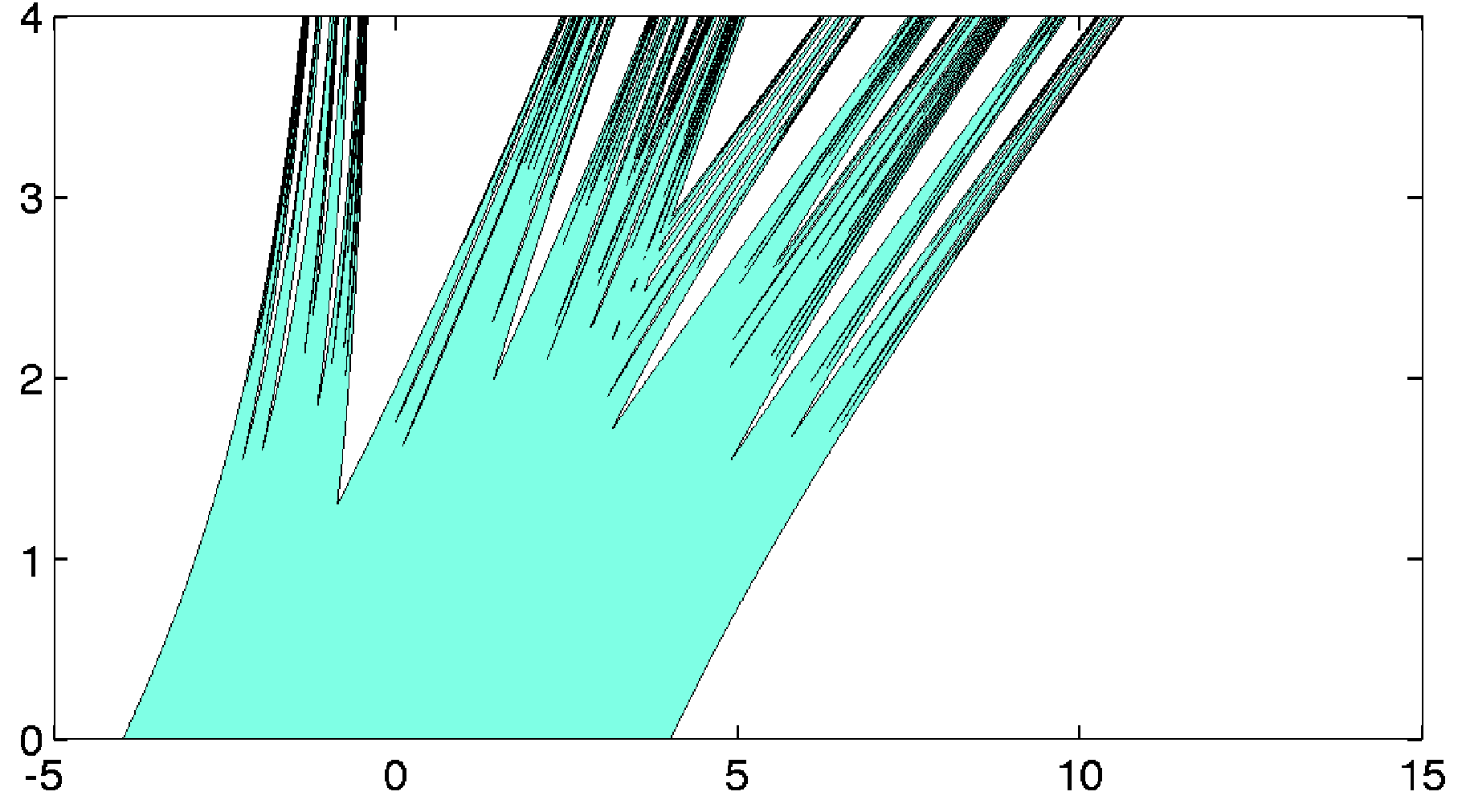}\hspace{10pt}
   \includegraphics[scale=0.3]{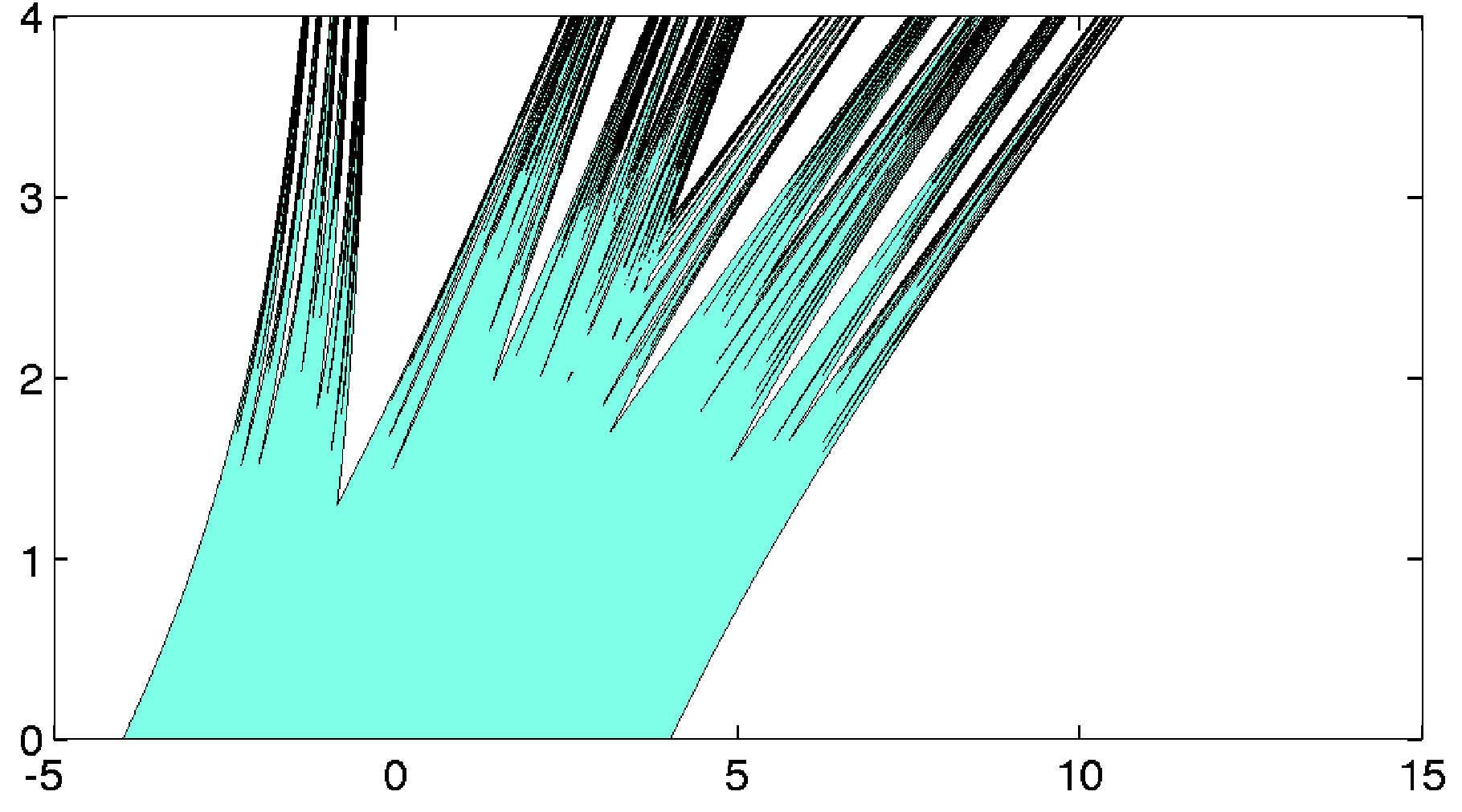}

\begin{picture}(0,0)
\put(-39,24){$k=7$} \put(133,24){$k=8$}
\put(-39,117){$k=5$} \put(133,117){$k=6$}
\put(-39,211){$k=3$} \put(133,211){$k=4$}
\put(-39,304){$k=1$} \put(133,304){$k=2$}
\put(-173,337){\rotatebox{90}{\footnotesize $\lambda$}}
\put(-173,243){\rotatebox{90}{\footnotesize $\lambda$}}
\put(-173,150){\rotatebox{90}{\footnotesize $\lambda$}}
\put(-173, 56){\rotatebox{90}{\footnotesize $\lambda$}}
\put(-1,337){\rotatebox{90}{\footnotesize $\lambda$}}
\put(-1,243){\rotatebox{90}{\footnotesize $\lambda$}}
\put(-1,150){\rotatebox{90}{\footnotesize $\lambda$}}
\put(-1, 56){\rotatebox{90}{\footnotesize $\lambda$}}
\put(-89,288){\footnotesize $E$} \put(82,288){\footnotesize $E$}
\put(-89,194){\footnotesize $E$} \put(82,194){\footnotesize $E$}
\put(-89,101){\footnotesize $E$} \put(82,101){\footnotesize $E$}
\put(-89,7){\footnotesize $E$} \put(82,7){\footnotesize $E$}
\end{picture}
\end{center}

\vspace*{-.5em}
\caption{\label{fig:2d}
Approximations $\Sigma_{k,\lambda}+\Sigma_{k,\lambda}$ of the spectrum of the square Fibonacci operator, as a function of $\lambda$. For $k=8$ and $\lambda=4$, $\Sigma_{k,\lambda}+\Sigma_{k,\lambda}$ is the union of 311 disjoint intervals.
}
\end{figure}

\begin{figure}
\begin{center}
   \includegraphics[scale=0.3]{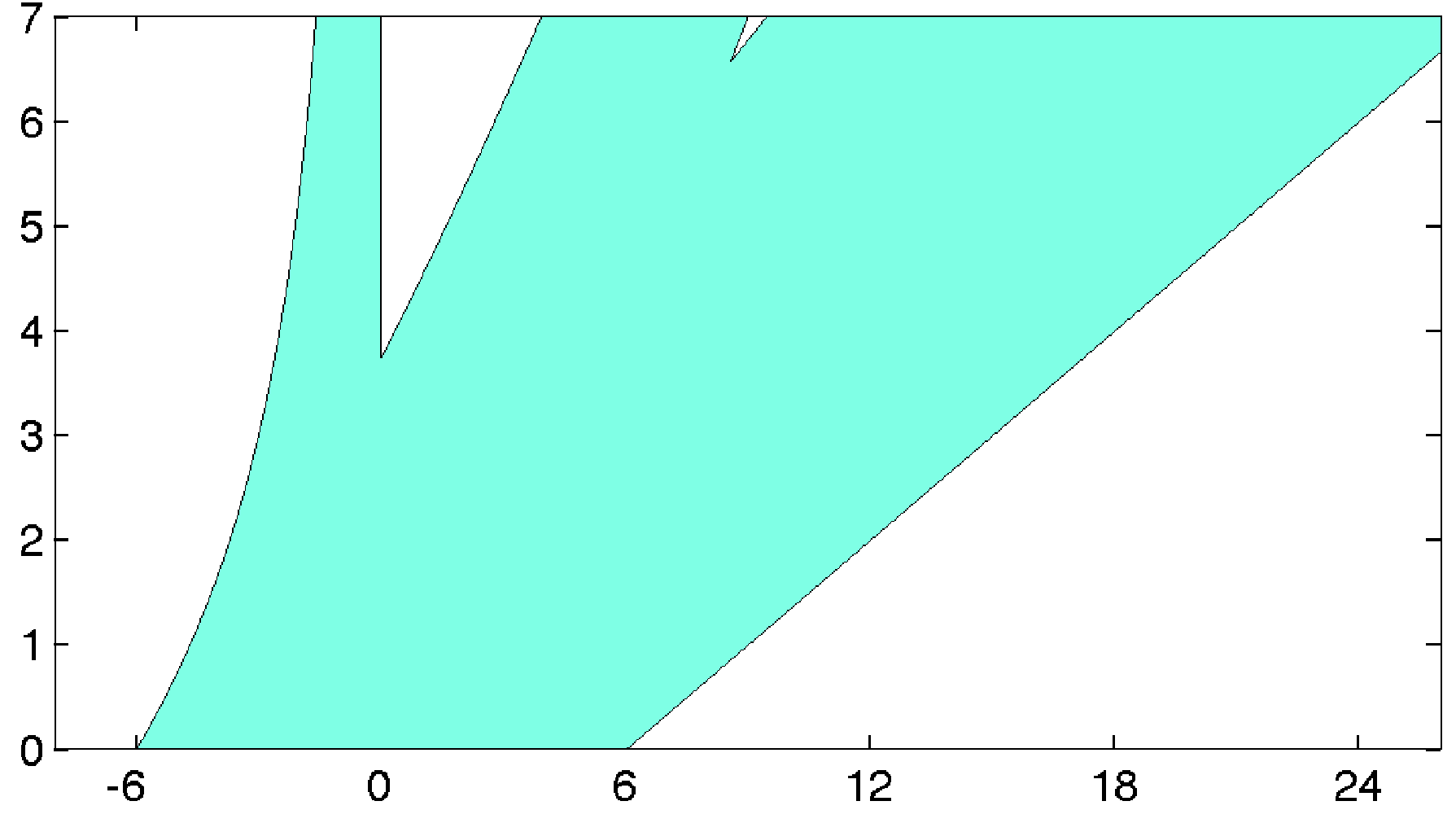}\hspace{10pt}
   \includegraphics[scale=0.3]{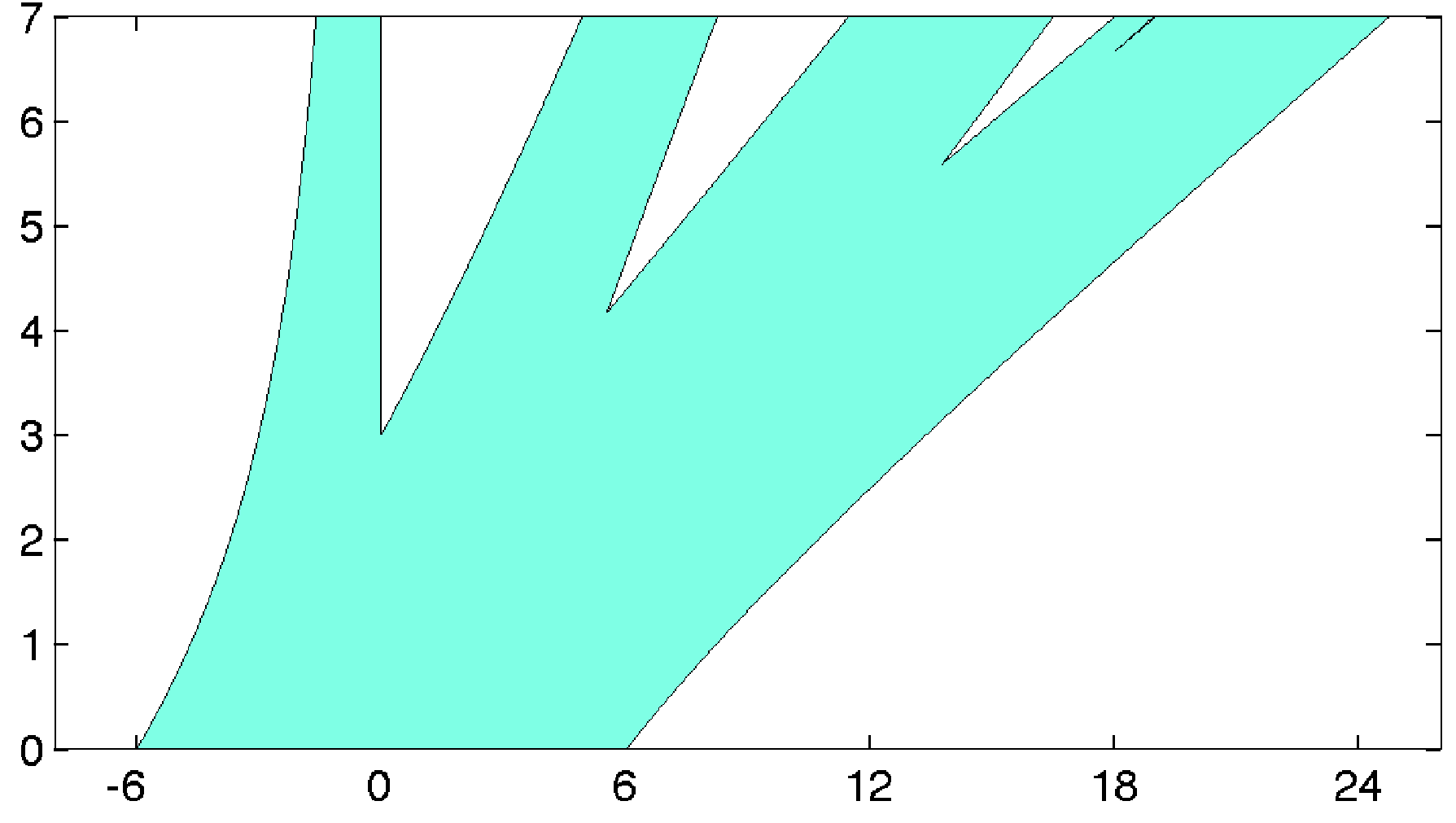}

\vspace*{.5em}
   \includegraphics[scale=0.3]{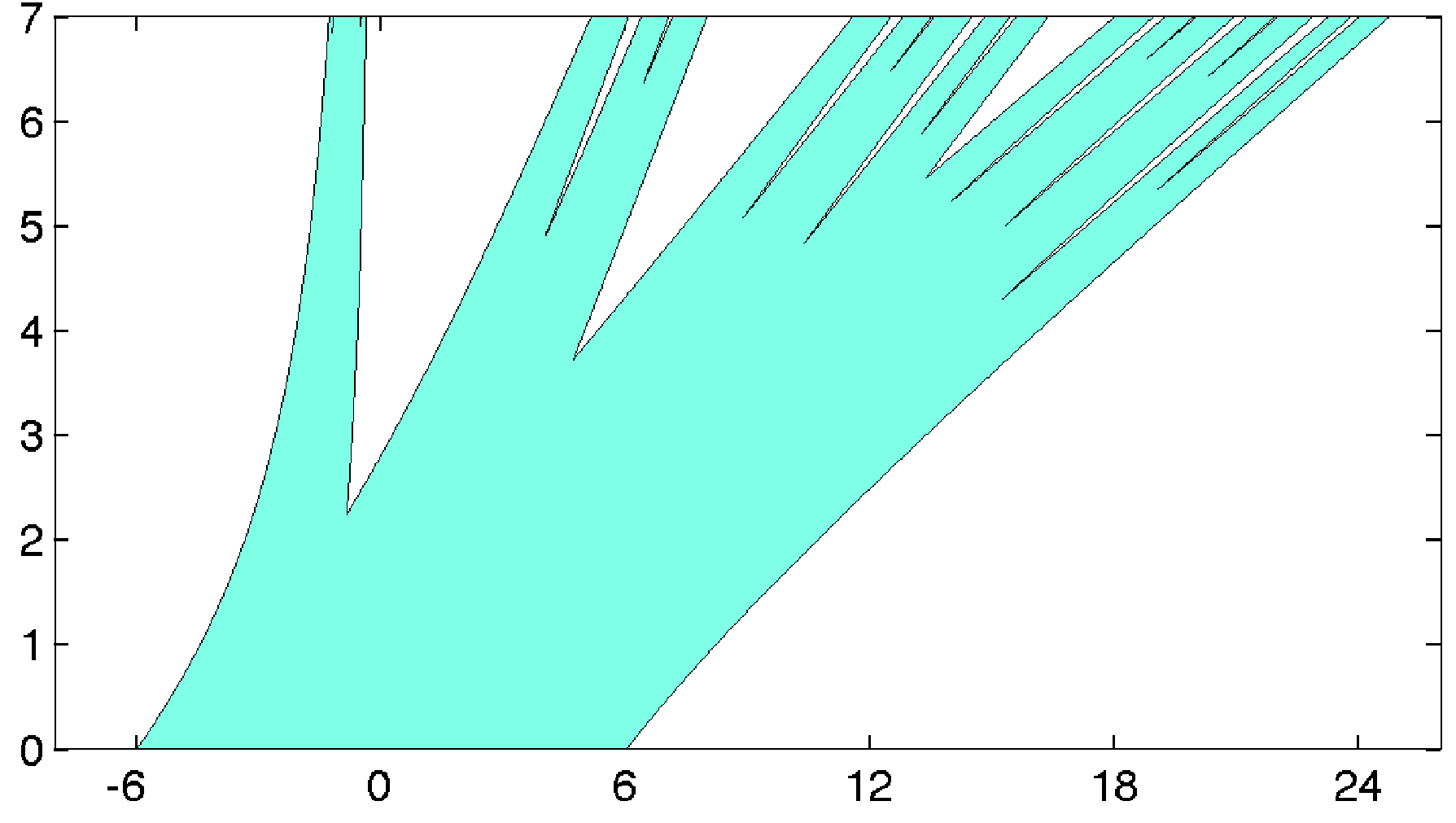}\hspace{10pt}
   \includegraphics[scale=0.3]{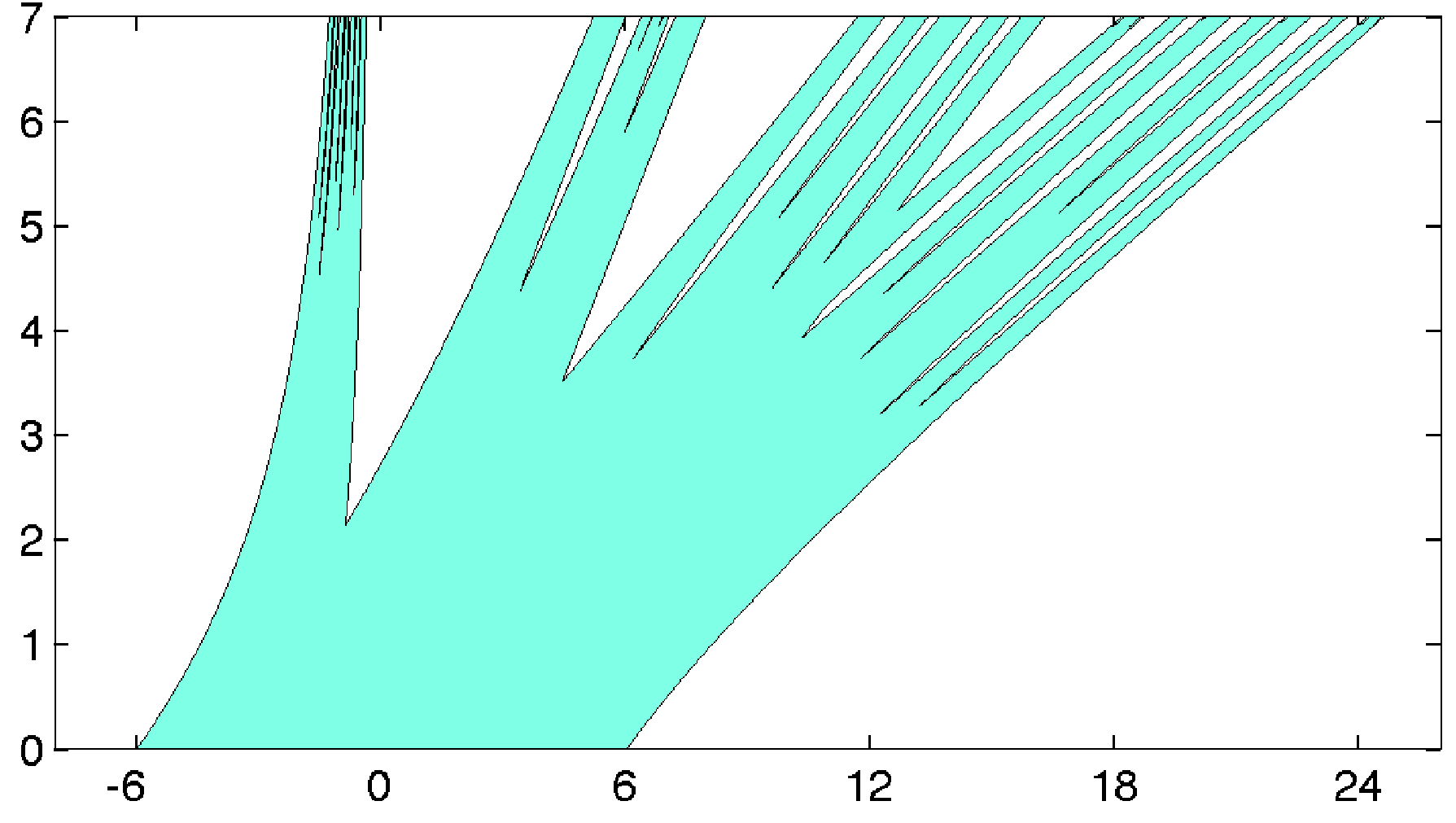}

\vspace*{.5em}
   \includegraphics[scale=0.3]{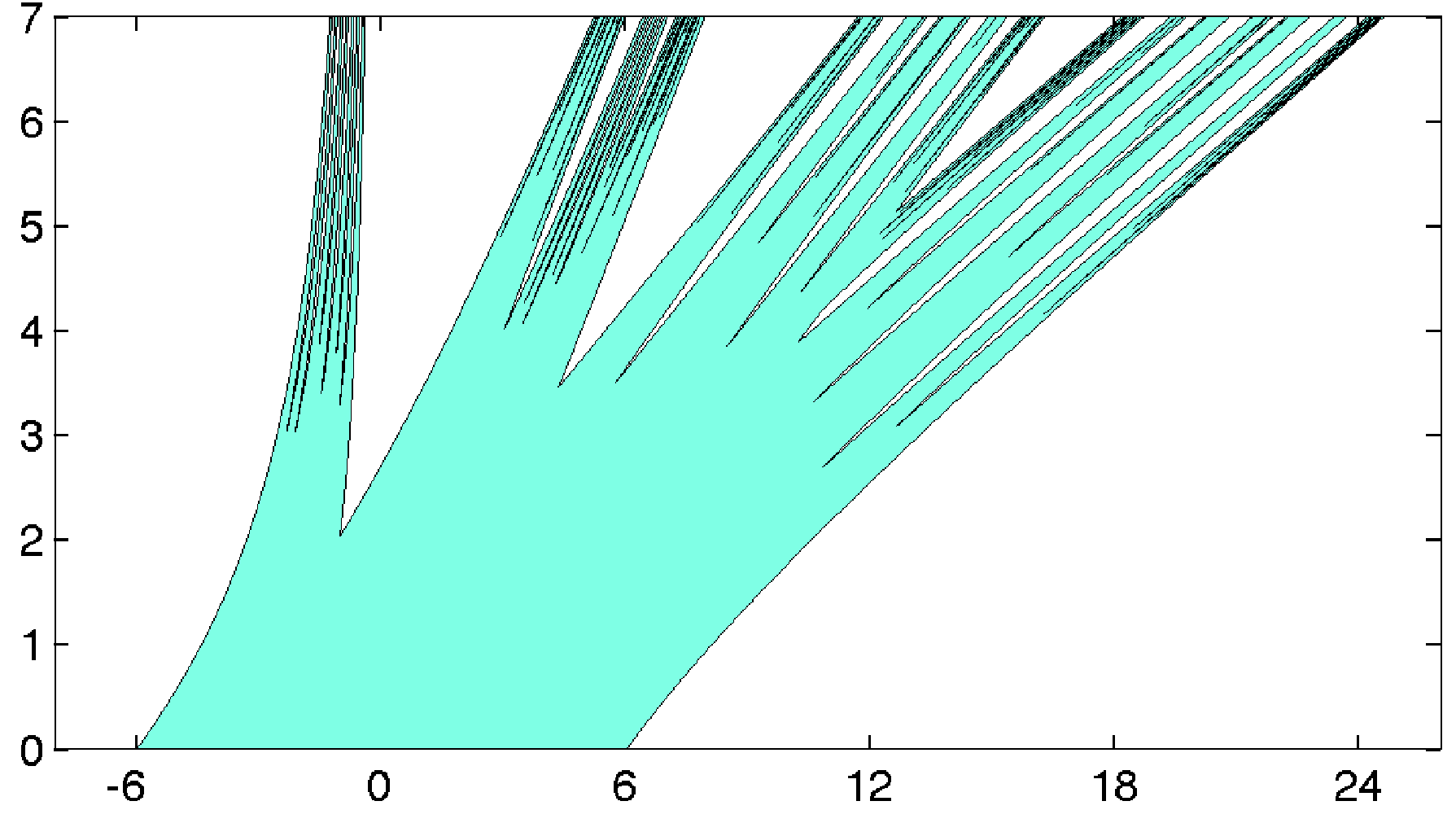}\hspace*{10pt}
   \includegraphics[scale=0.3]{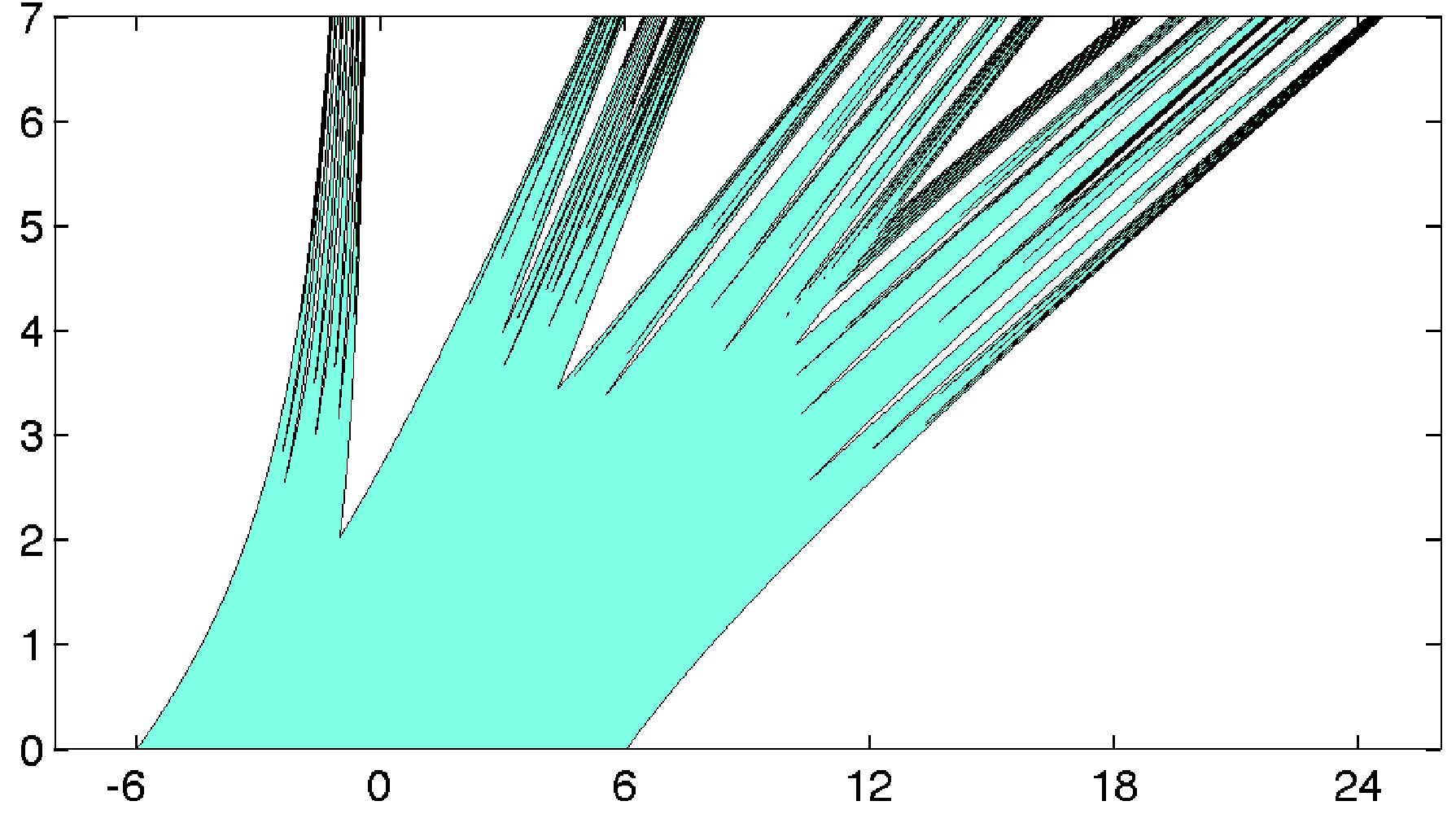}

\vspace*{.5em}
   \includegraphics[scale=0.3]{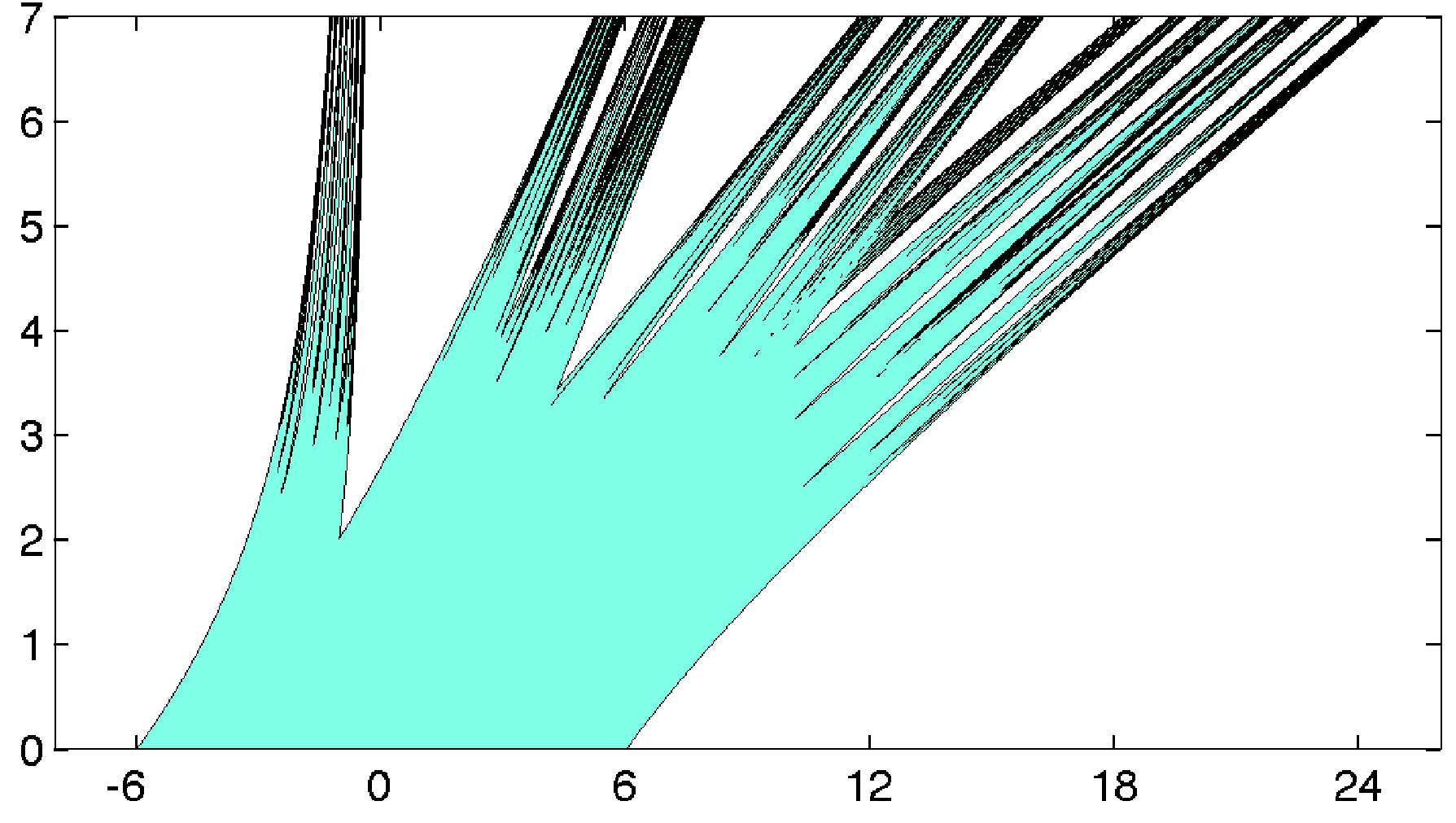}

\begin{picture}(0,0)
\put(49,24){$k=7$}
\put(-36,117){$k=5$} \put(133,117){$k=6$}
\put(-36,211){$k=3$} \put(133,211){$k=4$}
\put(-36,304){$k=1$} \put(133,304){$k=2$}
\put(-173,337){\rotatebox{90}{\footnotesize $\lambda$}}
\put(-173,243){\rotatebox{90}{\footnotesize $\lambda$}}
\put(-173,150){\rotatebox{90}{\footnotesize $\lambda$}}
\put(-1,337){\rotatebox{90}{\footnotesize $\lambda$}}
\put(-1,243){\rotatebox{90}{\footnotesize $\lambda$}}
\put(-1,150){\rotatebox{90}{\footnotesize $\lambda$}}
\put(-87, 56){\rotatebox{90}{\footnotesize $\lambda$}}
\put(-89,289){\footnotesize $E$} \put(82,289){\footnotesize $E$}
\put(-89,195){\footnotesize $E$} \put(82,195){\footnotesize $E$}
\put(-89,102){\footnotesize $E$} \put(82,102){\footnotesize $E$}
\put(-5,8){\footnotesize $E$}
\end{picture}
\end{center}

\vspace*{-.5em}
\caption{\label{fig:3d}
Approximations $\Sigma_{k,\lambda} + \Sigma_{k,\lambda} + \Sigma_{k,\lambda}$
of the spectrum of the cubic Fibonacci operator, as a function of $\lambda$.
For $k=7$ and $\lambda=7$, $\Sigma_{k,\lambda}+\Sigma_{k,\lambda}+\Sigma_{k,\lambda}$
is the union of 482 disjoint intervals.
}
\end{figure}
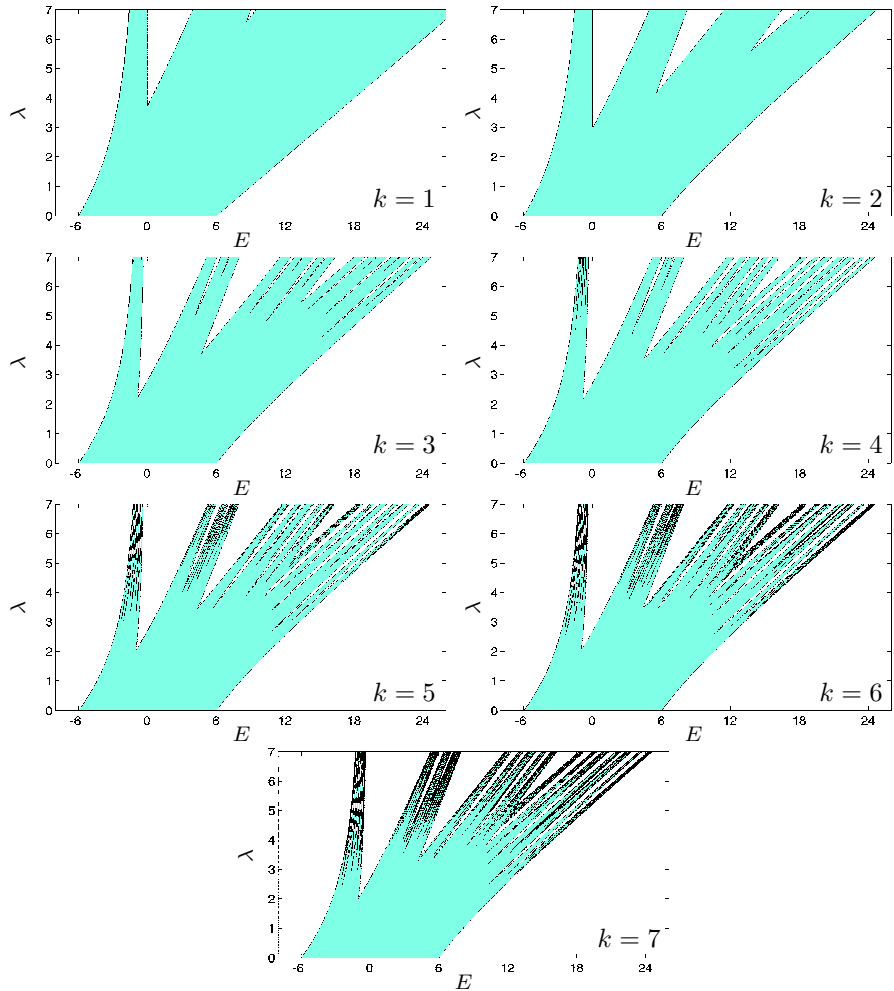

\begin{figure}
\begin{center}
   \includegraphics[scale=0.5]{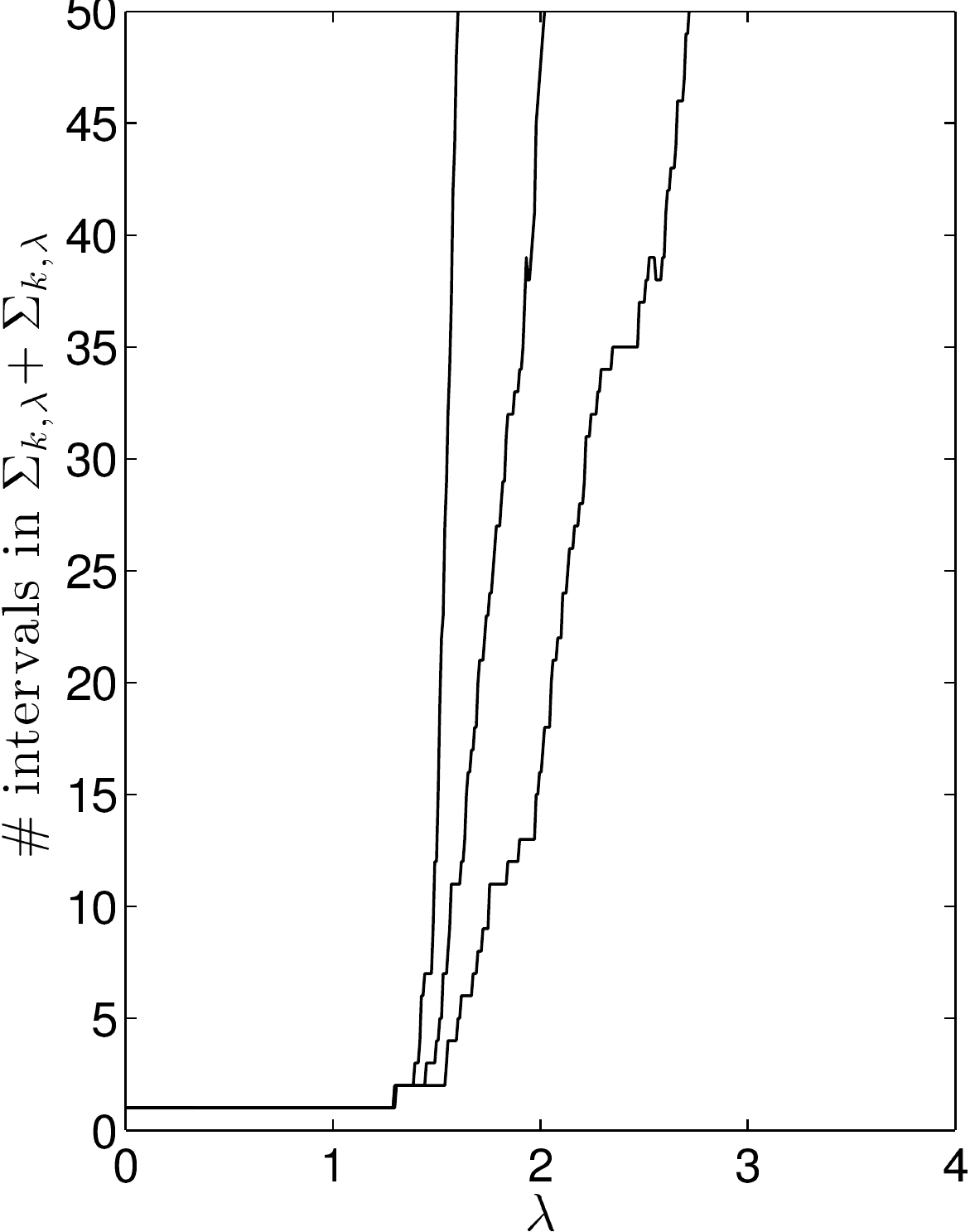}\quad
   \includegraphics[scale=0.5]{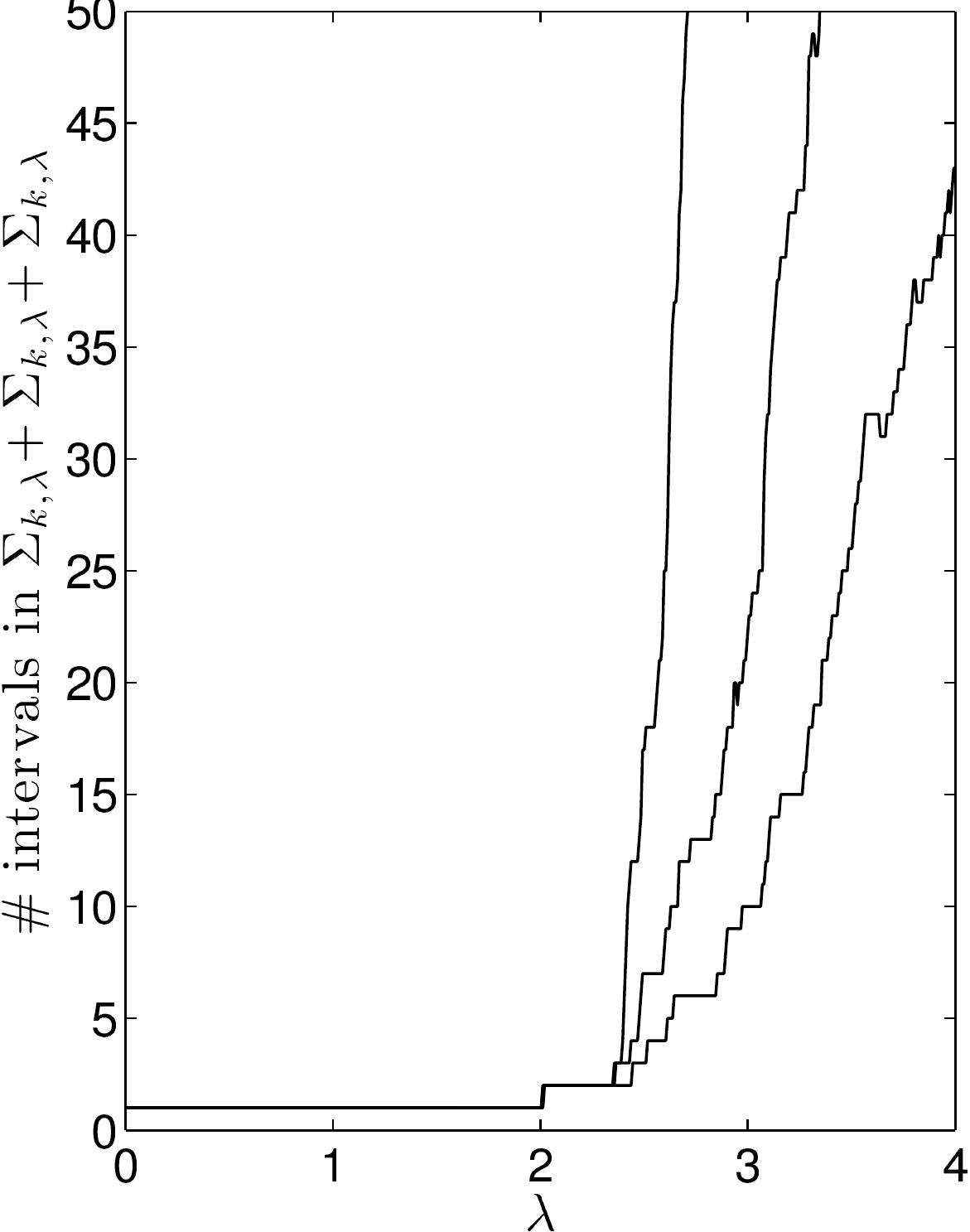}
\begin{picture}(0,0)
   \put(-281,156){\rotatebox{88}{\footnotesize{$k=13$}}}
   \put(-267.5,156){\rotatebox{88}{\footnotesize{$k=9$}}}
   \put(-248,156){\rotatebox{80}{\footnotesize{$k=7$}}}
   \put(-64,156){\rotatebox{88}{\footnotesize{$k=13$}}}
   \put(-45,156){\rotatebox{88}{\footnotesize{$k=9$}}}
   \put(-23,156){\rotatebox{80}{\footnotesize{$k=7$}}}
\end{picture}
\end{center}
\vspace*{-1em}
\caption{\label{fig:int_plot}
Number of intervals in the spectral approximations $\Sigma_{k,\lambda}+\Sigma_{k,\lambda}$
and $\Sigma_{k,\lambda}+\Sigma_{k,\lambda}+\Sigma_{k,\lambda}$, as a function of $\lambda$.
}
\end{figure}
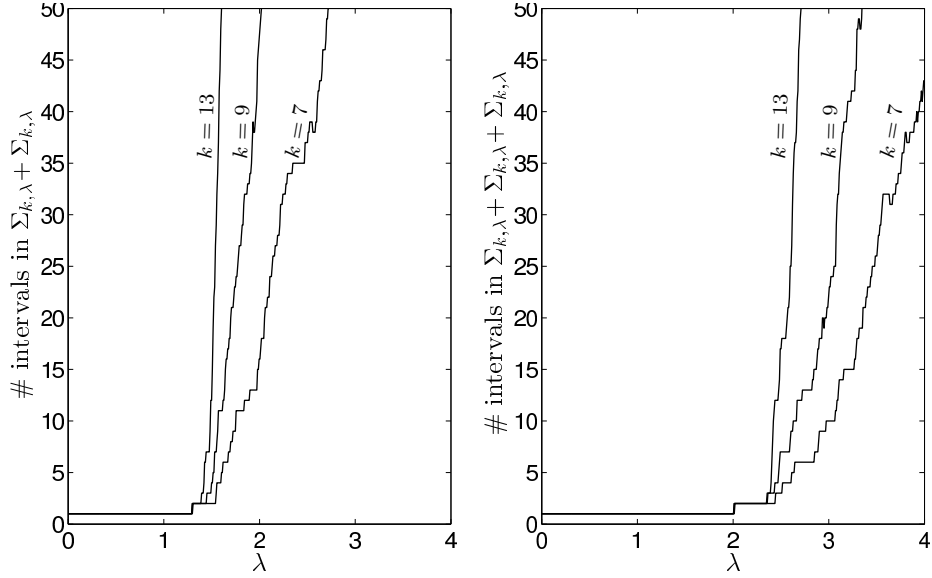

\begin{figure}
\begin{center}
   \hspace*{-3em}\includegraphics[scale=0.5]{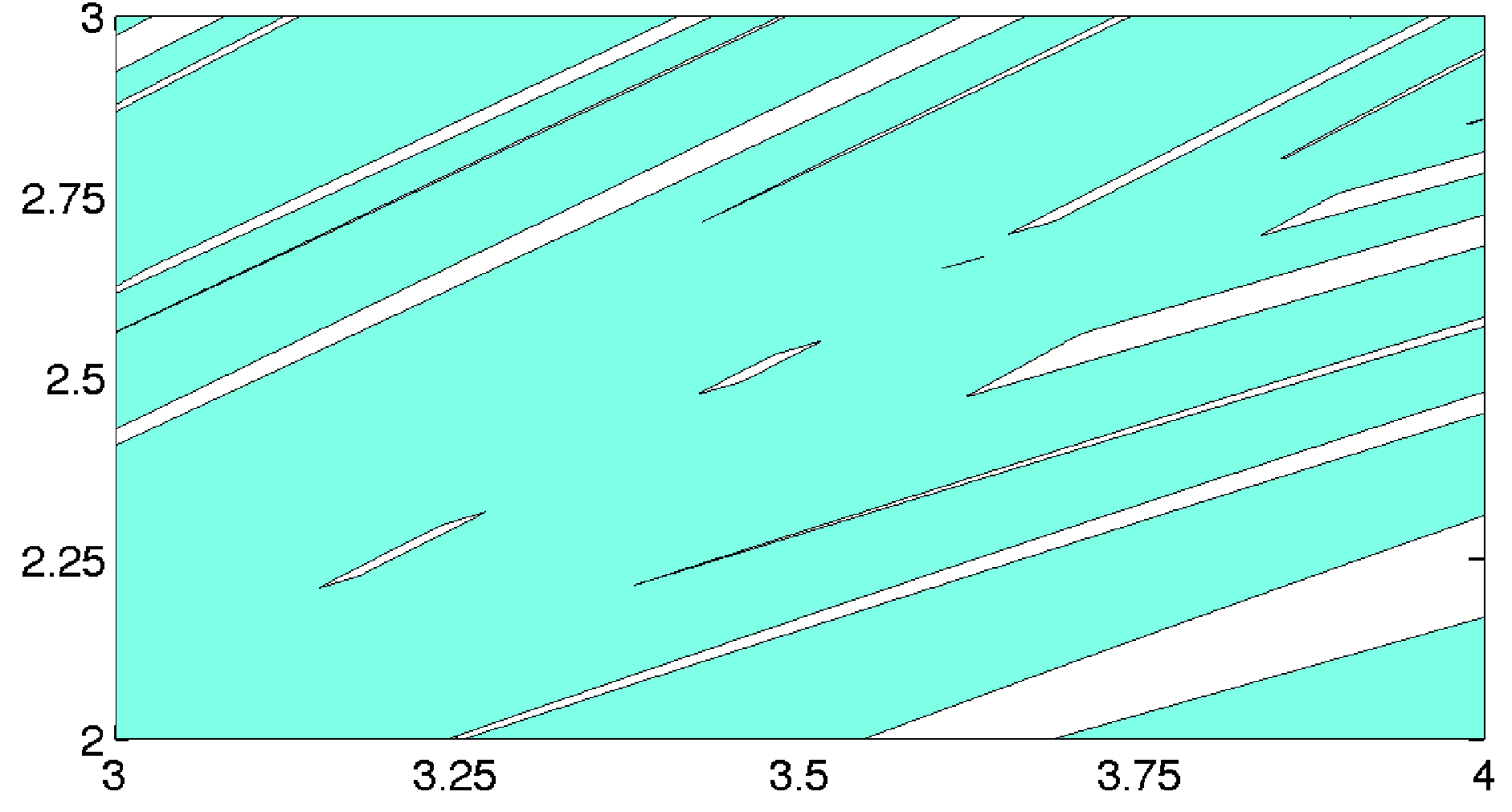}

\vspace*{1em}
   \hspace*{-3em}\includegraphics[scale=0.5]{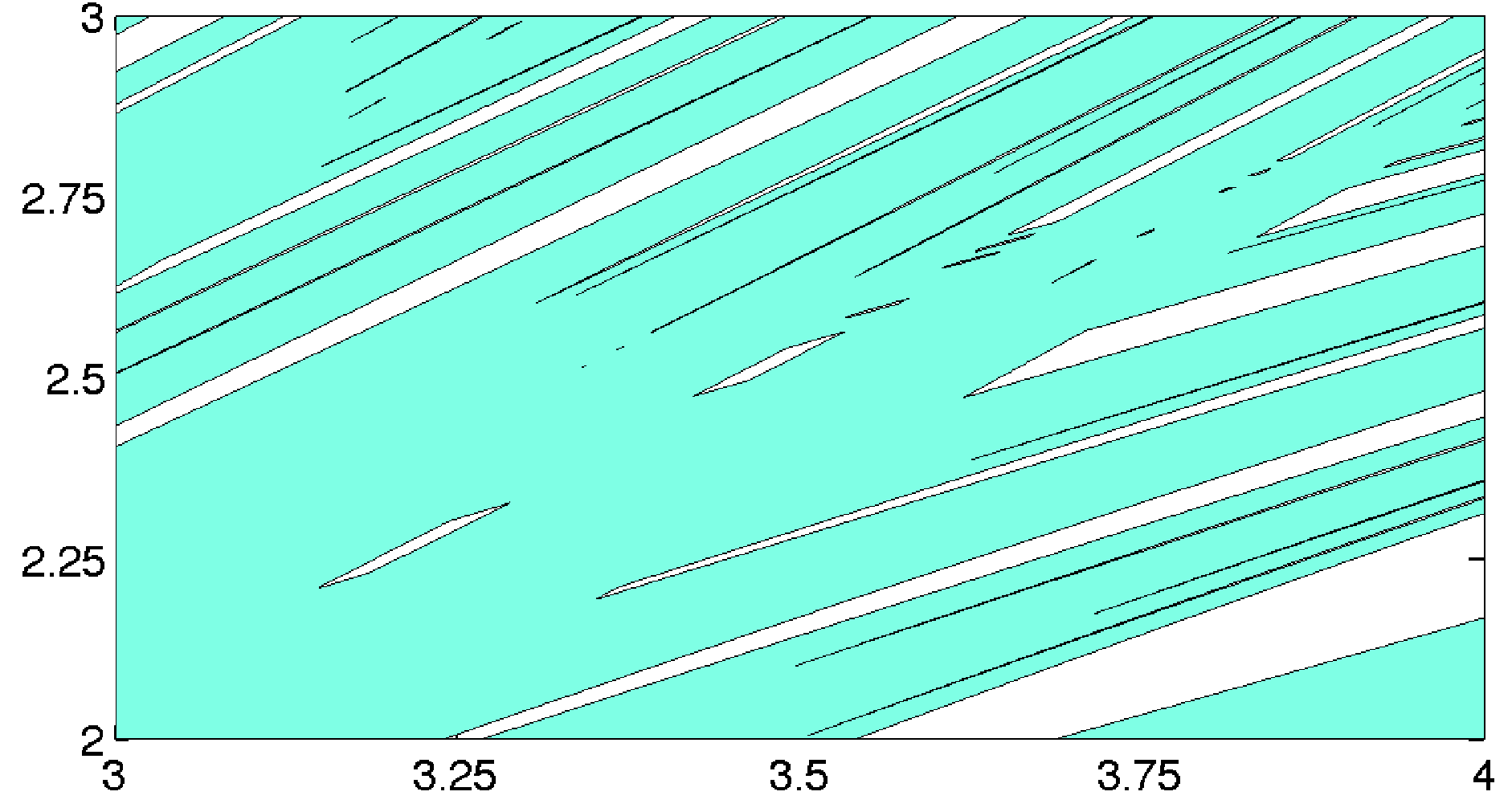}

\begin{picture}(0,0)
\put(126,188){$k=7$}
\put(126,29){$k=8$}
\put(-156,244){\rotatebox{90}{$\lambda$}}
\put(-156,87){\rotatebox{90}{$\lambda$}}
\put(82,7){$E$}
\put(82,164){$E$}
\end{picture}
\end{center}

\vspace*{-1.5em}
\caption{\label{fig:2dzoom}
Approximations $\Sigma_{k,\lambda}+\Sigma_{k,\lambda}$ of the square
Fibonacci spectrum $\Sigma_\lambda+\Sigma_\lambda$ as in Figure~\ref{fig:2d},
magnified to show the opening and closing of gaps as $\lambda$ increases.
How this structure affects $\Sigma_\lambda+\Sigma_\lambda$ is not presently understood.
}
\end{figure}

As described in Section~\ref{sec:2d3d}, the estimates $\Sigma_{k,\lambda}$ for the one-dimensional Fibonacci spectrum can readily be translated into approximations for the square and cubic cases, as investigated by Even-Dar Mandel and Lifshitz~\cite{EL06}. As described in Theorem~\ref{t.4}, $\Sigma_\lambda$ need not be a Cantor set, especially for small coupling constants.  This behavior is apparent in Figures~\ref{fig:2d} and~\ref{fig:3d}, which illustrate $\Sigma_{k,\lambda}+\Sigma_{k,\lambda}$ and $\Sigma_{k,\lambda} + \Sigma_{k,\lambda} + \Sigma_{k,\lambda}$ for various values of $k$ and $\lambda$. For a finite range of small $\lambda$ values, the spectra comprise intervals that branch into a greater number of intervals as $k$ and $\lambda$ increase. Figure~\ref{fig:int_plot} shows the growth in the number of intervals present in these approximations as a function of $\lambda$ for three different values of $k$. This plot makes evident rapid (but not always monotone) growth in the number of intervals with $\lambda$. Figure~\ref{fig:2dzoom} illustrates the opening and closing of gaps for the square problem, revealing an intriguing structure for finite $k$. How does this structure develop as $k$ increases, and, indeed, is it reflected in $\Sigma_\lambda+\Sigma_\lambda$? At present these questions remain open.

Tables~\ref{tbl:2d} and~\ref{tbl:3d} investigate the square and cubic
spectral estimates more precisely,
giving the values of $\lambda$ where multiple intervals first emerge.
These results confirm and sharpen the observation of Even-Dar Mandel and Lifshitz~\cite{EL06}
that $\Sigma_{k,\lambda}+\Sigma_{k,\lambda}$ transitions from one to two
intervals near $\lambda=1.3$, while $\Sigma_{k,\lambda}+\Sigma_{k,\lambda}+\Sigma_{k,\lambda}$
makes the same transition near $\lambda=2$.
For these finite values of $k$, it is apparent that $\Sigma_{k,\lambda}+\Sigma_{k,\lambda}$
and $\Sigma_{k,\lambda}+\Sigma_{k,\lambda} + \Sigma_{k,\lambda}$ both
transition to two intervals, then three intervals, and so on.
What do these calculations suggest about the limit $k\to\infty$?
For example, is the $\lambda$ value at which $\Sigma_{k,\lambda}+\Sigma_{k,\lambda}$
transitions from two to three intervals converging?
Is there a finite span of $\lambda$ values for which
$\Sigma_{k,\lambda}+\Sigma_{k,\lambda}$ persists as the union of two intervals
as $k\to\infty$, or does $\Sigma_\lambda + \Sigma_\lambda$ transition from one
interval directly to a Cantorval or Cantor set?
(See Problems~\ref{prob:cantorval} and~\ref{prob:2d} below.)

\begin{table}
\caption{\label{tbl:2d}
Estimates of $\lambda^*$, the $\lambda$ value for which the thickness of
$\Sigma_{k,\lambda}$ equals one, along with
$\lambda_{k,m}$, the coupling constant
where $\Sigma_{k,\lambda}+\Sigma_{k,\lambda}$
splits from $m$ to $m+1$ intervals, for $m=1,\ldots, 4$.}

\begin{center}
\begin{tabular}{r | r | r | r | r | r}
\multicolumn{1}{c|}{$k$} &
\multicolumn{1}{c|}{$\lambda_k^*$} &
\multicolumn{1}{c|}{$\lambda_{k,1}$} &
\multicolumn{1}{c|}{$\lambda_{k,2}$} &
\multicolumn{1}{c|}{$\lambda_{k,3}$} &
\multicolumn{1}{c}{$\lambda_{k,4}$} \\[.25em] \hline
 & & & & &  \\[-.75em]
 6  &  1.313172936 &  1.313172936 &    1.624865906 &    1.649775155 &    1.708521471 \\
 7  &  1.298964798 &  1.298964798 &    1.543759898 &    1.548912772 &    1.596682038 \\
 8  &  1.296218739 &  1.296218739 &    1.494856217 &    1.514291562 &    1.520122025 \\
 9  &  1.294303086 &  1.294303086 &    1.445808095 &    1.492410878 &    1.512965310 \\
 10 &  1.293935333 &  1.293935333 &    1.442778219 &    1.446787662 &    1.472813609 \\
 11 &  1.293679331 &  1.293679331 &    1.430901095 &    1.436192692 &    1.437915282 \\
 12 &  1.293630242 &  1.293630242 &    1.402035016 &    1.415460742 &    1.426586813 \\
 13 &  1.290031553 &  1.293596081 &    1.392730451 &    1.412863780 &    1.419815054 \\
 14 &  1.288819456 &  1.293589532 &    1.382510414 &    1.404399139 &    1.408704405 \\
 15 &  1.287431935 &  1.293584975 &    1.380466052 &    1.399646887 &    1.400190389 \\
 16 &  1.287269802 &  1.293584102 &    1.380121550 &    1.388518687 &    1.397593470 \\
 17 &  1.287084388 &  1.293583494 &    1.379851608 &    1.387310733 &    1.395556145 \\
 18 &  1.287062735 &  1.293583377 &    1.379806139 &    1.385835331 &    1.393702258 \\
 19 &  1.287037977 \\
 20 &  1.287035086
\end{tabular}
\end{center}
\end{table}

\begin{table}
\caption{\label{tbl:3d}
Estimates of $\lambda_{k,m}$, the coupling constant
where $\Sigma_{k,\lambda}+\Sigma_{k,\lambda}+\Sigma_{k,\lambda}$
splits from $m$ to $m+1$ intervals, for $m=1,\ldots, 4$.}
\begin{center}
\begin{tabular}{r | r | r | r | r}
\multicolumn{1}{c|}{$k$} &
\multicolumn{1}{c|}{$\lambda_{k,1}$} &
\multicolumn{1}{c|}{$\lambda_{k,2}$} &
\multicolumn{1}{c|}{$\lambda_{k,3}$} &
\multicolumn{1}{c}{$\lambda_{k,4}$} \\[.25em] \hline
 & & & &  \\[-.75em]
 6 &    2.025741216 &    2.544063632 &    2.573539294 &    2.842670115 \\
 7 &    2.012664501 &    2.438240772 &    2.511570744 &    2.606841186 \\
 8 &    2.011113604 &    2.376933028 &    2.498126298 &    2.498926850 \\
 9 &    2.009524869 &    2.364541039 &    2.435665993 &    2.473875055 \\
10 &    2.009337409 &    2.357357667 &    2.412613336 &    2.421115367 \\
11 &    2.009145619 &    2.355932060 &    2.399696274 &    2.408616763 \\
12 &    2.009123008 &    2.355107791 &    2.392573154 &    2.401253561 \\
13 &    2.009099880 &    2.354944739 &    2.391094663 &    2.397036745 \\
14 &    2.009097154 &    2.354850520 &    2.390282080 &    2.393347062 \\
15 &    2.009094365 &    2.354831891 &    2.390113912 &    2.393329303 \\
16 &    2.009094036 &    2.354821128 &    2.390021550 &    2.393302392 \\
17 &    2.009093700 &    2.354819000 &    2.390002443 &    2.393300376\\
\end{tabular}
\end{center}
\end{table}

\begin{figure}
\begin{center}
\includegraphics[scale=.355]{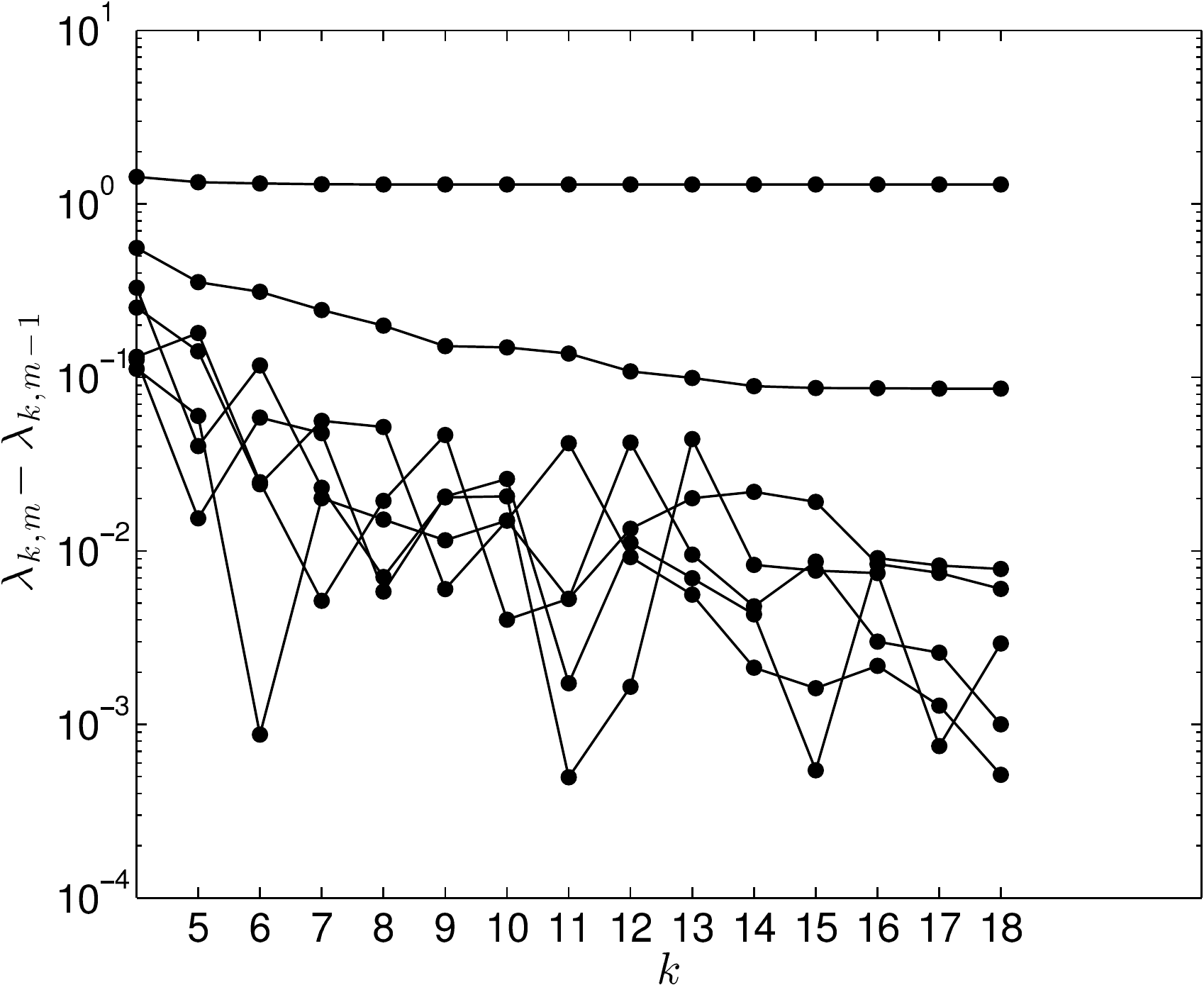} \hspace*{0em}
\includegraphics[scale=.355]{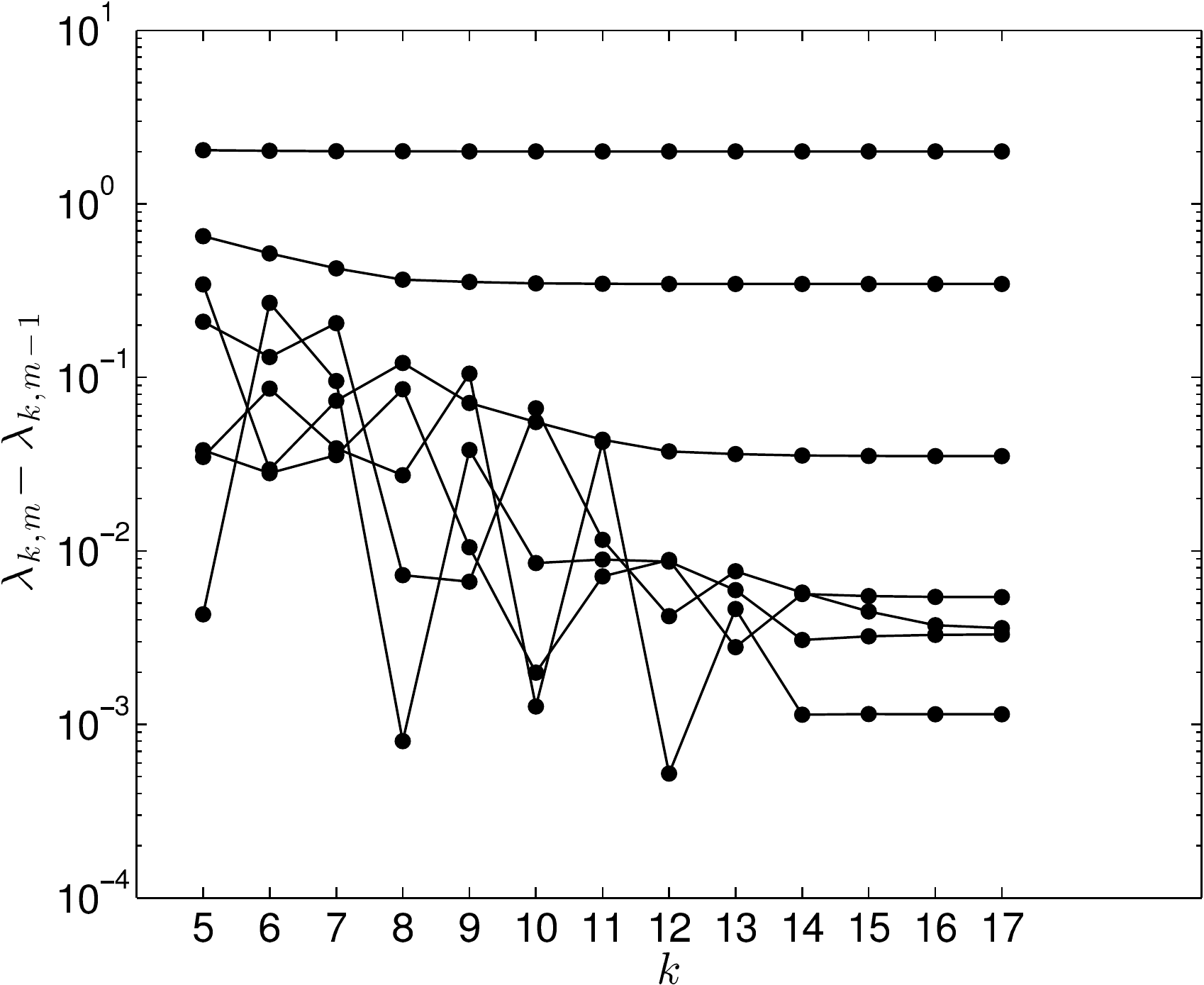}
\begin{picture}(0,0)
\put(-210.5,115){\footnotesize $m=1$}
\put(-210.5,85){\footnotesize $m=2$}
\put(-210.5,61){\footnotesize $m=4$}
\put(-210.5,55){\footnotesize $m=3$}
\put(-210.5,48){\footnotesize $m=7$}
\put(-210.5,36){\footnotesize $m=6$}
\put(-210.5,28){\footnotesize $m=5$}
\put(-29.5,119){\footnotesize $m=1$}
\put(-29.5,100){\footnotesize $m=2$}
\put(-29.5,75){\footnotesize $m=3$}
\put(-29.5,58){\footnotesize $m=5$}
\put(-29.5,51.5){\footnotesize $m=6$}
\put(-29.5,45){\footnotesize $m=4$}
\put(-29.5,37){\footnotesize $m=7$}
\end{picture}
\end{center}

\vspace*{-1em}
\caption{\label{fig:trans_plot3}
The span of $\lambda$ values (i.e., $\lambda_{k,m}-\lambda_{k,m-1}$) for which $\Sigma_{k,\lambda}+\Sigma_{k,\lambda}$ (left)
and $\Sigma_{k,\lambda}+\Sigma_{k,\lambda}+\Sigma_{k,\lambda}$ (right) comprise $m$ intervals for $m=1,\ldots, 7$.
}
\end{figure}
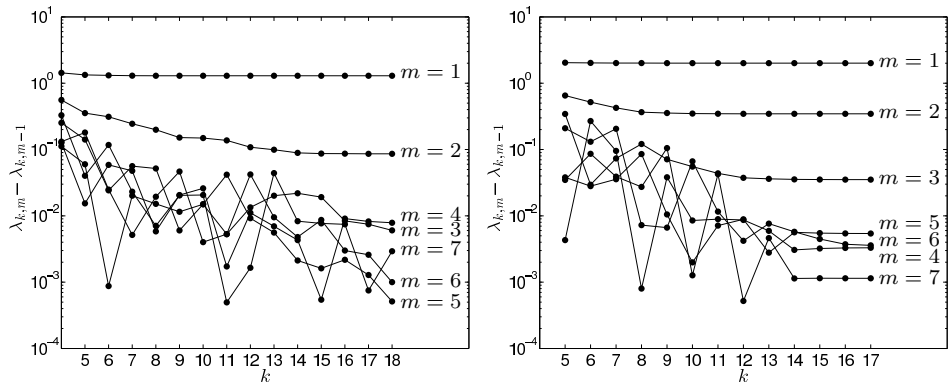

Let $\lambda_{k,m}$ denote the value of $\lambda$ at which $\Sigma_{k,\lambda}+\Sigma_{k,\lambda}$ (or $\Sigma_{k,\lambda}+\Sigma_{k,\lambda}+\Sigma_{k,\lambda}$) first splits from $m$ to $m+1$ intervals as $\lambda$ increases, with $\lambda_{k,0}=0$. (Our detailed computations suggest that, for small values of $m$, there is only one such point of transition; for larger numbers of intervals, gap closings complicate the picture, as seen in Figure~\ref{fig:2dzoom}.) Figure~\ref{fig:trans_plot3} plots $\lambda_{k,m}-\lambda_{k,m-1}$ as a function of $k$ for $m=1,\ldots, 7$ for the square and cubic Hamiltonians. Do the transition points converge as $k\to\infty$? First consider the plot on the left, for the square Hamiltonian. For $m=1$ and $m=2$, $\lambda_{k,m}$ appears to converge; however, the points of transition to $m\ge 3$ intervals do not show such consistency: it is unclear if these $\lambda_{k,m}$ values are converging. It may be that the coupling constants at which $\Sigma_{\lambda, k} + \Sigma_{\lambda,k}$ breaks into $m>3$ intervals are converging to the point at which the spectrum breaks into $m=3$ intervals as $k\to \infty$. Now consider the plot on the right of Figure~\ref{fig:trans_plot3}, for the cubic Hamiltonian. In contrast to the square case, these results suggest the $\lambda_{k,m}$ values converge to distinct points as $k\to\infty$ for all values $m=1,2,\ldots, 7$ shown, inviting the conjecture that there exist $\lambda$ values for which $\Sigma_\lambda+\Sigma_\lambda+\Sigma_\lambda$ is the union of $m$ disjoint intervals for all $m\ge 1$.

\section{Conjectures and Open Problems}\label{s.8}

In this final section we discuss various open problems that are suggested by the existing results and address generalizations, strengthenings, and related issues.

We begin with open problems for the Fibonacci Hamiltonian. The existing quantitative results concern estimates for dimensional properties of the spectrum, the density of states measure, and the spectral measures, as well as estimates for the transport exponents. In almost all cases, the asymptotic behavior is known in the regimes of small and large coupling. While the bounds we obtain are monotone, we would like to understand whether the quantities themselves have this property:

\begin{problem}
Are the various quantities we consider {\rm (}in particular, $\dim_H \Sigma_\lambda${\rm )} monotone in $\lambda$?
\end{problem}

The known estimates for the local scaling exponents and in particular the optimal H\"older exponent of the spectral measures (see \cite{DG2} and references therein) are clearly not optimal, and in particular do not identify their asymptotics in the extremal coupling regimes. For the density of states measure, which is an average of spectral measures, we have much better information. Can one find ways to find equally good estimates for spectral measures?

\begin{problem}
What can one say about the spectral measures? In particular, are their dimensional properties uniform across the hull and/or across the spectrum? Moreover, what are the asymptotics as $\lambda \downarrow 0$ and $\lambda \uparrow \infty$?
\end{problem}

We know that $\dim \Sigma_\lambda$ goes to one as $\lambda$ goes to zero. In addition, we would be interested in the following:

\begin{problem}
Does the right-derivative of $\dim \Sigma_\lambda$ exist at zero?
\end{problem}

If it does, due to Theorem \ref{t.1} it must be finite and non-zero.

\bigskip

Let us now turn to the higher-dimensional separable analogs of the Fibonacci Hamiltonian (e.g., the square or cubic Fibonacci Hamiltonian). Recall that the spectrum of such an operator is given by the sum of the one-dimensional spectra, which in turn are Cantor sets. Recall also that at sufficiently small coupling, these sum sets are intervals, while at sufficiently large coupling, they are Cantor sets as well. Concretely, this uses that if the thickness of a Cantor set $C$ is larger than 1, then $C+C$ is an interval by Theorem~\ref{t.sumofcantorsets} and, on the other hand, if the upper box counting dimension of $C$ is strictly less than $1/2$, then $C+C$ is a Cantor set. It is natural to ask what shape the higher-dimensional spectra have at intermediate coupling, that is, we wish to study how the transition from $C + C$ being an interval to being a Cantor set happens when the thickness of $C$ decreases.

\bdef
A compact set $C\subset \mathbb{R}^1$ is a {\rm Cantorval} if it has a dense interior {\rm (}i.e., $\overline{int(C)} = C${\rm )}, it has a continuum of connected components, and none of them is isolated.
\endef

Here is a general result on the occurrence of Cantorvals in the context of taking sums of Cantor sets \cite{MMR}:

\bthm\label{t.moreira}
There is an open set $\mathcal{U}$ in the space of dynamically defined Cantor sets such that for generic $C_1, C_2\in \mathcal{U}$, the sum $C_1+C_2$ is a Cantorval.
\ethm

Unfortunately, this result does not provide any specific and verifiable genericity conditions that would allow one to check that the sum of two given specific Cantor sets is indeed a Cantorval. Thus, for our purpose we need a solution to the following problem.

\begin{problem} \label{prob:cantorval}
Provide specific verifiable conditions on a Cantor set $C$ which imply that the sum $C+C$ is a Cantorval.
\end{problem}

Ideally, such a criterion would be applicable to the spectrum of the Fibonacci Hamiltonian and establish that, say, the spectrum of the square Fibonacci Hamiltonian is a Cantorval for intermediate values of the coupling constant $\lambda$. The next step would then be to study the transitions between the three regimes. We ask whether there are two sharp transitions; compare \cite{EL07} for closely related numerical evidence and discussion.

\begin{problem} \label{prob:2d}
Let $H^{(2)}_\lambda$ be the separable square Fibonacci Hamiltonian. Prove that there are values $0 < \lambda' < \lambda'' < \infty$ such that for $\lambda \in (0,\lambda')$, the spectrum $\sigma(H_\lambda)$ is an interval {\rm (}or a finite union of intervals{\rm )}, for $\lambda \in (\lambda',\lambda'')$, it is a Cantorval, and for $\lambda \in (\lambda'',\infty)$, it is a Cantor set.
\end{problem}

Notice that this will provide an example of a ({\it topologically!}\kern1pt) new structure of the spectrum for ``natural'' potentials.

\bigskip

Moving on from the Fibonacci case, which has a description via a substitution rule as well as via a simple quasi-periodic expression, there are two natural choices of a more general setting.

For a different choice of the underlying substitution rule, one always has an associated trace map. However, our understanding of the dynamics of such a trace map is in general far more limited than the one in the Fibonacci case. As a consequence, outside of the Fibonacci case there is a scarcity of quantitative results for dimensional issues (such as the dimension of the spectrum, the dimension of the density of states measure, or the dimension of the spectral measures). For example, here is a simple open problem that is currently completely out of reach:

\begin{problem}
Study other trace maps {\rm (}e.g., period doubling and Thue-Morse{\rm )}; in particular, find the asymptotics of  the Hausdorff dimension of the spectrum as the coupling constant tends to zero and infinity. 
\end{problem}

The other natural generalization of the Fibonacci potential is to replace the golden ratio in its quasi-periodic description by a general irrational number. Thus, we discuss some open problems for Sturmian potentials next.

Let us say that two Cantor sets $C_1$ and $C_2$ on $\mathbb{R}^1$ are {\it diffeomorphic} if there are neighborhoods $U_1(C_1)$, $U_2(C_2)$, and a diffeomorphism $f:U_1\to U_2$ such that $f(C_1)=C_2$.

\begin{problem}
Suppose that $\alpha=[a_1, a_2, \ldots]$ and $\beta=[b_1, b_2, \ldots]$ are such that for some $k \in \mathbb{Z}$ and all large enough $i \in \Z_+$ we have $b_{i+k}=a_i$. Prove that in this case, the Sturmian spectra $\Sigma_{\lambda,\alpha}$ and $\Sigma_{\lambda,\beta}$ are diffeomorphic.
\end{problem}

Notice also that due to the ergodicity of the Gauss map, a solution of this problem would also imply that the following long standing conjecture is correct:

\begin{problem}
For any fixed $\lambda > 0$, the dimension $\dim_H \Sigma_{\lambda,\alpha}$ is almost everywhere constant in $\alpha$.
\end{problem}

Finally, let us emphasize that most of the questions related to  higher dimensional models (described in Section \ref{s.3}) are completely open. So we formulate an extremely general problem:

\begin{problem}
Study spectral properties {\rm (}e.g., the shape of the spectrum and the type of the spectral measures{\rm )} and transport properties of higher dimensional operators; for example, study these questions for the particular case of the Laplacian on the graph associated with a Penrose tiling.
\end{problem}

\end{document}